\documentclass[longauth]{aa} 
\usepackage{graphicx}
\usepackage{txfonts}
\usepackage{amsmath,amsfonts,amssymb} 
\usepackage{multicol,caption}
\usepackage{lipsum}
\usepackage{booktabs}
\usepackage{threeparttable}
\usepackage{multirow}
\usepackage{here}
\usepackage{lscape}

\bibpunct{(}{)}{;}{a}{}{,}
\usepackage[colorlinks=true,urlcolor=blue,linkcolor=blue,citecolor=blue]{hyperref}

\newcommand{\TESS}{\emph{TESS}}

\usepackage{ragged2e}

\usepackage{tikz,xcolor,hyperref}
\definecolor{lime}{HTML}{A6CE39}
\DeclareRobustCommand{\orcidicon}{%
	\hspace{-1.5mm}
	\begin{tikzpicture}
	\draw[lime, fill=lime] (0,0) 
	circle [radius=0.16] 
	node[white] {{\fontfamily{qag}\selectfont \tiny ID}};
	\draw[white, fill=white] (-0.0625,0.095) 
	circle [radius=0.007];
	\end{tikzpicture}
	\hspace{-2.5mm}
}
\foreach \x in {A, ..., Z}{%
	\expandafter\xdef\csname orcid\x\endcsname{\noexpand\href{https://orcid.org/\csname orcidauthor\x\endcsname}{\noexpand\orcidicon}}
}

\foreach \x in {a, ..., z}{%
	\expandafter\xdef\csname orcid\x\endcsname{\noexpand\href{https://orcid.org/\csname orcidauthor\x\endcsname}{\noexpand\orcidicon}}
}

\begin{document}

\title{Three short-period Earth-sized planets around M dwarfs discovered by TESS: TOI-5720\,b, TOI-6008\,b, and TOI-6086\,b}
		
\author{
K. Barkaoui\orcidA{}\inst{\ref{astro_liege},\ref{MIT},\ref{IAC_Laguna}} \thanks{E-mail: \color{blue}khalid.barkaoui@uliege.be}
 \and R.~P.~Schwarz\orcidB{}\inst{\ref{Harvard_USA}} 
 \and N.~Narita\orcidG{}\inst{\ref{Univ_tokyo},\ref{Astro_tokyo},\ref{IAC_Laguna}} 
  \and P.~Mistry\orcidf{}\inst{\ref{National_Ins_Tech_India}}      
 \and C.~Magliano\orcidm{}\inst{\ref{Dip_Fisica_Ita},\ref{INFN_Ita},\ref{INAF_Ita}} 
 \and T.~Hirano\orcidQ{}\inst{\ref{Astro_tokyo},\ref{National_Astro_Obs_Tokyo}} 
 \and M.~Maity\orcidg{}\inst{\ref{Presi_univ_India}}      
 \and A.J.~Burgasser\orcidH{}\inst{\ref{ucsd}}    
 \and B.V.~Rackham\orcidN{}\inst{\ref{MIT},\ref{Dep_Kavli}}  
 \and F.~Murgas\inst{\ref{IAC_Laguna},\ref{Univ_LaLaguna}}    
 \and F.J.~Pozuelos\orcidD{}\inst{\ref{iaa}}   
 \and K.~G.~Stassun\orcido{}\inst{\ref{uvander}}    
 \and M.E.~Everett\orcida{}\inst{\ref{Opt_Inf_Ast_USA}}   
 \and D.R.~Ciardi\orcidj{}\inst{\ref{NASA_Exo_Sc_Ins_USA}} 
 \and C.~Lamman\orcidr\inst{\ref{Harvard_USA}}    
\and E.~K.~Pass\orcidt{}\inst{\ref{Harvard_USA}}  
\and A.~Bieryla\orcidu{}\inst{\ref{Harvard_USA}}
 \and C.~Aganze\orcidL{}\inst{\ref{ucsd2},\ref{stanford}}    
 \and E.~Esparza-Borges\orcide{}\inst{\ref{IAC_Laguna}}  
 \and K.A.~Collins\orcidE{}\inst{\ref{Harvard_USA}} 
 \and G.~Covone\orcidn{}\inst{\ref{Dip_Fisica_Ita},\ref{INFN_Ita},\ref{INAF_Ita}} 
 \and J.~de~Leon\orcidS{}\inst{\ref{Multi_tokyo}}  
 \and M.~D\'evora-Pajares\orcidl{}\inst{\ref{ugr}}  
 \and J.~de~Wit\inst{\ref{MIT}}      
 \and Izuru Fukuda\orcidY{}\inst{\ref{Multi_tokyo}}    
 \and A.~Fukui\orcidO{}\inst{\ref{Univ_tokyo},\ref{IAC_Laguna}}    
 \and R.~Gerasimov\orcidI{}\inst{\ref{ucsd2},\ref{notredame}}    
 \and M.~Gillon\orcidC{}\inst{\ref{astro_liege}}     
 \and Y.~Hayashi\orcidR{}\inst{\ref{Multi_tokyo}}      
 \and S.B.~Howell\orcidi{}\inst{\ref{Ames_NASA}}  
 \and M.~Ikoma\orcidX{}\inst{\ref{Division_Scien_Tokyo}}     
 \and K.~Ikuta\orcidV{}\inst{\ref{Multi_tokyo}} 
 \and J.M.~Jenkins\orcidV{}\inst{\ref{Ames_NASA}} 
 \and P.R.~Karpoor\orcidK{}\inst{\ref{ucsd}}    
 \and Y.~Kawai\orcidU{}\inst{\ref{Multi_tokyo}}   
 \and T.~Kimura\inst{\ref{Division_Scien_Tokyo}}  
 \and T.~Kotani\inst{\ref{Astro_tokyo},\ref{National_Astro_Obs_Tokyo},\ref{Astro_Sc_Prog_SOKENDAI_Tokoy}}    
 \and D.W.~Latham\orcidh{}\inst{\ref{Harvard_USA}} 
\and M.~Mori\orcidT{}\inst{\ref{Multi_tokyo}}     
 \and E.~Pall\'{e}\inst{\ref{IAC_Laguna},\ref{Univ_LaLaguna}}   
 \and H.~Parviainen\orcidd{}\inst{\ref{Univ_LaLaguna},\ref{IAC_Laguna}}   
 \and Y.G.~Patel\orcidb{}\inst{\ref{Opt_Inf_Ast_USA}}   
 \and G.~Ricker\inst{\ref{Kavli_MIT}} 
 \and H.M.~Relles\orcidp{}\inst{\ref{Harvard_USA}} 
 \and A.~Shporer\orcidF{}\inst{\ref{Dep_Kavli}}  
 \and S.~Seager\orcidq{}\inst{\ref{UCSDiego},\ref{Univ_LaLaguna},\ref{Univ_of_Maryl}} 
 \and E.~Softich\orcidJ{}\inst{\ref{ucsd}}    
 \and G.~Srdoc\inst{\ref{Kotiza_Obs}}  
 \and M.~Tamura\orcidZ{}\inst{\ref{Dep_Ast_UnivTokyo_Hongo},\ref{Astro_tokyo},\ref{National_Astro_Obs_Tokyo}}  
 \and C.A.~Theissen\orcidM{}\inst{\ref{ucsd}}    
 \and J.D.~Twicken\inst{\ref{Ames_NASA},\ref{Colorado_Boulder}}   
 \and R.~Vanderspek\inst{\ref{Kavli_MIT}}  
 \and N.~Watanabe\orcidW{}\inst{\ref{Multi_tokyo}}   
 \and C.N.~Watkins\orcidP{}\inst{\ref{Harvard_USA}}     
 \and J.N.~Winn\inst{\ref{Astro_Prin}}   
 \and B.~Wohler\orcids{}\inst{\ref{SETI_Inst_USA},\ref{Ames_NASA}}    
 }
	
\institute{
	Astrobiology Research Unit, Universit\'e de Li\`ege, All\'ee du 6 Ao\^ut 19C, B-4000 Li\`ege, Belgium \label{astro_liege}
	\and 
    Department of Earth, Atmospheric and Planetary Science, Massachusetts Institute of Technology, 77 Massachusetts Avenue, Cambridge, MA 02139, USA \label{MIT}
	\and 
    Instituto de Astrof\'isica de Canarias (IAC), Calle V\'ia L\'actea s/n, 38200, La Laguna, Tenerife, Spain \label{IAC_Laguna}
    \and 
    Center for Astrophysics \textbar  Harvard \& Smithsonian, 60 Garden St, Cambridge, MA 02138, USA \label{Harvard_USA}
    \and 
    Komaba Institute for Science, The University of Tokyo, 3-8-1 Komaba, Meguro, Tokyo 153-8902, Japan \label{Univ_tokyo}
    \and 
    Astrobiology Center, 2-21-1 Osawa, Mitaka, Tokyo 181-8588, Japan \label{Astro_tokyo}
    \and Department of Physics, Sardar Vallabhbhai National Institute of Technology, Surat-395007, Gujarat, India 
    \label{National_Ins_Tech_India}
    \and Dipartimento di Fisica "Ettore Pancini", Università di Napoli Federico II, Napoli, Italy \label{Dip_Fisica_Ita}
    \and INFN, Sezione di Napoli, Complesso Universitario di Monte S. Angelo, Via Cintia Edificio 6, 80126 Napoli, Italy \label{INFN_Ita}
    \and INAF - Osservatorio Astronomico di Capodimonte, via Moiariello 16, 80131 Napoli, Italy \label{INAF_Ita}
    \and National Astronomical Observatory, 2-21-1 Osawa, Mitaka-shi, Tokyo 181-8588, Japan \label{National_Astro_Obs_Tokyo}
    \and Department of Physics, Presidency University, Kolkata-700073, West Bengal, India \label{Presi_univ_India}
    \and
    Department of Astronomy \& Astrophysics, University of California San Diego, La Jolla, CA 92093, USA\label{ucsd}
    \and 
    Department of Physics and Kavli Institute for Astrophysics and Space Research, Massachusetts Institute of Technology, Cambridge, MA 02139, USA \label{Dep_Kavli}
    \and 
    Departamento de Astrof\'isica, Universidad de La Laguna (ULL), E-38206 La Laguna, Tenerife, Spain \label{Univ_LaLaguna}
    \and Instituto de Astrof\'isica de Andaluc\'ia (IAA-CSIC), Glorieta de la Astronom\'ia s/n, 18008 Granada, Spain \label{iaa}
    \and
    Department of Physics and Astronomy, Vanderbilt University, Nashville, TN 37235, USA\label{uvander}
    \and NSF’s National Optical-Infrared Astronomy Research Laboratory, 950 N. Cherry Ave., Tucson, AZ 85719, USA \label{Opt_Inf_Ast_USA}
    \and NASA Exoplanet Science Institute-Caltech/IPAC, Pasadena, CA 91125, USA \label{NASA_Exo_Sc_Ins_USA}
    \and  Department of Physics, Center for Astrophysics \& Space Sciences, University of California San Diego, La Jolla, CA 92093, USA\label{ucsd2}
    \and
    Department of Physics \& Astronomy, Stanford University, Stanford, CA 94305, USA\label{stanford}
    \and Department of Multi-Disciplinary Sciences, Graduate School of Arts and Sciences, The University of Tokyo, 3-8-1 Komaba, Meguro, Tokyo 153-8902, Japan \label{Multi_tokyo}
    \and Dpt. Física Teórica y del Cosmos, Universidad de Granada, Campus de Fuentenueva s/n, 18071 Granada, Spain \label{ugr}
    \and
    Department of Physics and Astronomy, University of Notre Dame, South Bend, IN 46556, USA\label{notredame}
    \and NASA Ames Research Center, Moffett Field, CA 94035, USA \label{Ames_NASA}
    \and Division of Science, National Astronomical Observatory of Japan, 2-21-1 Osawa, Mitaka, Tokyo 181-8588, Japan \label{Division_Scien_Tokyo}
    \and Astronomical Science Program, Graduate University for Advanced Studies, SOKENDAI, 2-21-1, Osawa, Mitaka, Tokyo, 181-8588, Japan \label{Astro_Sc_Prog_SOKENDAI_Tokoy}
    \and Department of Physics and Kavli Institute for Astrophysics and Space Research, Massachusetts Institute of Technology, Cambridge, MA 02139, USA \label{Kavli_MIT}
    \and Center for Astrophysics and Space Sciences, UC San Diego, UCSD Mail Code 0424, 9500 Gilman Drive, La Jolla, CA 92093-0424, USA \label{UCSDiego}
    \and Department of Astronomy, University of Maryland, College Park, MD 20742, USA \label{Univ_of_Maryl}
    \and Kotizarovci Observatory, Sarsoni 90, 51216 Viskovo, Croatia \label{Kotiza_Obs}
    \and Department of Astronomy, University of Tokyo, 7-3-1 Hongo, Bunkyo-ku, Tokyo 113-0033, Japan \label{Dep_Ast_UnivTokyo_Hongo}
    \and Department of Astrophysical and Planetary Sciences, University
of Colorado Boulder, Boulder, CO 80309, USA \label{Colorado_Boulder}
    \and Department of Astrophysical Sciences, Princeton University, Princeton, NJ 08544, USA \label{Astro_Prin}
    \and SETI Institute, Mountain View, CA 94043 USA \label{SETI_Inst_USA}
 }
	
\date{Received/accepted}
\titlerunning{TOI-5720\,b, TOI-6008\,b, TOI-6086\,b}\authorrunning{K. Barkaoui et al.}	 

\abstract{
One of the main goals of the NASA Transiting Exoplanet Survey Satellite (TESS) mission  is the discovery of Earth-like  planets around  nearby M-dwarf stars.
We present the discovery and validation of three new short-period Earth-sized planets orbiting nearby M dwarfs: TOI-5720\,b, TOI-6008\,b, and TOI-6086\,b. We  combined TESS data, ground-based multicolor light curves, ground-based optical and near-infrared spectroscopy, and  Subaru/IRD radial velocity data to validate the planetary candidates and constrain the  physical parameters of the systems.
In addition, we used  archival images, high-resolution imaging, and statistical validation techniques to support the planetary validation.
TOI-5720\,b is an Earth-sized planet with a radius of $R_p = 1.09 \pm 0.07~R_\oplus$. It orbits a nearby (36~pc) M2.5 host with an orbital period of $P = 1.4344555 \pm  0.0000036 $~days.  It has an equilibrium temperature of $T_{\rm eq} = 708 \pm 19$~K (assuming a null albedo) and an incident flux of $S_p = 41.7 \pm 4.5~S_\oplus$. 
TOI-6008\,b is a short-period planet of $ P = 0.8574347 \pm 0.0000424$~day. It has a radius of $R_p = 1.03 \pm 0.05~R_\oplus$, an equilibrium temperature of $T_{\rm eq} = 707 \pm 19$~K, and an incident flux of $S_p = 41.5 \pm 4.5~S_\oplus$. The host star (TOI-6008) is a nearby (23~pc) M5 with an effective temperature of $T_{\rm eff} = 3075 \pm 75~K$. Based on the radial velocity measurements collected with Subaru/IRD, we set a $3\sigma$ upper limit of $M_p < 4M_\oplus$, thus ruling out a star or brown dwarf as the transiting companion.
TOI-6086\,b orbits its nearby (32~pc) M3 host star ($T_{\rm eff} = 3200 \pm 75$~K) every $1.3888725 \pm 0.0000827$~days and has a radius of $ R_p = 1.18 \pm 0.07~R_\oplus$, an equilibrium temperature of $T_{\rm eq} =  634 \pm 16$~K, and an incident flux of $S_p = 26.8 \pm 2.7~S_\oplus$. 
Additional high-precision radial velocity measurements are needed to derive the planetary masses and bulk densities and to search for additional planets in the systems. 
Moreover, short-period Earth-sized planets orbiting around nearby M dwarfs  are suitable targets for an atmospheric characterization with the James Webb Space Telescope through transmission and emission spectroscopy and phase-curve photometry.
}


\maketitle

\section{Introduction}

M dwarfs are stars with effective temperatures $<3800$~K (e.g. \citet{Reid_1995AJ,Nutzman_2008,Kaltenegger_2009,Winters_2015}). They are the most common stars in our Galaxy \citep{henry1994solar,Kirkpatrick_1999ApJ}. M-dwarf stars are excellent targets for a search for transiting Earth-like planets. These systems offer a rare opportunity for exploring their physical, orbital, and atmospheric parameters  due to their low masses, small sizes and low luminosities.
The combination of the high-equilibrium temperature of the planet, low mass, small radius, and infrared brightness of the host star make the short-period planets around M dwarfs exciting targets for a detailed atmospheric characterization using  emission and transmission spectroscopy and phase-curve photometry \citep{Zieba_2023Natur,Greene_2023Natur}.
These planets with especially short periods (of about one day) may provide  some additional information about the formation, evolution, and interior structure of exoplanetary systems \citep{Winn_2018NewAR}.
Moreover, they are also excellent systems based on which the slope of the radius valley can be derived (see, e.g., \cite{Fulton_2017AJ}), and  star-planet interactions, such as the tidal dissipation' effect within their host stars (see, e.g., \cite{Aller_2020AandA}) can be determined.

In this context, we present the discovery and validation of three short-period Earth-sized \emph{TESS} planets around the nearby M dwarfs TOI-5720, TOI-6008, and TOI-6086.  By combining  \emph{TESS} data, and ground-based photometric and spectroscopic measurements, we find that TOI-5720\,b, TOI-6008\,b and TOI-6086\,b have radii of $R_p = 1.09 \pm 0.07~R_\oplus$, $R_p = 1.03 \pm 0.05~R_\oplus$ and $R_p = 1.18 \pm 0.07~R_\oplus$ with orbital periods of $P = 1.4344555 \pm  0.0000036$, $P = 0.8574347 \pm 0.0000424$, and $P = 1.3888725 \pm 0.0000827$~days, respectively.

The paper is organized as follows. Section~\ref{photometric_observation} describes the \emph{TESS} and ground-based observations.
Section~\ref{stellar_carac} describes the stellar characterization of the host stars using spectral energy distributions (SEDs) and spectroscopic observations.
Section~\ref{Validate_planet} presents the validation of the transit signals in the light curves. The planet searches and detection limits from the TESS photometry are presented in Section~\ref{search}.
Section~\ref{Global_modelling} presents our global analysis of all available photometric and spectroscopic datasets of the planetary systems,  allowing us to derive the physical parameters of the systems. Our discussion and conclusion are presented in Section~\ref{Result_discuss}.

\section{Observations and data reduction} \label{photometric_observation}

\subsection{\TESS~ photometry}\label{tess_phot}
\begin{figure*}[!ht]
	\includegraphics[scale=0.32]{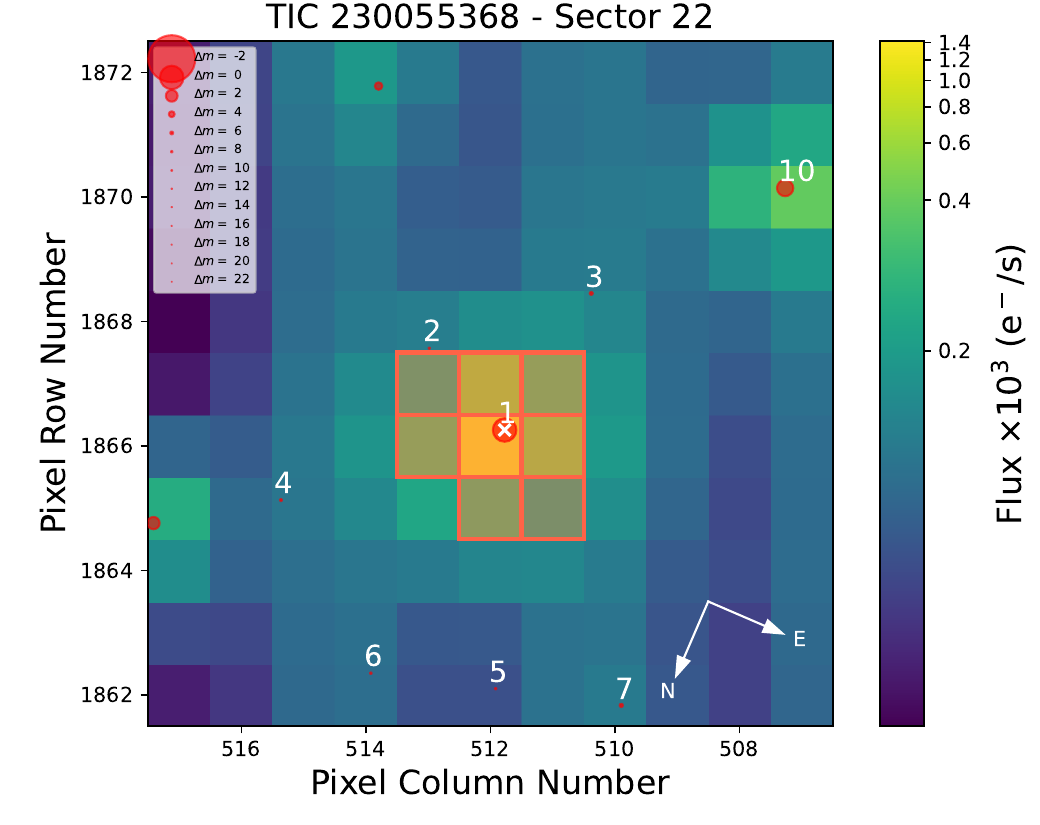}
    \includegraphics[scale=0.32]{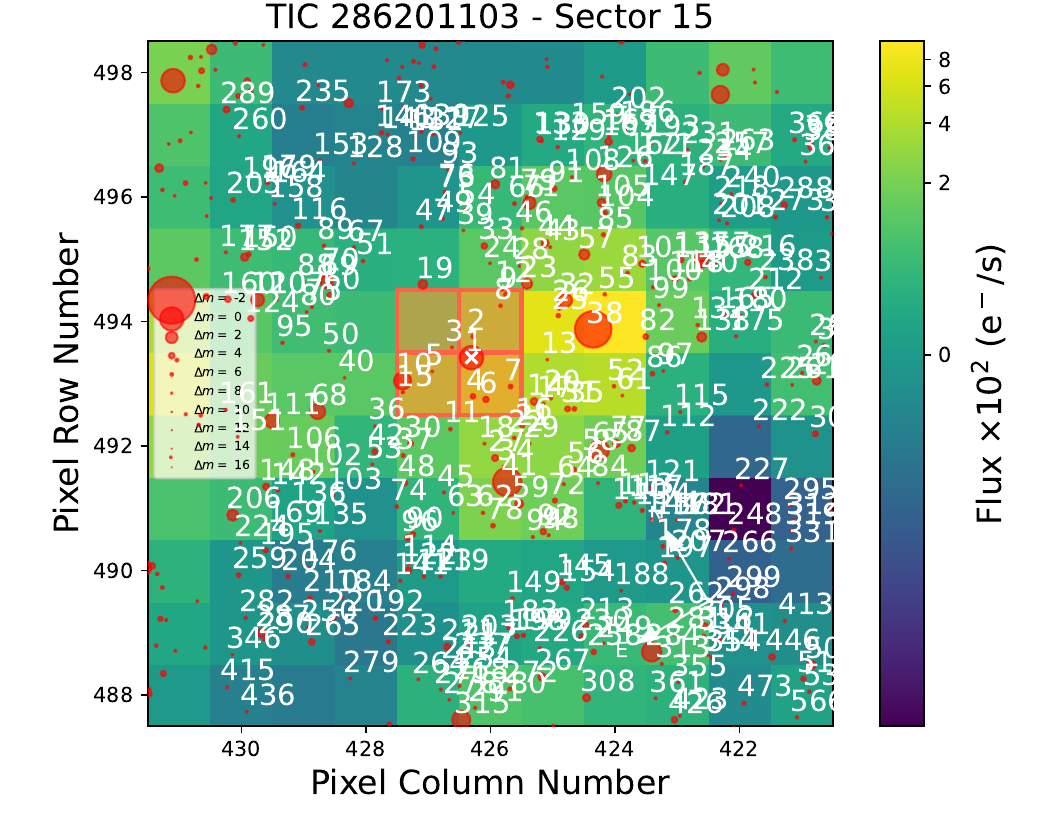}
    \centering\includegraphics[scale=0.32]{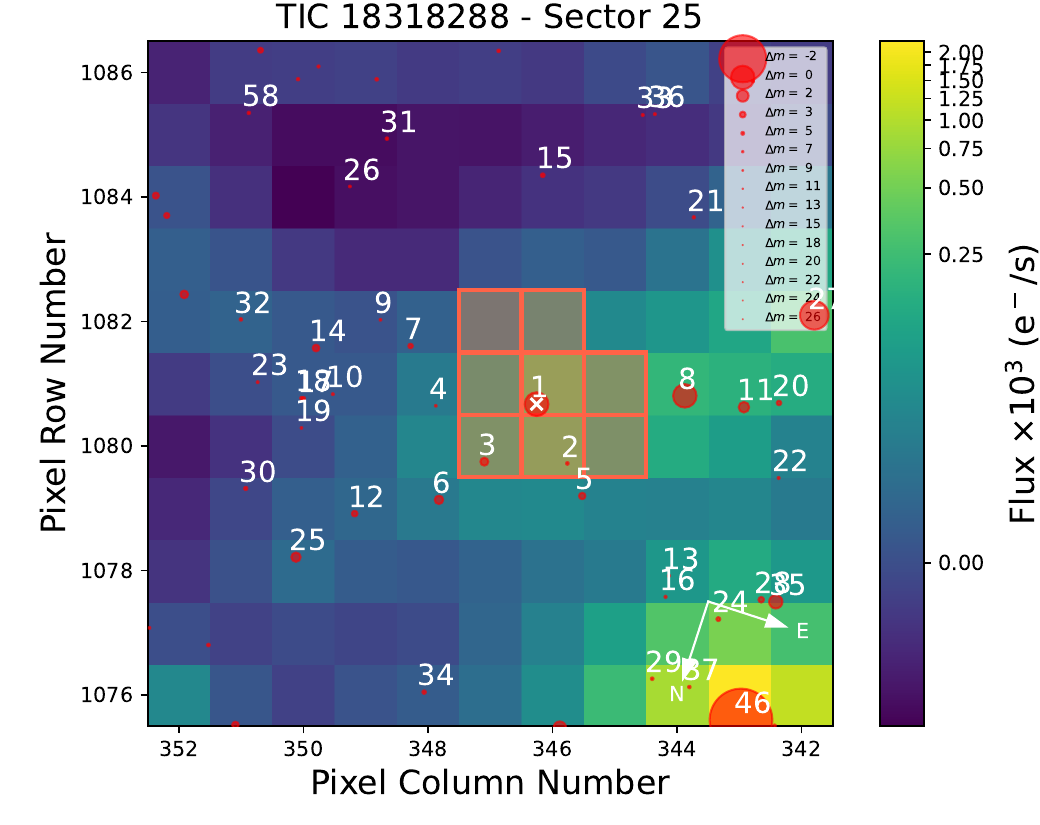}
	\caption{TESS target pixel file images of TOI-5720 (left panel) TOI-6008 (middle panel) and TOI-6086 (right panel) observed in sectors 22, 15 and 25, respectively. The images are made by the {\tt tpfplotter} package \citep{Aller_2020AandA}.  Red dots and red shaded polygons show the positions of Gaia DR3 sources and photometric apertures used to extract the photometric measurements, respectively.}
	\label{Target_pixel}
\end{figure*}

\begin{figure*}[!ht]
	\centering
	   \includegraphics[scale=0.27]{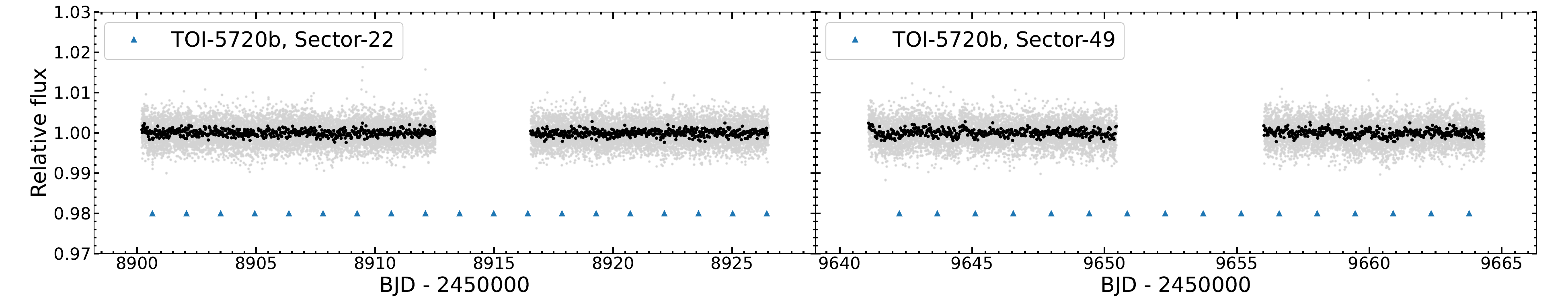}
    \includegraphics[scale=0.27]{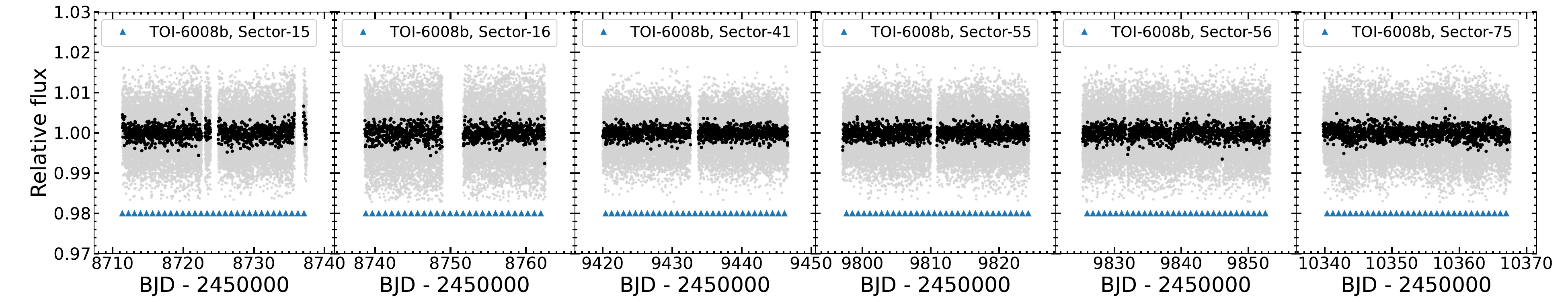}
    \includegraphics[scale=0.27]{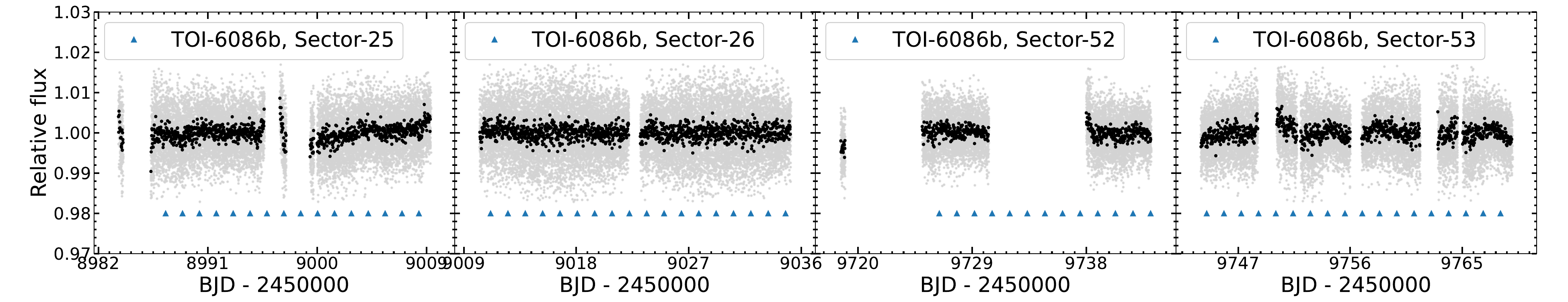}
	\caption{TESS PDC-SAP flux extracted from the 2-minute cadence data of TOI-5720 (top panel), TOI-6008 (middle panel), and TOI-6086 (bottom panel). TOI-5720 was observed in sectors 22 and 49, TOI-6008 was observed in sectors 15, 16, 41, 55, 56 and 75, and TOI-6086 was observed in sectors 25, 26, 52 and 53. The light grey points show the 2-minute cadence data, and the black points shows the flux in 30-minutes bins. The transit locations of TOI-5720\,b, TOI-6008\,b, and TOI-6086\,b are shown with the blue triangles.}
	\label{TESS_LCs_BJD}
\end{figure*}

\begin{figure*}[!ht]
	   \includegraphics[scale=0.3]{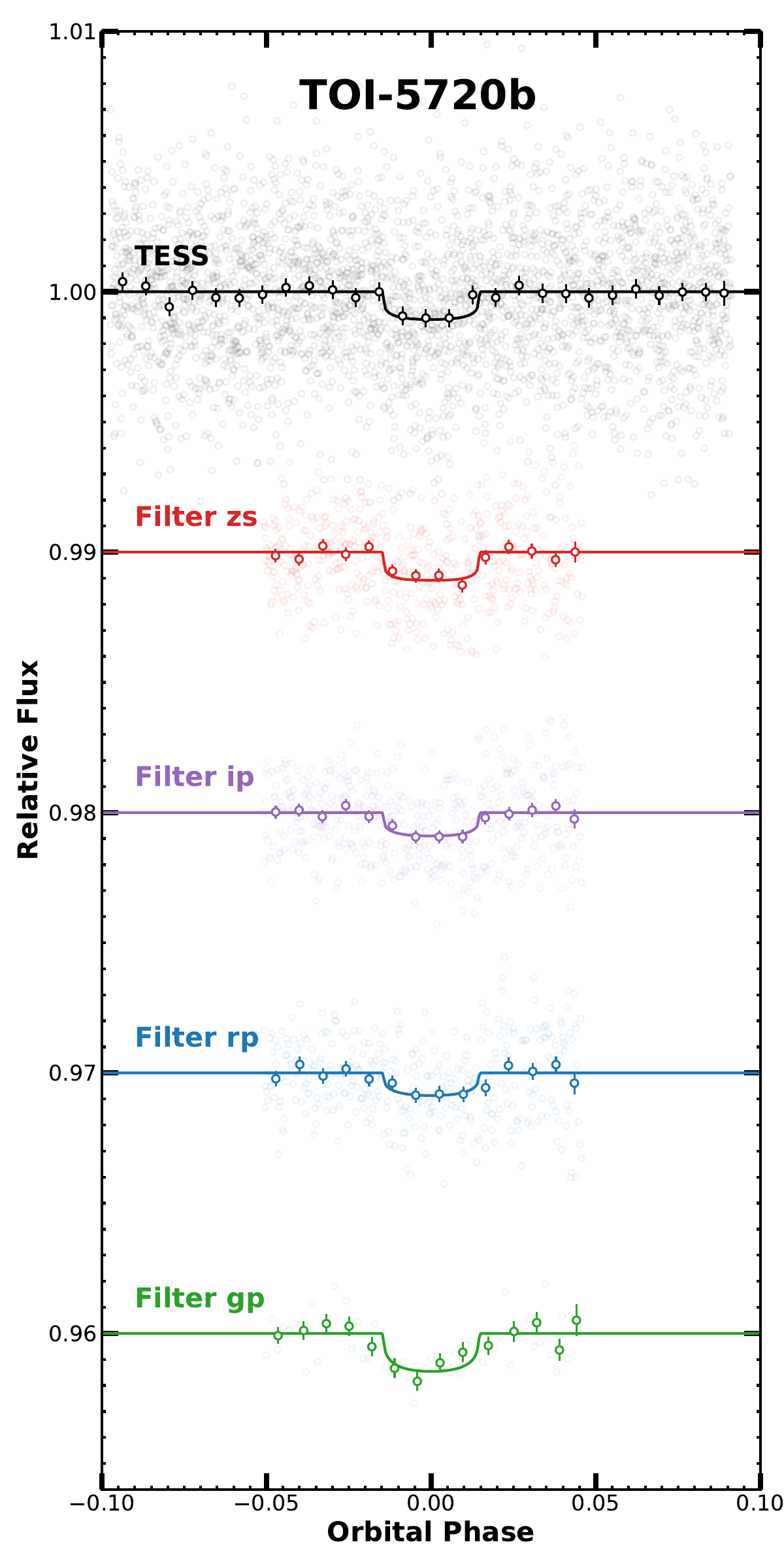}
     \includegraphics[scale=0.3]{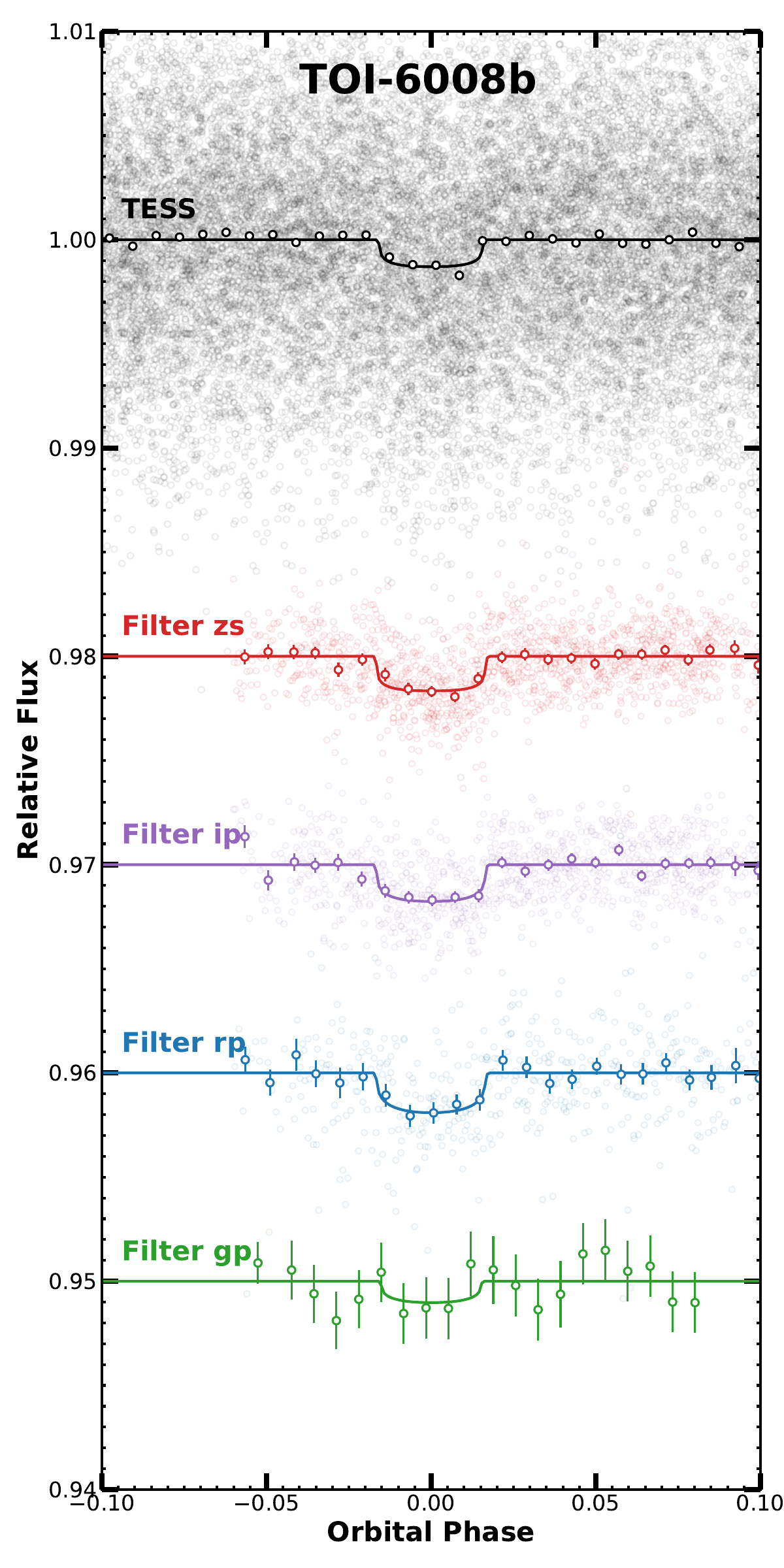}
     \includegraphics[scale=0.3]{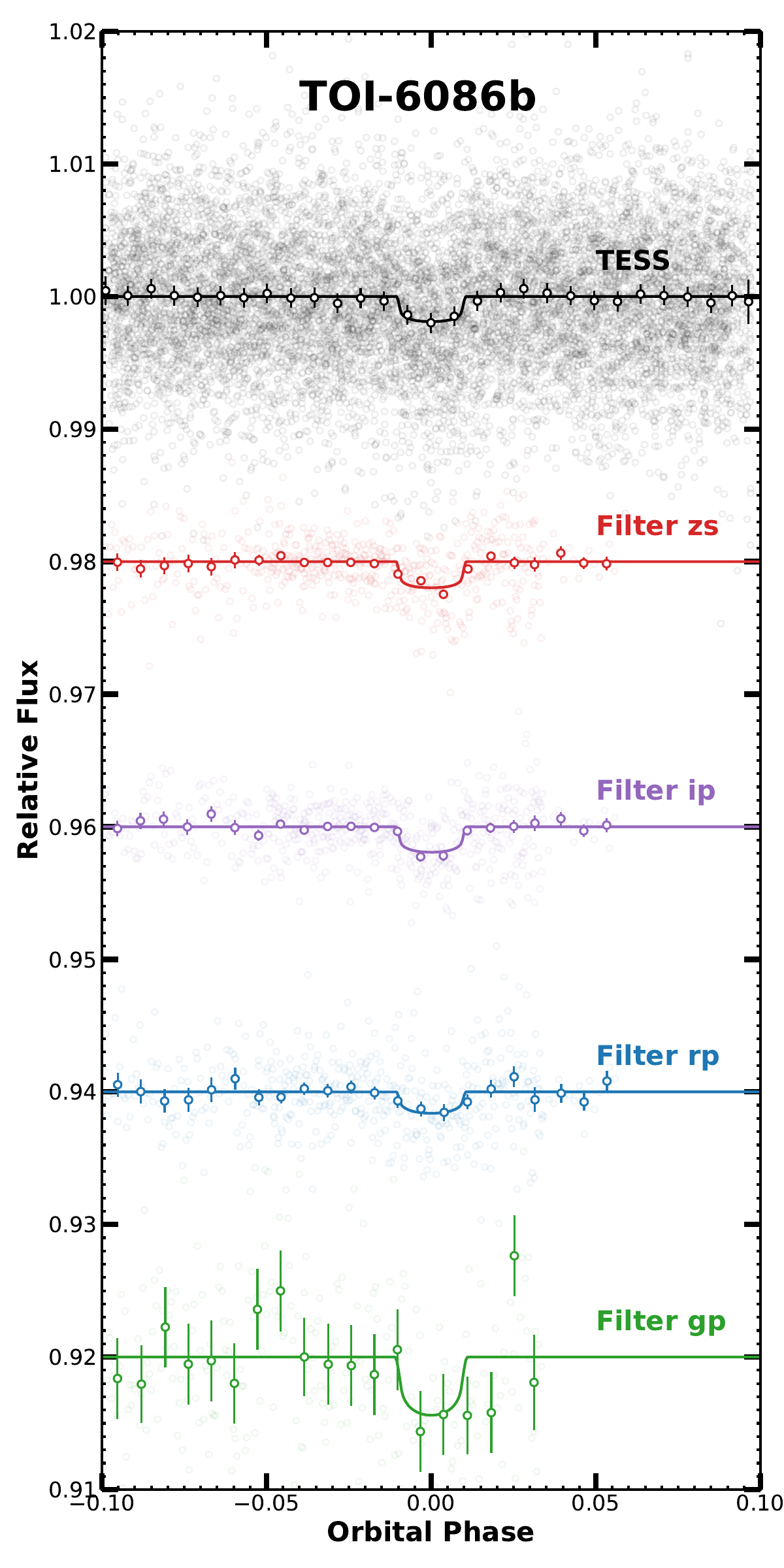}
	\caption{TESS and ground-based photometric data of TOI-5720\,b (left panel), TOI-6008\,b (middle panel) and TOI-6086\,b (right panel). Top plots show the TESS phased-folded, and other plots show the ground-based phased-folded by filter with best-fitting transit model is superimposed. The light curves are shifted along the y-axis for clarity.}
	\label{TOI_6008_5720_TESS_LC}
\end{figure*}

The host stars TIC\,230055368 (TOI-5720), TIC\,286201103 (TOI-6008) and  TIC\,18318288 (TOI-6086) were observed by the TESS \citep{Ricker_2015JATIS_TESS} mission with a cadence of 2-minutes. 
TOI-5720 was observed in sectors 22 (2020 February 18 to 2020 March 18) and 49 (2022 February 26 to 2022 March 26).
TOI-6008 was observed in sectors 15 (2019 August 15 to 2019 September 11), 16 (2019 September 11 to 2019 October 07), 41 (2021 July 23 to 2021 August 20), 55 (2022 August 05 to 2022 September 01), 56 (2022 September 01 to 2022 September 30) and 75 (2024 January 30 to 2024  February 26).  
TOI-6086 was observed in sectors 25 (2020 May 13 to 2020 June 08), 26 (2020 June 08 to 2020 July 04), 52 (2022 May 18 to 2022 June 13) and 53 (2022 June 13 to 2022 July 09).
To model the TESS data, we retrieved the presearch data conditioning light curves (PDC-SAP; \citet{Stumpe_2012PASP,Smith_2012PASP,Stumpe_2014}) constructed by the TESS Science Processing Operations Center (SPOC; \citet{SPOC_Jenkins_2016SPIE}) at the Ames Research Center from the Mikulski Archive for Space Telescopes.
The PDC-SAP light curves were calibrated for any instrument systematics and crowding effects.  The TESS field of view for each target and  photometric apertures used with the location of nearby Gaia DR3 sources \citep{Gaia_Collaboration_2021AandA} are shown in \autoref{Target_pixel}.  Figures \ref{TESS_LCs_BJD}, \ref{TOI_6008_5720_TESS_LC}  and \ref{TESS_LCs} show the TESS photometric data for each target.

\subsection{Ground-based photometric follow-up}
We performed ground-based observation follow-up of TOI-5720, TOI-6008 and TOI-6086 as part of the  TESS follow-up observing program (TFOP) in order  to ({\it i}) rule out the nearby eclipsing binaries (NEBs) as potential sources of the TESS detection, ({\it ii}) confirm the transit event on the target star, ({\it iii}) check the chromaticity of the transit depth measurement, and ({\it iv}) refine the transit ephemerides.
 We used the {\tt TESS transit finder} tool \citep{jensen2013} in order to schedule the photometric follow-up observations. All ground-based photometric observations are summarized in Table~\ref{obs_table}.

\subsubsection{LCOGT-2.0m MuSCAT3}
We used the Las Cumbres Observatory Global Telescope (LCOGT; \cite{Brown_2013}) 2.0m Faulkes Telescope North located at Haleakala Observatory in Hawaii to observe  TOI-5720\,b, TOI-6008\,b and TOI-6086\,b. 
The telescope is equipped with the MuSCAT3 multiband imager \citep{Narita_2020SPIE11447E}. 
One full transit of TOI-5720\,b was observed on UTC May 12 2023, three full transits of TOI-6008\,b were observed on UTC May 7, 13 and 31 2023, and two full transits of TOI-6086\,b were observed on UTC July 6 and 31 2023. All observations were carried out simultaneously with the Sloan-$g'$, -$r'$, -$i'$ and Pan-STARRS-$z_\mathrm{s}$ filters. The stellar fluxes were extracted using uncontaminated photometric apertures (see Table~\ref{obs_table}).

The science images were calibrated using the standard LCOGT {\tt BANZAI} pipeline \citep{McCully_2018SPIE10707E}, and aperture photometry was performed using {\tt AstroImageJ}\footnote{{\tt AstroImageJ:}~\url{https://www.astro.louisville.edu/software/astroimagej/}} \citep{Collins_2017} software.

\subsubsection{MuSCAT2}
We observed one full transit of TOI-6008\,b on UTC July 27 2023 and one full transit of TOI-6086\,b on UTC July 29 2023 with the 1.52m Telescopio Carlos Sanchez (TCS) at the Teide Observatory in Spain. 
The telescope is equipped with the MuSCAT2 multicolor imager \citep{Narita2018}. 
Both  transits were observed simultaneously in the Sloan-$g$, -$r$, -$i$, and Pan-STARRS-$z_\mathrm{s}$.
The stellar fluxes  were extracted using an uncontaminated photometric aperture (see Table~\ref{obs_table}). 
The data calibration and aperture photometry were performed using the MuSCAT2 photometry pipeline \citep{Parviainen2020}. 

\subsubsection{LCOGT-1.0m}
Two full transits of TOI-6086\,b were observed with  LCOGT-1m0 network on UTC March 10 and 20 2023 in the Sloan-$i'$ filter. The telescopes are equipped with $4096 \times 4096$ SINISTRO cameras with a pixel scale of $0.389\arcsec$ per pixel and an  FOV of $26' \times 26'$. 
The science images were calibrated by the standard LCOGT {\tt BANZAI} pipeline, and aperture photometry  was performed using the software {\tt AstroImageJ}. Both  observations are used to check nearby eclipsing binaries, but they were not included in our global analysis because of the low signal-to-noise ratio (S/N).

\begin{table*}[!ht]
 \begin{center}
 {\renewcommand{\arraystretch}{1.4}
 \begin{tabular}{l c c c c c c cccccc}
 \toprule
Planet & Date (UT) & Filter & Telescope & Exptime  &  FWHM & Aperture & Baseline & $\beta_w$ & $\beta_r$ & $CF$ & Comment \\ 
       &           &        &           & [second] & [arcsec] & [arcsec] \\
 \hline
             & May 12 2023 & $z_\mathrm{s}$ & MuSCAT3 & 17  & 3.3 & 4.5 & $p(t^2)$ & 1.26 & 1.08 & 1.36 & Full \\
 TOI-5720\,b & May 12 2023 & $i'$ & MuSCAT3 & 19 &  3.3 & 4.5 & $p(t^2 + b^1)$ & 1.22 & 1.04 & 1.27 & Full \\
             & May 12 2023 & $r'$ & MuSCAT3 & 25 &  2.9 & 4.5 & $p(t^2 + b^1)$ & 1.26 & 1.09 & 1.37 & Full \\
             & May 12 2023 & $g'$ & MuSCAT3 & 209 &  2.6 & 4.5 & $p(t^2 + a^1)$ & 0.92 & 1.07 & 0.99 & Full \\
 \hline
             & May 07 2023 & $z_\mathrm{s}$ & MuSCAT3 & 20 & 2.1 & 3.2 & $p(t^2)$ & 1.13 & 1.45 & 1.64 & Full \\
             & May 07 2023 & $i'$ & MuSCAT3 & 30 &  2.2 & 3.2 & $p(t^2)$ & 1.13 & 1.20 & 1.36 & Full \\
             & May 07 2023 & $r'$ & MuSCAT3 & 88 &  2.3 & 3.2 & $p(t^2)$ & 1.24 & 1.22 & 1.52 & Full \\
 
             & May 13 2023 & $z_\mathrm{s}$ & MuSCAT3 & 20 &  2.9 & 4.5 & $p(t^2 + b^1)$  & 1.17 & 1.23 & 1.44 & Full \\
             & May 13 2023 & $i'$ & MuSCAT3 & 30 &  3.5 & 4.8 & $p(t^2)$ & 1.16 & 1.30 & 1.51 & Full \\
             & May 13 2023 & $r'$ & MuSCAT3 & 88 &  3.5 & 5.0 & $p(t^2 + a^2)$ & 1.11 & 1.02 & 1.13 & Full \\
 TOI-6008\,b & May 13 2023 & $g'$ & MuSCAT3 & 500 &  3.4 & 5.0 & $p(t^2)$ & 0.85 & 1.23 & 1.04 & Full \\
 
             & May 31 2023 & $z_\mathrm{s}$ & MuSCAT3 & 20 &  2.6 & 4.0 & $p(x^2)$ & 1.15 & 2.46 & 2.82 & Full \\
             & May 31 2023 & $i'$ & MuSCAT3 & 30 & 2.9 & 4.2 & $p(b^2)$ & 1.09 & 1.51 & 1.64 & Full \\
             & May 31 2023 & $r'$ & MuSCAT3 & 88 & 2.7 & 4.2 & $p(t^2)$ & 1.05 & 1.19 & 1.26 & Full \\
 
             & July 27 2023 & $z_\mathrm{s}$ & MuSCAT2 & 30 & 2.1 & 4.3 & $p(t^2)$ & 1.02 & 1.31 & 1.34 & Full \\
             & July 27 2023 & $i'$ & MuSCAT2 & 30 & 2.4 & 4.4 & $p(t^2)$ & 1.01 & 1.08 & 1.09 & Full \\
             & July 27 2023 & $r'$ & MuSCAT2 & 88 & 2.5 & 4.5 & $p(t^2)$ & 0.98 & 1.18 & 1.18 & Full \\
             & July 27 2023 & $g'$ & MuSCAT2 & 30 & 2.7 & 4.5 & $p(t^2)$ & 1.00 & 1.04 & 1.04 & Full \\
 \hline
             & July 06 2023 & $z_\mathrm{s}$ & MuSCAT3 & 35 & 2.4 & 4.8 & $p(t^2)$ & 0.97 & 1.00 & 0.97 & Full \\
             & July 06 2023 & $i'$ & MuSCAT3 & 43 & 2.3 & 4.8 & $p(t^2)$ & 0.95 & 1.00 & 0.95 & Full \\
             & July 06 2023 & $r'$ & MuSCAT3 & 63 & 2.5 & 4.8 & $p(t^2 + b^1)$ & 1.04 & 1.00 & 1.04 & Full \\
  
             & July 31 2023 & $z_\mathrm{s}$ & MuSCAT3 & 35 & 4.1 & 6.9 & $p(t^2)$  & 1.33 & 1.22 & 1.49 & Full \\
             & July 31 2023 & $i'$ & MuSCAT3 & 43 & 4.3 & 6.9 & $p(t^2)$ & 1.16 & 1.04 & 1.21 & Full \\
  TOI-6086\,b & July 31 2023 & $r'$ & MuSCAT3 & 63 & 4.3 & 6.9 & $p(t^2)$ & 1.01 & 1.14 & 1.15 & Full \\
  
             & July 29 2023 & $z_\mathrm{s}$ & MuSCAT2 & 20 & 2.1 & 10.9  & $p(t^2)$ & 0.95 & 1.00 & 0.95 & Full \\
             & July 29 2023 & $i'$ & MuSCAT2 & 20 & 2.1 & 10.9 & $p(t^2)$ & 1.02 & 1.09 & 1.011 & Full \\
             & July 29 2023 & $r'$ & MuSCAT2 & 40 & 2.5 & 10.9 & $p(t^2)$ & 1.00 & 1.09 & 1.09 & Full \\
             & July 29 2023 & $g'$ & MuSCAT2 & 25 & 2.8 & 10.9 & $p(t^2)$ & 1.00 & 1.36 & 1.36 & Full \\
\hline
 \end{tabular}}
 \caption{Observational  parameters for TOI-5720\,b, TOI-6008\,b, and TOI-6086\,b:  Date of the observation, filter, telescope, exposure time(s), and FWHM of the point-spread function and photometric aperture. The analysis parameters are the selected baseline model, red noise $\beta_r$, white noise $\beta_w$ and $CF$ (Correction Factor). The baseline model is shown as a polynomial $p()$ of systematic parameters, with $t$ is the time, $x$ and $y$ are the positions of the star on the detector, $a$ is the airmass, $b$ is the background, and $f$ is the full with at half maximum of the PSF.}
 \label{obs_table}
 \end{center}
\end{table*}


\subsubsection{Shane/Kast optical spectroscopy} 
\label{Sec_Shane_Kast}
\begin{figure}[!ht]
    \centering
    \includegraphics[width=\columnwidth, keepaspectratio]{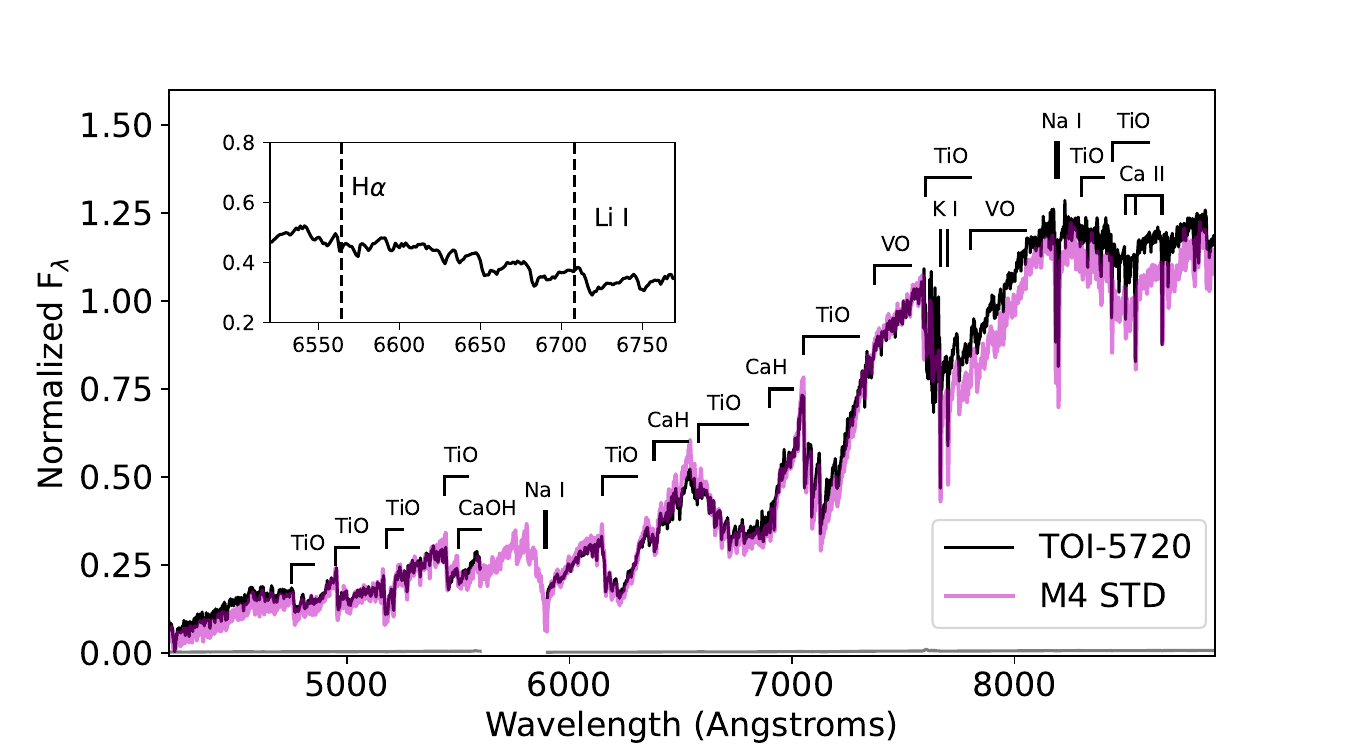}
    \includegraphics[width=\columnwidth, keepaspectratio]{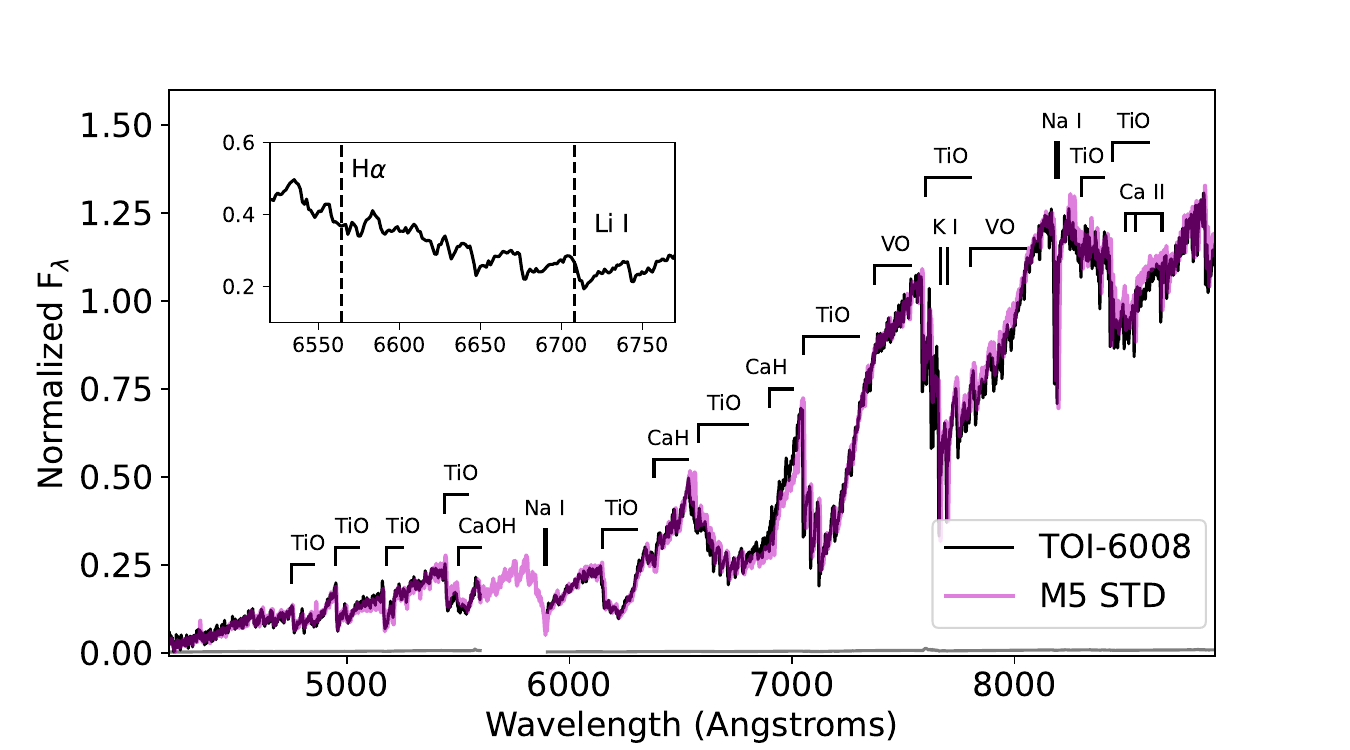}
    \includegraphics[width=\columnwidth, keepaspectratio]{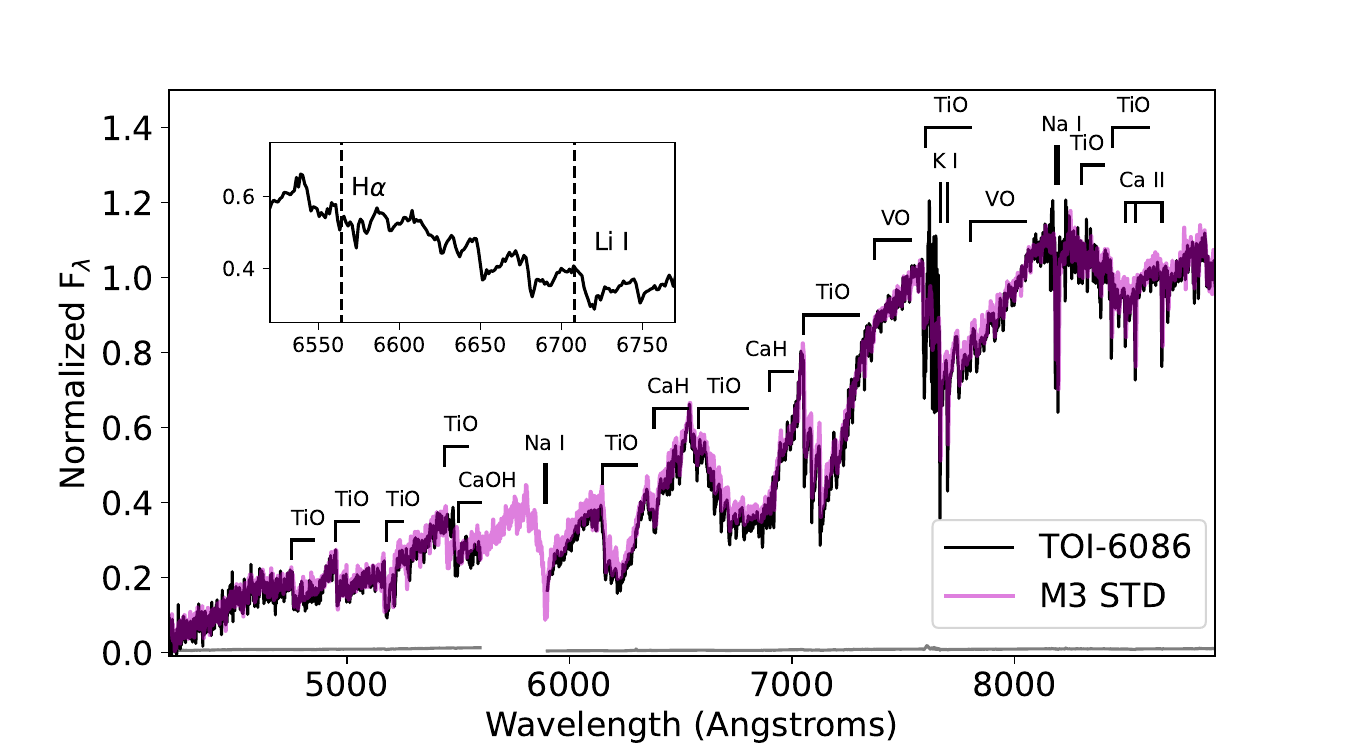}
    \caption{
        Shane/Kast blue and red optical spectra of TOI-5720 (top), TOI-6008 (middle), and TOI-6086 (bottom).
        The target spectra (black lines) are normalized at 7400~{\AA} and compared to their best-fit SDSS standard template (magenta lines)
        from \citet{2007AJ....133..531B}.
        The gaps in the target spectra between 5600~{\AA} and 5900~{\AA} correspond to the dichroic split between the Kast blue and red channels.
        Key spectral features in the 4200--8900~{\AA} region are labeled.
        The inset box highlights the region around the 6563~{\AA} H$\alpha$ and 6708~{\AA} Li~I lines, neither of which are detected in any of the sources.
}
    \label{fig_Shane_kast}
\end{figure}

All three host stars were observed with the Kast Double Spectrograph \citep{kastspectrograph} mounted on the 3m Shane Telescope at Lick Observatory.
TOI-5720 and TOI-6008 were observed on UT June 2, 2023 in hazy to clear conditions with a seeing of 1$\arcsec$;
TOI-6086 was observed on UT August 27, 2023 in hazy conditions with a seeing of 1$\farcs$2.
For all three sources, we used the 1$\arcsec$ wide slit, with the 600/7500 grating in the red channel, 
providing a wavelength coverage of 5900--9000~{\AA}  at an average resolution of $\lambda/\Delta\lambda$ = 1950,
and the 600/4310 grism in the blue channel, providing a wavelength coverage 3600--5600~{\AA}  at an average resolution of $\lambda/\Delta\lambda$ = 1500. The channels were split by the D57 dichroic with a crossover wavelength of 5700~{\AA}.
For all sources, we obtained one blue exposure and two red exposures, for total intergration times of 1000~s, 600~s, and 600~s for TOI-5720, TOI-6008, and TOI-6086, respectively.
We also observed a nearby G2\,V star for each source to  calibrate the telluric absorption, a spectrophotomeric flux standard star on each night to calibrate the flux (Fiege 34 on June 2, BD 28+4211 on August 27; \citealt{1992PASP..104..533H,1994PASP..106..566H}), and flat-field and HeNeAr arc lamps at the start of each night to calibrate pixel response and wavelength. The data reduction was performed using  the kastredux package\footnote{kastredux:~\url{https://github.com/aburgasser/kastredux}.}, which included image reduction (pixel response calibration and bad pixel masking), boxcar extraction with median background subtraction of the spectra, wavelength calibration using the arc lamps, flux calibration using the flux standard, and correction for telluric absorption in the red channel using the G2\,V stars (The blue data were not corrected for telluric absorption). 
The final spectra have median signal-to-noise ratios of 194/64, 133/35, 100/26 in the red/blue channels for TOI-5720, TOI-6008, and TOI-6086, respectively (see \autoref{fig_Shane_kast}).

\subsubsection{IRTF/SpeX spectroscopy}
\label{Sec_IRTF_SpeX}
\begin{figure}[!ht]
    \centering
    \includegraphics[width=\columnwidth, height=0.45\textheight, keepaspectratio]{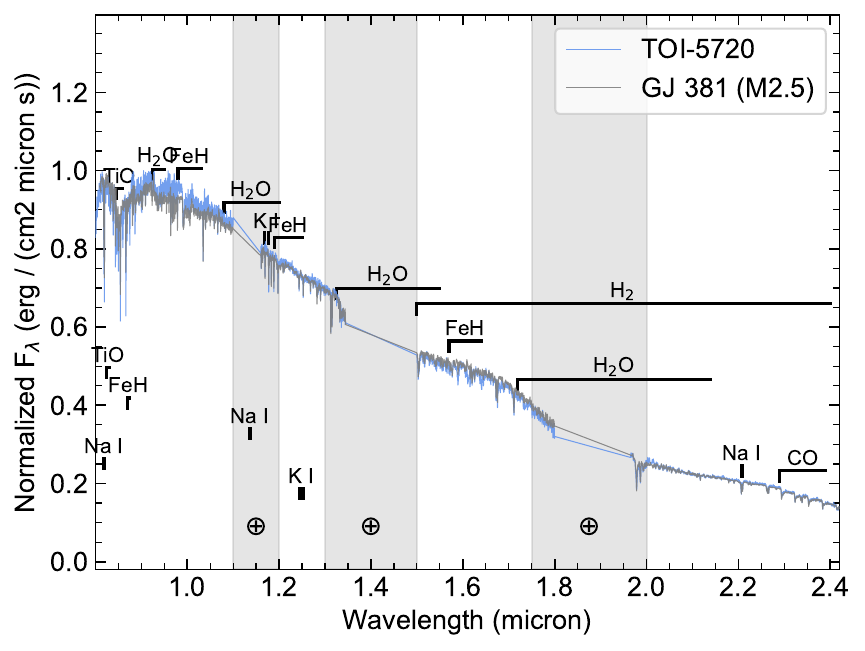}\\
    \includegraphics[width=\columnwidth, height=0.45\textheight, keepaspectratio]{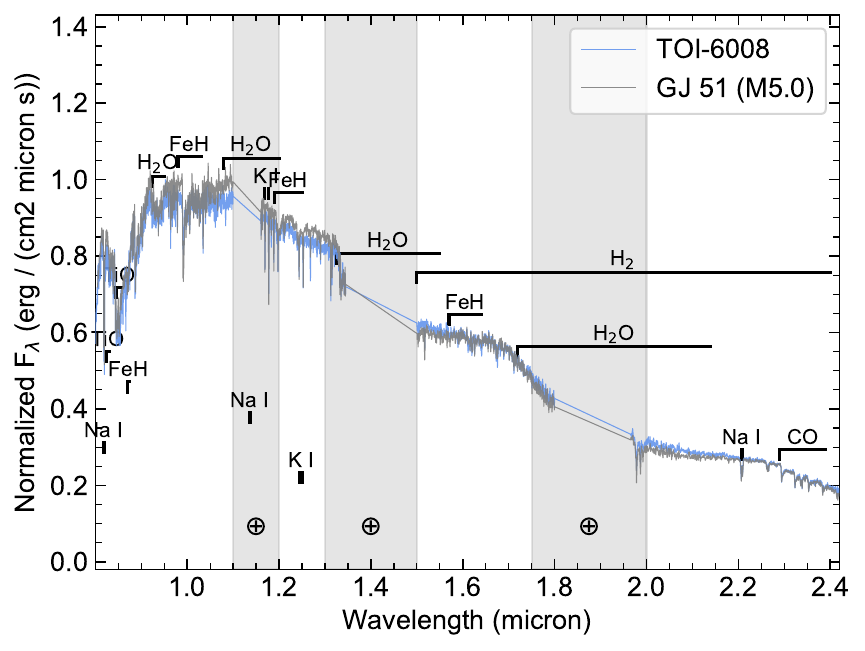}\\
    \includegraphics[width=\columnwidth, height=0.45\textheight, keepaspectratio]{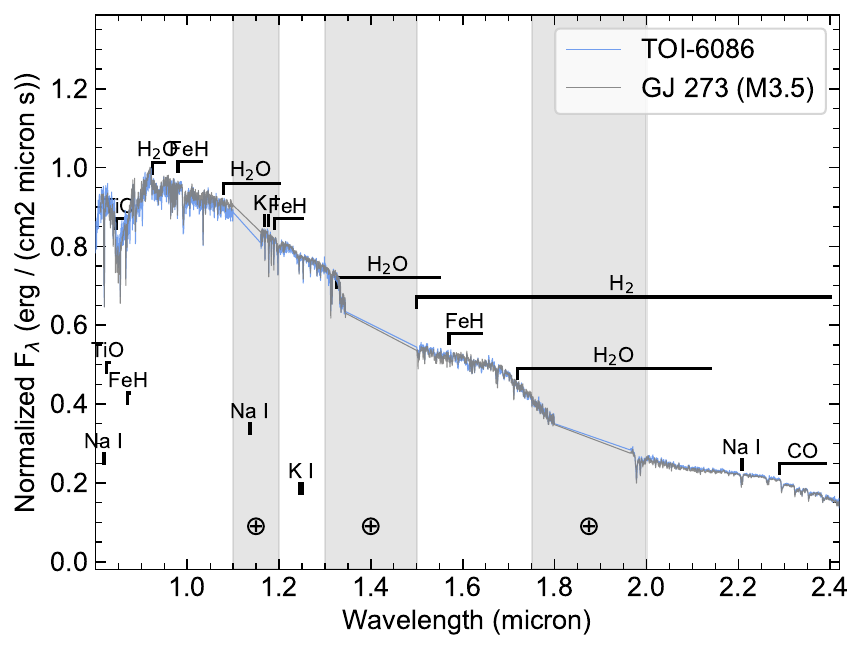}
    \caption{
        SpeX SXD spectra of TOI-5720 (top), TOI-6008 (middle), and TOI-6086 (bottom).
        The target spectra (blue) are shown alongside the spectrum of the best-fit standard star (gray).
        Regions of high telluric absorption are shaded, and strong spectral features of M dwarfs are highlighted.
}
    \label{fig_IRTF_spex}
\end{figure}

We used the SpeX spectrograph \citep{Rayner2003} on the 3.2m-NASA Infrared Telescope Facility (IRTF) to collect medium-resolution near-infrared spectra of TOI-6086 on UT 2023 May 4 and of TOI-5720 and TOI-6008 on UT 2023 June 5. Thin cirrus clouds were present on both nights.
Seeing was 1$\farcs$0 on the May night and 0$\farcs$6 on the June night. On both nights, we used the short-wavelength cross-dispersed (SXD) mode with the $0\farcs3 \times 15''$ slit aligned to the parallactic angle, providing spectra covering 0.80--2.42\,$\mu$m with a resolving power of $R{\sim}2000$ and an average of 2.5\,pixels per resolution element.
Nodding in an ABBA pattern, we collected six integrations of TOI-5720 and TOI-6008 and eight integrations of TOI-6086 with integration times of 70.0\,s, 70.0\,s, and 149.7\,s, respectively.
After each science observation, we collected a set of standard SXD flat-field and arc-lamp exposures and six integrations of A0\,V standard stars.
We reduced the data with Spextool v4.1 \citep{Cushing2004}, following the standard approach outlined in the Spextool User  Manual\footnote{Available at \url{http://irtfweb.ifa.hawaii.edu/~spex/observer/}.}.
The final spectra have median signal-to-noise ratios per pixel of 124, 109, and 133 for TOI-5720, TOI-6008, and TOI-6086, respectively (see \autoref{fig_IRTF_spex}).

\subsection{Spectroscopy}  \label{stellar_spectro}

\subsubsection{Spectroscopy from Subaru/IRD} 
We performed high-resolution spectroscopy for TOI-6008 with the InfraRed Doppler (IRD) instrument \citep{2012SPIE.8446E..1TT, 2018SPIE10702E..11K} mounted at the Subaru 8.2m telescope. IRD can obtain a near-infrared spectrum between from 950 nm to 1740 nm at one exposure with a spectral resolution of $R\approx 70,000$. Between UT 2023 July 31 and August 6, a total of 15 frames were collected for TOI-6008, for which integration times were set to 1200--1500~seconds. We also injected the laser-frequency comb (LFC) into a secondary fiber so that we were able to trace the temporal variation of the instrumental profile. 

\begin{figure}[!ht]
    \centering
    \includegraphics[scale=0.215]{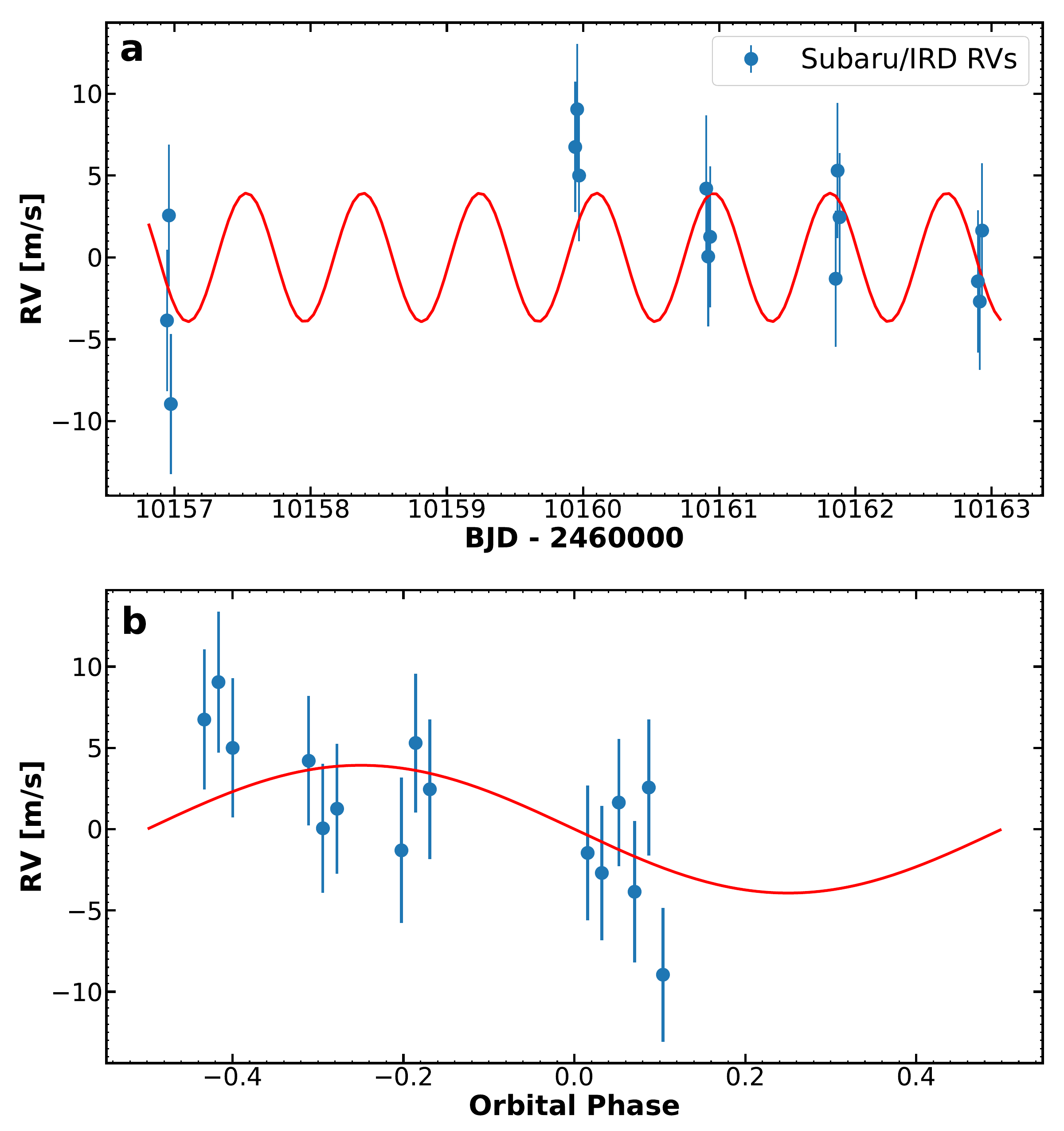}
    \caption{Radial velocities obtained with the Subaru-8.2m/IRD spectrograph for TOI-6008. 
    {\bf a}, RVs observations vs. time.
    {\bf b}, Phase-folded RV observations. The best-fitting model is superimposed in red.
    }
    \label{RVs_curve_TOI6008}
\end{figure}

IRD data were reduced following the reduction procedure in \citet{2020PASJ...72...93H}, where custom reduction codes as well as \texttt{IRAF} \citep{1993ASPC...52..173T} were used. For each frame, one-dimensional stellar and LFC spectra were separately extracted by this reduction. The reduced stellar spectra typically have (S/N) of 60--70 per pixel at 1000 nm. These spectra were then subjected to the radial velocity (RV) measurements with the standard RV pipeline for IRD \citep{2020PASJ...72...93H}, and we extracted relative RVs for individual frames using the template-matching technique. 
The resulting RV measurements and curve are presented in \autoref{table_IRD_RVs_TOI6008} and \autoref{RVs_curve_TOI6008}, respectively; the RV internal errors returned by the pipeline were typically $\sim 3$~m~s$^{-1}$. 

\begin{table}[!ht]
\centering
	{\renewcommand{\arraystretch}{1.1}
		\begin{tabular}{lcc}
			\hline
			BJD-TDB     & RV [km/s] & $\sigma_{RV}$ [km/s]       \\
			\hline\hline
            2460156.9444130 	&  6.337714	&  0.00332 \\
            2460156.9586420 	&  6.344117	&  0.00310 \\
            2460156.9728711 	&  6.332584	&  0.00301 \\
            2460159.9418833 	&  6.345601	&  0.00327 \\
            2460159.9561104 	&  6.347892	&  0.00330 \\
            2460159.9703249 	&  6.343831	&  0.00323 \\
            2460160.9039142 	&  6.342193	&  0.00282 \\
            2460160.9181436 	&  6.338026	&  0.00282 \\
            2460160.9323684 	&  6.339218	&  0.00286 \\
            2460161.8541920 	&  6.335824	&  0.00348 \\
            2460161.8684328 	&  6.342417	&  0.00321 \\
            2460161.8826727 	&  6.339555	&  0.00325 \\
            2460162.8982452 	&  6.334722	&  0.00306 \\
            2460162.9124742 	&  6.333475	&  0.00303 \\
            2460162.9294896 	&  6.337795	&  0.00274 \\
			\hline
		\end{tabular}}
		\caption{Radial velocity measurements for TOI-6008 obtained from  the Subar-8.2m/IRD  spectrograph.}
		\label{table_IRD_RVs_TOI6008}
\end{table}

\subsubsection{Spectroscopy from TRES} 
\label{TRES_spectr}
 Two reconnaissance spectra of TOI-5720 were obtained on UT December 19 and 24, 2022 using the Tillinghast reflector Echelle Spectrograph (TRES; \citet{gaborthesis_TrES}) mounted on the 1.5m Tillinghast Reflector telescope at the Fred Lawrence Whipple Observatory (FLWO) in Arizona. TRES is a fiber-fed spectrograph that observes in the 390--910\,nm range with a resolving power of R=44,000. Spectra were extracted using the methods described in \citet{buchhave2010}.

For the two TRES observations of TOI-5720, we measured the radial velocities using a pipeline optimized for mid- to late-M dwarfs (see, e.g., \citet{Pass_2023} for further details). In brief, this pipeline performs a cross correlation between the observed spectra and an empirical mid- to late-M dwarf template using six TRES echelle orders in the range of 6400--7850\,\AA. The pipeline also produces carefully calibrated radial velocity uncertainties, taking into account the signal-to-noise ratio of the observed spectra, template mismatch, rotational broadening, and the long-term stability of the spectrograph and errors in the barycentric correction. The two resulting RV measurements are $35.692 \pm 0.031$~km/s on UT December 19 and $35.684 \pm 0.024$~km/s for UT December 24, and the upper limit on the projected rotational velocity is found to be $v\sin(i) < 3.4$~km/s.

\subsection{High-resolution imaging using Palomar-5m0/PHARO}

We assessed the possibility of a blend of nearby stellar companions as part of our standard process for validating transiting exoplanets, which can dilute the transit light curves and underestimated planetary radii measurements \citep{Ciardi_2015ApJ}. We observed TOI-5720, TOI-6008, and TOI-6086 with WIYN 3.5~m/NESSI, Palomar-5m0/PHARO, and Robo-AO high-resolution imaging. Despite the combination of Gaia catalog and high-resolution imaging, there is no indication of any additional close stellar companions.

\begin{figure}[!ht]
    \centering
    \includegraphics[width=\columnwidth, height=0.4\textheight, keepaspectratio]{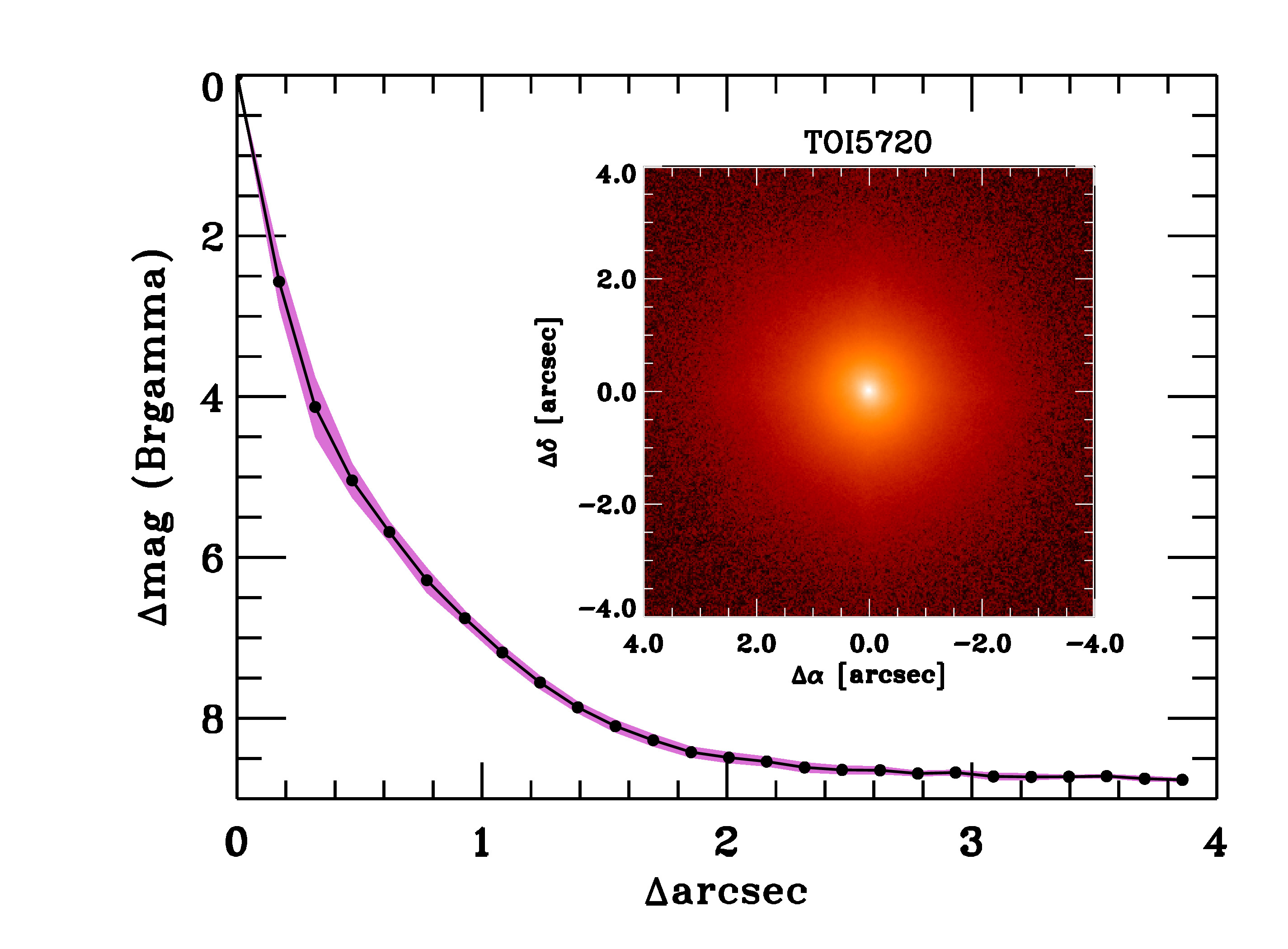}
    \includegraphics[width=\columnwidth, height=0.4\textheight, keepaspectratio]{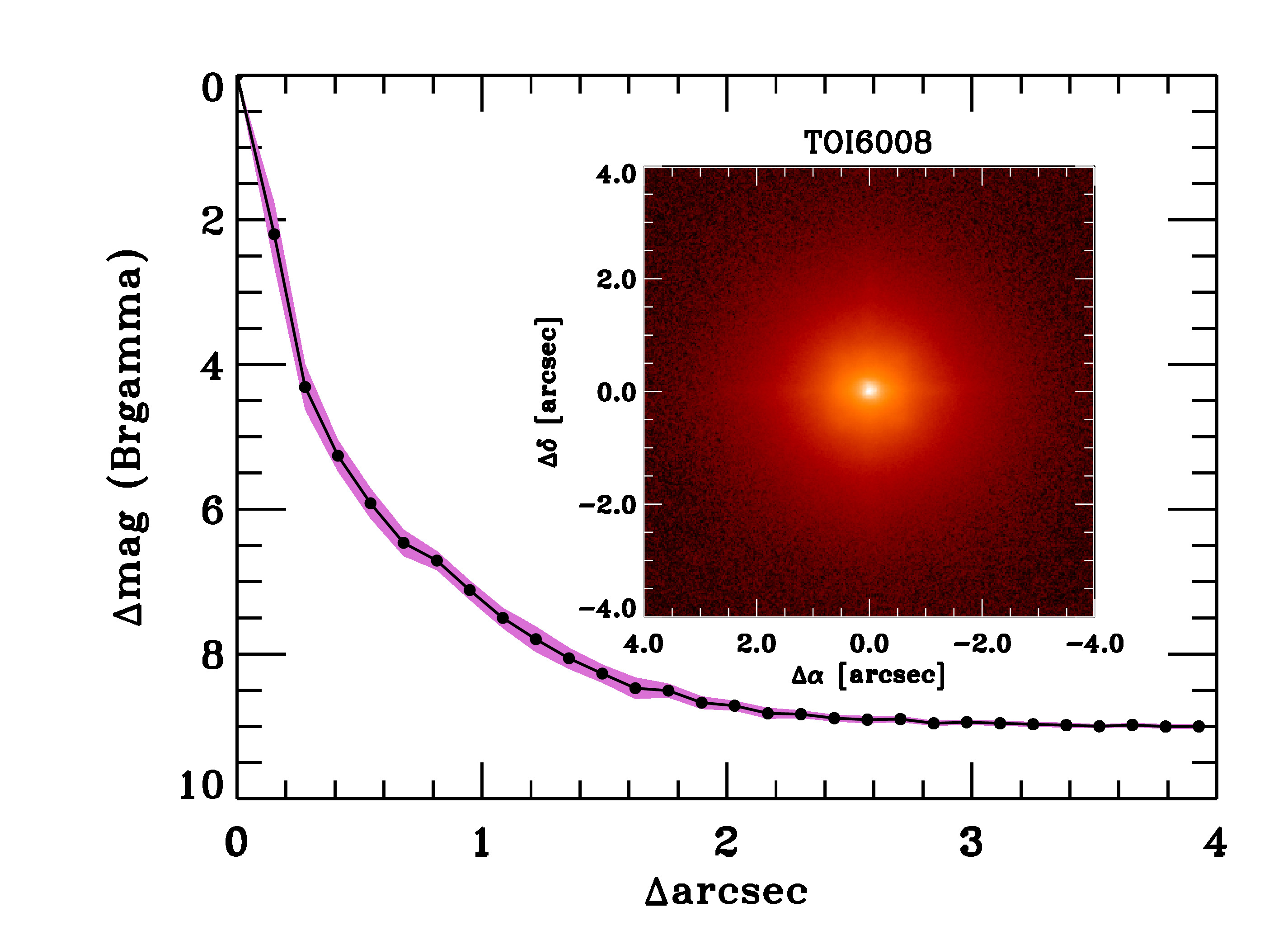}
    \includegraphics[width=\columnwidth, height=0.4\textheight, keepaspectratio]{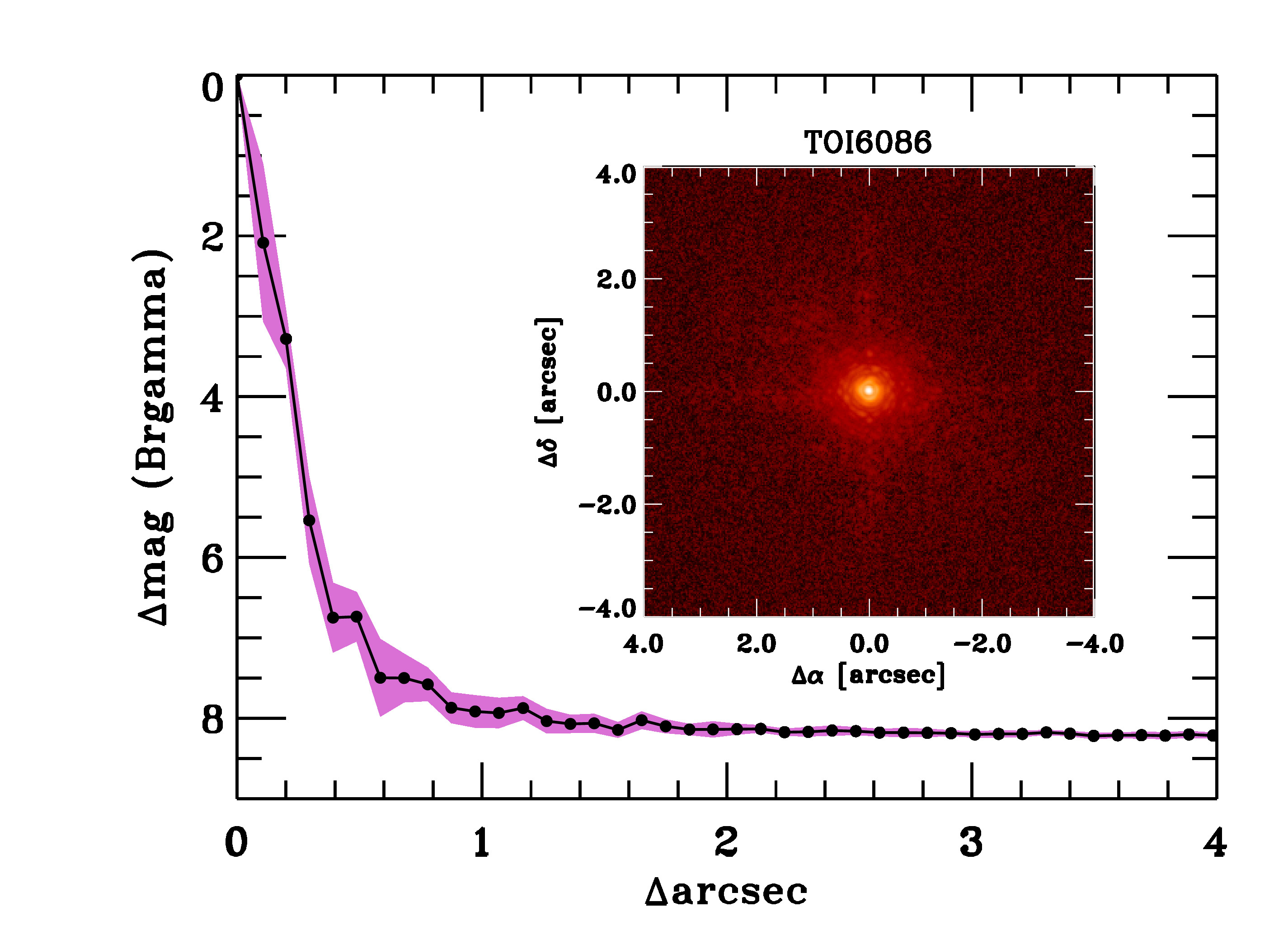}
    \caption{High resolution imaging of TOI-5720 (top panel), TOI-6008 (middle pane), and TOI-6086 (bottom panel) obtained with the Palomar/PHARO Speckle Imager on UT 2023 June 7 and 8. No close stellar companion sources were detected near the targets.}
    \label{tois_palomar_pharo}
\end{figure}

We observed TOI-5720, TOI-6008, and TOI-6086 with the PHARO instrument \citep{Hayward_2001} behind the natural guide star AO system P3K \citep{Dekany_2013} on UT 2023 June 7 and 8 in the narrow-band $Br-\gamma$ filter ($\lambda_0 = 2.1686; \Delta \lambda = 0.0326$~$\mu$m). PHARO is equipped with a 1024$\times$1024 HgCdTe HAWAII detector, with a pixel scale of 0.025~arcsec, resulting in a total FOV of 25~arcsec.
Each dither position was observed three times for a total of 15 frames  with an exposure time of 9.9s for TOI-5720, 2.8s for TOI-6008, and 9.9s for TOI-6086. The total on-source time was 148.5s for TOI-5720, 42.5s for TOI-6008, and 148.5s for TOI-6086. The science data were calibrated by sky-subtraction and flat-fielding. Further details of the observations from  Palomar-5m0/PHARO are available in \cite{Hayward_2001}. No companion sources were detected for TOI-5720, TOI-6008, and TOI-6086 in the PHARO data (see \autoref{tois_palomar_pharo}).


\section{Stellar characterization} \label{stellar_carac}

\begin{table*}
	{\renewcommand{\arraystretch}{1.34}
		\begin{tabular}{lcccc}
			\hline
			\hline
			\multicolumn{3}{c}{  Star information}   \\
			\hline
			\hline
			Parameter & TOI-5720 &  TOI-6008 & TOI-6086 & Source   \\
			\hline
			{\bf Identifying information:} & & \\
			TIC            & 230055368 & 286201103  & 18318288 & \\
			GAIA DR3 ID  & 3994047048130880384 & 2168280502430898944  &  4597447518944594176 &\\
			2MASS ID      & J11211567+2516255 & J20414744+4938482  &  J17412905+3018135 & \\
			&&&& \\
			{\bf Parallax  and distance:} &   \\
			RA [J2000]     &  11:21:15.26 & 20:41:47.70  & 17:41:28.90  & (1) \\
			Dec [J2000]    & +25:16:23.86 &  +49:38:48.84  & +30:18:13.59 & (1)\\
			Plx [$mas$] & $27.878 \pm 0.022$ &   $ 43.44 \pm 0.016$ & $ 31.7665 \pm 0.0142$ & (1)\\
            $\mu_{RA}$ [mas yr$^{-1}$] & $-350.0406 \pm 0.0233$ & $158.334 \pm 0.019$ & $-110.609 \pm 0.013$ & (1) \\
            $\mu_{Dec}$ [mas yr$^{-1}$] & $-91.5117 \pm 0.0248$ &  $44.342 \pm 0.0122$ & $2.127 \pm 0.016$ & (1) \\
			Distance [pc]  & $35.87 \pm 0.03$  &  $ 23.02 \pm 0.01$ & $31.47 \pm 0.014$ & (1)\\
			&&&& \\
			{\bf Photometric properties:} & \\
			TESS$_{\rm mag}$           & $11.619 \pm 0.007$ &  $12.294 \pm  0.008$  & $12.411 \pm 0.007$ & (2)  \\
			$V_{\rm mag}$ [UCAC4]       & $13.847 \pm 0.043$  & $15.97 \pm 0.20$   &  $14.98 \pm 0.06$ & (3) \\
			$B_{\rm mag}$ [UCAC4]       & $15.469 \pm 0.083$  &  $16.824$  & $16.78 \pm 0.06$ & (3) \\
			$J_{\rm mag}$ [2MASS]       & $10.123 \pm 0.022$ &   $10.416 \pm 0.024$  &  $10.82 \pm 0.02$ & (4) \\
			$H_{\rm mag}$ [2MASS]       & $9.485 \pm 0.019$ &  $9.830 \pm 0.030$  &  $10.25 \pm 0.02$ & (4) \\
			$K_{\rm mag}$ [2MASS]       & $9.288 \pm 0.015$ &  $9.541 \pm 0.018$  &  $9.99 \pm 0.02$ & (4) \\			
			$G_{\rm mag}$ [Gaia DR3]    &  $15.594 \pm 0.001$ & $13.712 \pm 0.001$  &  $13.797 \pm 0.001$ & (1)  \\
			$W1_{\rm mag}$ [WISE]       & $ 9.134 \pm 0.023 $ & $9.334 \pm 0.022$ &  $9.80 \pm 0.02$ & (5) \\
			$W2_{\rm mag}$ [WISE]       & $ 8.988 \pm 0.020 $ &  $9.183 \pm 0.02$  & $9.62 \pm 0.02$  & (5) \\
			$W3_{\rm mag}$ [WISE]       & $ 8.892 \pm 0.029 $ &  $9.234 \pm 0.053$  & $9.53 \pm 0.04$ & (5)\\
            $W4_{\rm mag}$ [WISE]       & $ 8.835 \pm 0.438 $ &  $9.128$  &  $8.87 \pm 0.46$ &(5)\\
			&&&& \\
			\multicolumn{2}{l}{\bf Spectroscopic and derived parameters}   \\
			Optical spectral type             &  M3.5~$\pm$~1.0 & M5.0  & M3.0 & this work\\
			Near-infrared spectral type       &  M3.5~$\pm$~1.0     & M5.0  & M3.5 & this work\\
			SED $T_{\rm eff}$ [K]             & $ 3325 \pm 75$     &  $ 3075 \pm 75$   & $ 3200 \pm 75$ & this work\\
			SED $\log g_\star$ [dex]          & $ 5.01 \pm 0.04$   &  $ 5.01 \pm 0.02$ & $ 5.01 \pm 0.02$ & this work\\
			Optical [Fe/H] [dex]              &  $+0.03 \pm 0.20$  & $ \gtrsim+0.3$    &  $ +0.11 \pm 0.20$ & this work\\
			Near-infrared [Fe/H] [dex]        & $-0.19 \pm 0.12$   & $ +0.21 \pm 0.12$ &  $ -0.20 \pm 0.12$ & this work\\
			SED [Fe/H] [dex]                  & $ 0.0 \pm 0.3$     & $ +0.3 \pm 0.3$   & $ -0.2 \pm 0.3$ & this work\\
			  $v\sin(i)$  [km/s]                &  $<3.4$    &   -   &   -    &    this work \\
            $M_\star$  [$M_\odot$]            & $ 0.383 \pm 0.019$  & $ 0.230 \pm 0.011$ & $ 0.254 \pm 0.013$ & this work\\
			$R_\star$  [$R_\odot$]            & $ 0.383 \pm 0.018$  & $ 0.242 \pm 0.013$ &  $ 0.259 \pm 0.013$ & this work\\
            $\rho_\star$ [$\rho_\odot$]       & $ 6.82 \pm 1.02 $   & $ 16.22 \pm 2.73 $ & $ 14.62 \pm 2.3 $   & this work \\
            $F_{\rm bol} \times 10^{-10}$  [erg s$^{-1}$ cm$^{-2}$] & $4.027 \pm 0.094 $  &  $2.839 \pm 0.101 $  &  $2.051 \pm 0.072 $  &   this work\\ 
			$L_{\rm bol}$  [L$_\odot$]        & $ 0.01615 \pm 0.00019$  & $0.004691 \pm 0.000082$ &   $0.00634 \pm 0.00011$ &  this work\\
			$Av$ [mag]    & $0.00 \pm 0.02$   & $0.00 \pm 0.02$   &  $0.00 \pm 0.02$ &  this work\\
			Age (H$\alpha$)  [Gyr]            &  $\gtrsim$5  &  $\gtrsim$3 &  $\gtrsim$2 &  this work\\
			\hline
	\end{tabular} }
	\caption{Spectroscopy, astrometry and photometry stellar properties of the host stars TOI-5720, TOI-6008, and TOI-6086. 
	{\bf (1)} Gaia EDR3; \cite{Gaia_Collaboration_2021AandA}; 
	{\bf (2)} \emph{TESS} Input Catalog; \cite{Stassun_2018AJ_TESS_Catalog}; 
	{\bf (3)} UCAC4; \cite{Zacharias_2012yCat.1322};
	{\bf (4)} 2MASS; \cite{Skrutskie_2006AJ_2MASS};
	{\bf (5)} WISE; \cite{Cutri_2014yCat.2328}. 
	\label{stellarpar}}
\end{table*}

\begin{table*}
	\begin{center}
		{\renewcommand{\arraystretch}{1.2}
				\resizebox{0.93\textwidth}{!}{
			\begin{tabular}{llllc}
				\hline
			        Parameter &  TOI-5720\,b  & TOI-6008\,b  &   TOI-6086\,b  \\
           \hline
           \multicolumn{2}{l}{\it Planet/star area ratio}  \\
				\hline
            $(R_p/R_\star)^2_{\rm TESS}$ [ppm]      &  $ 725^{+188}_{-186} $ &   $ 1517^{-120}_{+122} $ &   $ 1531^{-370}_{+376} $  \\
            $(R_p/R_\star)^2_{{\rm Sloan}-z'}$ [ppm] &  $ 989^{+156}_{-155} $ & $ 1431^{-150}_{+151} $   &   $ 1783^{-261}_{+269} $     \\
			$(R_p/R_\star)^2_{{\rm Sloan}-i'}$ [ppm] &  $ 723^{+193}_{-190}$  & $ 1591^{-186}_{+188} $   & $ 1969^{-275}_{+286} $       \\
            $(R_p/R_\star)^2_{{\rm Sloan}-r'}$ [ppm] &  $635^{+218}_{-223}$   & $ 1420^{-292}_{+301} $   &  $ 1639^{-401}_{+405} $        \\
			$(R_p/R_\star)^2_{{\rm Sloan}-g'}$ [ppm] &  $724^{+188}_{-186}$   &  $ 970^{-610}_{+746} $   &  $ 2951^{-1260}_{+1902} $    \\
            \hline
            Orbital period $P$ [days]             &  $1.4344555 \pm 0.0000036$  & $0.8574347 \pm 0.0000424$ &  $ 1.3888725 \pm 0.0000827$ \\ 
            Transit-timing $T_0$  &  $ 10076.902501 \pm 0.00067$   &   $10078.016214 \pm 0.000311$  & $10131.969501 \pm 0.000567$    \\
            $[BJD_{\rm TDB} - 2450000]$   &  &&   \\
            Orbital semi-major axis $a$ [AU]      &  $0.018288 \pm 0.000279$   &  $ 0.010777 \pm 0.000162$  &  $ 0.01541 \pm 0.00024$    \\
            Impact parameter $b$ [$R_\star$]      &   $0.14 \pm 0.09 $   &  $0.45 \pm 0.08 $   &    $0.608 \pm 0.051$   \\
			Transit duration $W$ [min]            &   $ 61 \pm 2$    & $ 39 \pm 1$ &   $ 42 \pm 2$   \\
			Scaled semi-major axis  $a/R_\star$          &  $10.97 \pm 0.29$   &  $ 9.44 \pm 0.21$    &  $ 12.75 \pm 0.26$   \\
			Orbital inclination $i$ [deg]         &  $89.3 \pm 0.6$   &   $ 87.2 \pm 0.5$     &  $ 87.3 \pm 0.3$  \\
			Radius ratio $R_\oplus /R_\star $ &  $ 0.02796 \pm 0.00158$ &  $ 0.03867 \pm 0.00109$ & $ 0.04173 \pm 0.00220 $\\
            Radius $R_p$ [$R_\oplus $]              &  $1.09 \pm 0.07$   &  $ 1.03 \pm 0.05$    &   $ 1.18 \pm 0.07$  \\
			Irradiation $S_p$ [$S_\oplus$]          &  $41.7 \pm 4.5$   &   $ 41.5 \pm 4.5$     &    $ 26.8 \pm 2.7$ \\
            $^a$ Equilibrium temperature $T_{\rm eq}$ [K]       &  $708 \pm 19$   &    $ 707 \pm 19$     &      $ 634 \pm 16 $     \\
			  $^b$ Mass $M_p$ [$M_\oplus$ ]      &  $ <4.3$   &    $1.1^{+3.3}_{-0.9}$       &   -    \\
			$^b$ RV semi-amplitude $K$ [$m/s$]     &  $<64$   &  $2.1^{+5.9}_{-1.8}$  &  -  \\ 
            $^c$ Predicted mass $M_p$ [$M_\oplus$ ]      &  $ 1.32^{+0.92}_{-0.47}$   &   $1.10^{+0.73}_{-0.36}$     &     $1.68^{+1.15}_{-0.56}$  \\
            $^c$ Predicted RV  semi-amplitude $K$ [$m/s$]      &  $ 1.4^{+1.0}_{-0.5}$  &   $2.0^{+1.3}_{-0.6}$     &     $2.4^{+1.6}_{-0.8}$ \\
            $^d$ Predicted TSM & $9.7^{+7.1}_{-4.1}$ & $18.8 ^{+12.6}_{-6.8}$  & $12.0 ^{+8.5}_{-4.7}$  \\
            $^d$ Predicted ESM & $3.4 \pm 0.5 $ & 6.4$\pm$0.7  & 4.2$\pm$0.7  \\
   \hline
		\end{tabular}}}
	\end{center}
	\caption{Planetary physical parameters determined from our global MCMC analysis  (medians and $1\sigma$ error bar) for TOI-5720\,b, TOI-6008\,b, and TOI-6086\,b. $^a$  The equilibrium temperature derived assuming a null albedo. $^b$ Planetary mass and RV semi-amplitude values derived from RVs measurements collected by Subaru/IRD and TrES spectrographs for TOI-6008\,b, and TOI-5720\,b, respectively. $^c$ Predicted mass and RV semi-amplitude values calculated using the mass-radius relationship from \cite{Chen_Kipping_2017ApJ}. $^d$ TSM (Transmission spectroscopy metric) and ESM (emission spectroscopy metric) values calculated from \cite{kem}.}
	\label{tois_mcmc_params}
\end{table*}

\begin{table*}
	\begin{center}
		{\renewcommand{\arraystretch}{1.2}
				\resizebox{0.93\textwidth}{!}{
			\begin{tabular}{llccc}
				\hline
				Parameter &  TOI-5720  & TOI-6008  &   TOI-6086  \\
				          &        &   & \\
				\hline
				Mean density,  $\rho_\star$ [$\rho_\odot$]   &  $8.64^{+0.77}_{-0.67} $  &  $15.41^{+1.02}_{-0.95} $   &      $14.41^{+0.96}_{-0.89} $     \\
				Stellar mass, $M_\star$ [$M_\odot$ ]      & $0.397 ^{+0.017}_{-0.018}$   &  $0.227^{+0.010}_{-0.011} $    & $0.253^{+0.012}_{-0.012} $           \\
				Stellar radius,  $R_\star$ [$R_\odot$]      & $0.358^{+0.011}_{-0.011}$   &  $0.245^{+0.007}_{-0.007} $  & $0.260^{+0.007}_{-0.008} $     \\
				Luminosity, $L_\star$ [$L_\odot$]       & $0.01402 _{-0.00145}^{+0.00162}$  & $0.00480^{+0.00058}_{-0.00053} $  & $0.00639^{+0.00077}_{-0.00068} $    \\
                Effective temperature, $T_{\rm eff}$ [K] & $ 3324^{+73}_{-76} $ & $3073^{+75}_{-76} $  & $3205^{+76}_{-77} $ \\ 
                \hline
                \multicolumn{2}{l}{\it Quadratic Limb-Darkening coefficients}  \\
                \hline
			    $u_{\rm 1,TESS}$ &  $0.26 \pm 0.01$   &  $0.41 \pm 0.03$  & $0.21 \pm 0.02$ \\
				$u_{\rm 2,TESS}$ &  $0.39 \pm 0.01$  &   $0.14 \pm 0.01$  & $0.42 \pm 0.01$\\
				$u_{\rm 1,{\rm Pan-STARRS-z_s}}$ &  $0.16 \pm 0.01$   &  $0.34 \pm 0.02$  & $0.15 \pm 0.01$ \\
				$u_{\rm 2,{\rm Pan-STARRS-z_s}}$ &  $0.42 \pm 0.01$   &  $0.13 \pm 0.01$  & $0.43 \pm 0.01$\\
				$u_{\rm 1,{\rm Sloan}-i'}$ &  $0.30 \pm 0.01$   &  $0.42 \pm 0.02$  & $0.29 \pm 0.01$ \\
				$u_{\rm 2,{\rm Sloan}-i'}$ &  $0.35 \pm 0.01$   &  $0.22 \pm 0.01$  & $0.36 \pm 0.01$\\
				$u_{\rm 1,{\rm Sloan}-r'}$ &  $0.42 \pm 0.01$  &   $0.57 \pm 0.02$  & $0.40 \pm 0.01$\\
				$u_{\rm 2,{\rm Sloan}-r'}$ &  $0.34 \pm 0.01$   &  $0.26 \pm 0.01$  & $0.36 \pm 0.01$\\
                $u_{\rm 1,{\rm Sloan}-g'}$ &  $0.41 \pm 0.01$  &  $0.61 \pm 0.02$   & $0.39 \pm 0.02$ \\
				$u_{\rm 2,{\rm Sloan}-g'}$ &  $0.37 \pm 0.01$   &  $0.26 \pm 0.01$  & $0.40 \pm 0.04$  \\
				\hline
		\end{tabular}}}
	\end{center}
	\caption{Derived stellar parameters ($\rho_\star$, $M_\star$, $R_\star$, $L_\star$, and $T_{\rm eff}$) and limb-darkening coefficients for TOI-5720, TOI-6008, and TOI-6086 based on our global analysis.}
	\label{tois_LDs_param}
\end{table*}

\subsection{Shane/Kast spectroscopic analysis}

The Shane/Kast optical spectra of TOI-5720, TOI-6008, and TOI-6086 are shown in \autoref{fig_Shane_kast}.
We compared these spectra to the SDSS M dwarf templates from \citet{2007AJ....133..531B} and found best matches to M4, M5, and M3 templates, respectively. The index-based classification relations of \citet{1997AJ....113..806G} and \citet{2003AJ....125.1598L} confirm the last two classifications, but yield a significantly earlier type of M2.5 for TOI-5720, consistent with its near-infrared classification (see next section).
As neither the M2 nor M3 templates provide superior fits to the optical spectrum of this source, we assign and uncertain optical classification to this source, warranting further follow-up. 
None of these spectra shows evidence of H~\textsc{i} emission in the H$\alpha$ (6563~{\AA}), H$\beta$ (4861~{\AA}), or H$\gamma$ (4340~{\AA}) lines, 
with H$\alpha$ 3$\sigma$ equivalent width limits of 0.06~{\AA}), 0.12~{\AA}, and 0.18~{\AA}.
The lack of magnetic emission indicates ages greater than $\sim$3~Gyr, $\sim$5~Gyr, and $\sim$2~Gyr for the three sources \citep{2008AJ....135..785W}. 
To estimate the metallicity, we measured the $\zeta$ index based on TiO and CaH features \citep{2007ApJ...669.1235L,2013AJ....145...52M}, finding $\zeta$ = 1.02$\pm$0.01, 1.25$\pm$0.01, and 1.08$\pm$0.01 for TOI-5720, TOI-6008, and TOI-6086, respectively. 
For TOI-5720 and TOI-6086, these values indicate near-solar metallicities, [Fe/H] = 0.03$\pm$0.20 and [Fe/H] = 0.11$\pm$0.20 based on the calibration of \citet{2013AJ....145...52M}. The considerably higher $\zeta$ value for TOI-6008, outside the applicable range of the \citet{2013AJ....145...52M} relations suggests a significantly supersolar metallicity [Fe/H]~$\gtrsim$~0.3 (see \autoref{stellarpar}).

\subsection{IRTF/SpeX spectroscopic analysis}
The SpeX SXD spectra of TOI-5720, TOI-6008, and TOI-6086 are shown in \autoref{fig_IRTF_spex}. Using the SpeX prism library analysis toolkit \citep[SPLAT,][]{splat}, we assigned spectral types by comparing the spectra to those of single-star spectral standards in the IRTF Spectral Library \citep{Cushing2005, Rayner2009}.
Based upon the best matches, we adopt spectral types of M2.5 $\pm$ 0.5, M5.0 $\pm$ 0.5, and M3.5 $\pm$ 0.5 for TOI-5720, TOI-6008, and TOI-6086, respectively.
We also used SPLAT to measure the equivalent widths of the $K$-band Na\,\textsc{i} and Ca\,\textsc{i} doublets and the H2O--K2 index \citep{Rojas-Ayala2012}.
Using the \citet{Mann2013} relation between these observables and stellar metallicity, and propagating uncertainties using a Monte Carlo approach (see \citealt{Delrez2022}), we estimate iron abundances of $\mathrm{[Fe/H]} = -0.19 \pm 0.12$ for TOI-5720, $\mathrm{[Fe/H]} = +0.21 \pm 0.12$ for TOI-6008, and $\mathrm{[Fe/H]} = -0.20 \pm 0.12$ for TOI-6086 (see \autoref{stellarpar}).

\subsection{Spectral energy distribution analysis and evolutionary models} \label{SED_evol_analys}
\begin{figure*}[!ht]
	\centering
	\includegraphics[scale=0.34]{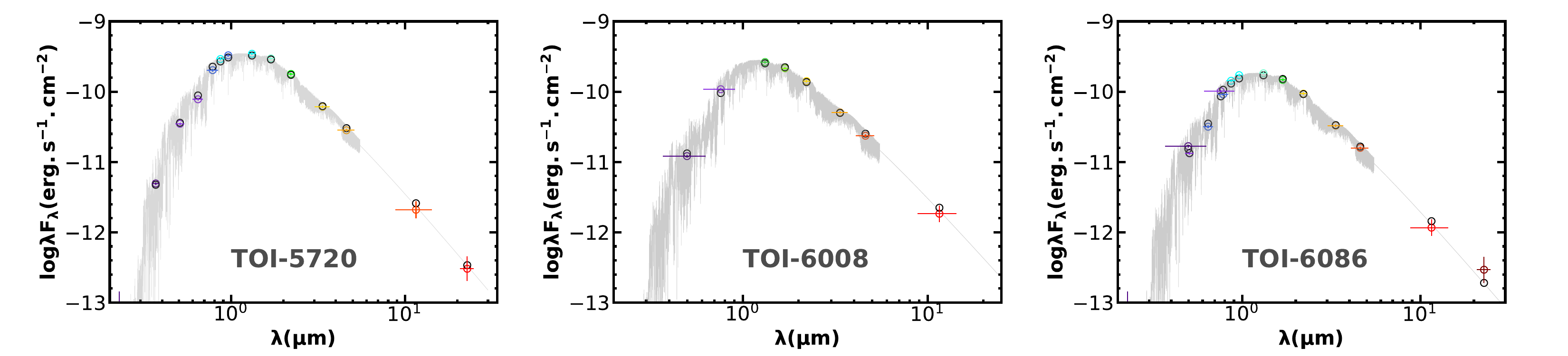}
	\caption{Spectral energy distribution (SED) fit of TOI-5720 (left), TOI-6008 (middle) and TOI-6086 (right). The gray curves are the best-fitting PHOENIX atmosphere models, the colored symbols with error bars are the observed fluxes (horizontal bars represent the effective bandpasses), and black symbols are the model fluxes.}
	\label{SED_plots}
\end{figure*}

 We conducted an independent analysis of the broadband  spectral energy distribution (SED) of the stars together with the {\it Gaia\/} DR3 parallax \citep[with no systematic offset applied; see, e.g.,][]{StassunTorres:2021} to derive an empirical measurement of the stellar radii,  following the same method as described in \citet{Stassun:2016,Stassun:2017,Stassun:2018}. We pulled the  {\it 2MASS} $JHK_S$ magnitudes,  the {\it WISE} W1--W4 magnitudes,  the {\it Pan-STARRS} $grizy$ magnitudes, and the {\it Gaia} $G_{\rm BP} G_{\rm RP}$ magnitudes. Together, the available photometry spans the full stellar SED over the wavelength range 0.4--20~$\mu$m (see \autoref{SED_plots}).  

 We performed a fit using PHOENIX stellar atmosphere models \citep{Husser:2013}, with the free parameters being the metallicity ([Fe/H]) and effective temperature ($T_{\rm eff}$). We set null extinction  due to the close proximity of the stars to Earth. The resulting fits (\autoref{SED_plots}) have a best-fit $T_{\rm eff} = 3325 \pm 75$~K, $T_{\rm eff} = 3075 \pm 75$~K, and $T_{\rm eff} = 3200 \pm 75$~K for TOI-5720, TOI-6008, and TOI-6086, respectively, and a best-fit [Fe/H] = $0.0 \pm 0.3$, $+0.3 \pm 0.3$, and $-0.2 \pm 0.3$ respectively, with a reduced $\chi^2$ of 1.8, 1.9, and 2.4 respectively (the latter star exhibits some UV excess that may indicate chromospheric activity). Integrating the model SED gives the bolometric flux at Earth, $F_{\rm bol} = 4.027 \pm 0.094 \times 10^{-10}$ erg~s$^{-1}$~cm$^{-2}$,  $F_{\rm bol} = 2.839 \pm 0.101 \times 10^{-10}$ erg~s$^{-1}$~cm$^{-2}$, and $F_{\rm bol} = 2.051 \pm 0.072 \times 10^{-10}$ erg~s$^{-1}$~cm$^{-2}$, respectively. Taking the $F_{\rm bol}$ together with the {\it Gaia\/} parallax directly gives the bolometric luminosity  as $L_{\rm bol} = 0.01615 \pm 0.00019$~L$_\odot$, $L_{\rm bol} = 0.004691 \pm 0.000082$~L$_\odot$, and $L_{\rm bol} = 0.00634 \pm 0.00011$~L$_\odot$ respectively. 
The stellar radii then follow from the Stefan-Boltzmann relation as $R_\star = 0.383 \pm 0.018$~R$_\odot$, $R_\star = 0.242 \pm 0.013$~R$_\odot$, and $R_\star = 0.259 \pm 0.013$~R$_\odot$, respectively. Finally, we can estimate the stellar mass from the empirical  mass-luminosity relation of \citet{Mann:2019}, giving $M_\star = 0.383 \pm 0.019$~M$_\odot$, $M_\star = 0.230 \pm 0.011$~M$_\odot$, and $M_\star = 0.254 \pm 0.013$~M$_\odot$, respectively.

\section{Planet validation} \label{Validate_planet}

\subsection{Data validation report for \emph{TESS}}

As we described in Section~\ref{tess_phot}, the SPOC pipeline extracted the photometric measurements  for TOI-5720, TOI-6008 and TOI-6086 using a cadence of 2-minutes, yielding orbital periods of  $P=1.43446$~days, $P=0.85737$~day, and $P=1.38886$~days, and signal-to-noise ratios of  $S/N = 7.7$, $S/N = 14.9$, and $S/N = 8.1$, respectively.
The TOI vetting team reviewed the SPOC data validation reports \citep{jenkins2002,jenkins2010,Jenkins_2020TPSkdph} on  May 26 2022, Sept 16 2022, and Sept 15 2022 \citep{guerrero2021} for TOI-5720, TOI-6008, and TOI-6086, respectively. 
The  transit depths were found to be $1.03 \pm 0.16$~ppt (for TOI-5720.01), $2.05 \pm 0.27$~ppt (for TOI-6008.01), and $1.37 \pm 0.24$~ppt (for TOI-6086.01), corresponding to a planet radius of $R_p = 1.4 \pm 0.3\pm$~R$_\oplus$, $R_p=1.2 \pm 0.2\pm$~R$_\oplus$, and $1.0 \pm 1.7\pm$~R$_\oplus$ for TOI-5720.01, TOI-6008.01, and TOI-6086.01, respectively. 
All additional validation tests, including bootstrap, centroid offset, ghost, difference-imaging centroid tests, and a search for discrepancies between odd and even transit depths were successfully passed. 
These conclusive tests prompted us to conduct further high-precision photometric follow-up and analysis to validate the planet candidates.

\subsection{Vetting}
While the transit method is one of the most fruitful technique for discovering exoplanets, several scenarios other than a transiting exoplanet can produce periodic dips in the brightness of a star. Numerous false-positive scenarios exist in astronomical observations: {\it i}) stellar binary systems, where two stars sharing a common center might create a signal resembling a planetary transit when their sizes differ or when their eclipse is partial; {\it ii}) blended binary stars, wherein the target blends the light of an eclipsing binary obscuring the secondary eclipse and confusing the primary with a planetary transit; and {\it iii}) other misleading signals from astrophysical sources (stellar spots, pulsations, and rotation) or instrument-related factors (jitter noise, and momentum dumps). 

Thus, the first essential step toward the confirmation of a transiting exoplanetary signal is the vetting. In addition to the SPOC DV reports for TOI-5720.01, TOI-6008.01, and TOI-6086.01, we conducted a uniform vetting analysis through the tool called discovery and vetting of exoplanets (\texttt{DAVE}; \citealt{Kostov2019}), which allowed us to examine a transit event on a double-level analysis: the pixel level by a photocenter analysis, and the light-curve level by a flux time-series analysis. 
The \texttt{centroid} module generates a difference image by subtracting the overall in-transit image from the out-of-transit image. It determines the photocenter by fitting the TESS pixel response function to the image, which is performed for each detected transit. The average centroid position of the events is calculated, along with its statistical significance. This aids in the vetting process. An offset in the centroid position from the expected target star position flags the event as a false positive (FP). However, the challenging interpretation due to artifacts or low S/N in DAVE-produced difference images makes the photocenter analysis unreliable. In these cases, we classify the signal as a planet candidate. The \texttt{Modelshift} module produces a phase-folded light curve with the best-fit trapezoid transit model. Its primary goal is to determine whether the signal source is an eclipsing binary system. It showcases key transit features, including the average primary signal and other significant features. Additionally, it compares odd and even transits by measuring their statistical difference. It also evaluates the transit shape, offering a comprehensive assessment of the observed transit properties.

In the vetting analysis of planetary transit candidates, the reliability can be compromised when the S/N of the event is particularly low. This occurs because low S/N signals may be indistinguishable from noise or other sources of variability in the data, making it challenging to accurately identify genuine transit events. As a result, the vetting efforts may be less effective in confirming or ruling out potential candidates for these cases. TOI-5720.01, TOI-6008.01, and TOI-6086.01 are three low S/N planet candidates, with the first and the third being near the detection threshold (S/N=$7.1$, \citealt{Borucki2011b}). However, their transit depths in the phase-folded light curve are slightly above the noise level. Moreover, we did not flag any particular anomalies in the shape of the transit, nor did statistical significant evidence arise between the odd and even transits. 
In the \texttt{centroid} module is concerned, TOI-5720.01 passed the photocenter test without any concern. The same test was inconclusive for both TOI-6008.01 and TOI-6086.01 since the TESS aperture masks used to extract their light curves are crowded with contaminating stars, resulting in an unreliable overall centroid measurement. However, for a few single clean centroid measurements the light photocenters are consistent with the position of both TOI-6008 and TOI-6086, as we  discuss in Sect.\ref{sec:ancillary_info}.

\subsection{Ancillary information}
\label{sec:ancillary_info}
Since the three planet candidates have a low S/N, we conducted  analysis in addition to those provided by \texttt{DAVE} to make our vetting process more solid.
TESS has a large pixel scale, about $21''$/pixel, with a focus-limited PSF. In order to check whether any resolved or unresolved sources fell within the aperture mask, thus contaminating the photometric signal, we consulted stellar catalogs \citep{Wenger2000, GaiaCollaboration2021}. While the \texttt{centroids} module should shed light on the contamination flux due to resolved nearby stars, the same cannot be said for the unresolved sources within the same pixel of the star target. At this level, we can only flag the presence of a potentially contaminating source that dilutes the transit depth. The worst scenario is when the contaminating source is the origin of the transit event. We here stress that this is an intrinsic limit of this analysis that can be overcome by follow-up photometric campaigns (see Sect. \ref{follow_photometric}).
    
TOI-5720 has only a very faint (TESS magnitude $=18.162$) nearby target, TIC 903650250, which is $\sim 55$ arcsec apart and seven magnitudes dimmer.
    
On the other hand, TOI-6008 is located in a region of the sky that is overdense in stars ($82$ known targets in $1$ arcmin). In particular, $5$ stars fall within the same pixel of TOI-6008 even if the brightest star (TIC~286201079) is almost $5$ magnitudes fainter than TOI-6008. Moreover, the brightest star is TIC 286201082 (TESS magnitude $=11.57$), which is $41$ arcsec ($\sim 2$ pixels) away from our target and $\sim 1$ magnitude brighter than it. Thus, the centroid measurements are triggered by the photometric variability of TIC~286201082. Another star at a distance of $42$~arcsec, TIC~286201047 (TESS magnitude $=13$), contributes to make the photocenter measurements challenging. Even though most centroid measurements are unreliable, a few are not altered by nearby sources. For these, the centroid is aligned with the position of TOI-6008.
    
The photometry obtained from TOI-6086 observations is contaminated by the star light of TIC 18318284 (TESS magnitude $=13.25$), which is only one magnitude fainter than TOI-6086 and lies at a distance of $46$~arcsec. Moreover, the photocenter measurements are completely compromised due to the presence of a $120$~arcsec star (Gaia DR3 4597447931261454080) that is 2 magnitudes brighter than TOI-6086. In this case as well, a few clean centroid measurements do not show any flag for an offset.
    
We also decided to conduct a pixel level light-curve (PLL) analysis using the publicly available \texttt{LATTE}\footnote{LATTE: \url{https://github.com/noraeisner/LATTE}} pipeline developed by \cite{Eisner2022}. This test allowed us to inspect the light curve for each pixel in the field of view of the corresponding target pixel file to check whether the transit occurs in nearby pixels of the mask. None of the pixels within the aperture mask other than the pixel corresponding to TOI-6008 and TOI-6086 shows a reliable dip in the starlight. We repeated the same procedure for all the transits observed by TESS with the same conclusions.


Although the orbital path of TESS has been designed to maximize the sky coverage while reducing the number of obstructions, the \emph{TESS} full frame images can occasionally be contaminated primarily by the zodiacal light and scattered light from Solar System objects \citep{Gangestad2013,Sullivan2015}. Thus, during the observation of each TESS sector, the flux of the background can vary widely. If any unusual events in the background occurred at the transit time of the planet candidate, they could introduce stilted signals into the light curve, contaminating or, even worse, triggering the detected transit. In order to ensure that no anomaly contaminated the signal, we visually inspected a three-day-long section of the background flux centered at the time of transit. We found no evidence of background artifacts for TOI-5720.01, TOI-6008.01, and TOI-6086.01.

We found no known TESS momentum dumps in correspondence of the transits of TOI-5720.01, TOI-6008.01, and TOI-6086.01. We therefore rule out any possible instrumental systematic scenario.
    

\subsection{Archival imaging}
\label{sec:archival_imaging}

We used archival images of TOI-5720, TOI-6008, and TOI-6086  to exclude any background-unresolved stellar companion that could be blended with targets at these current positions. Objects like this might introduce the transits signals that we detected in our observations, and skew the results that we obtained from the global analysis. 

\begin{figure*}[!h]
	\centering
	\includegraphics[scale=0.25]{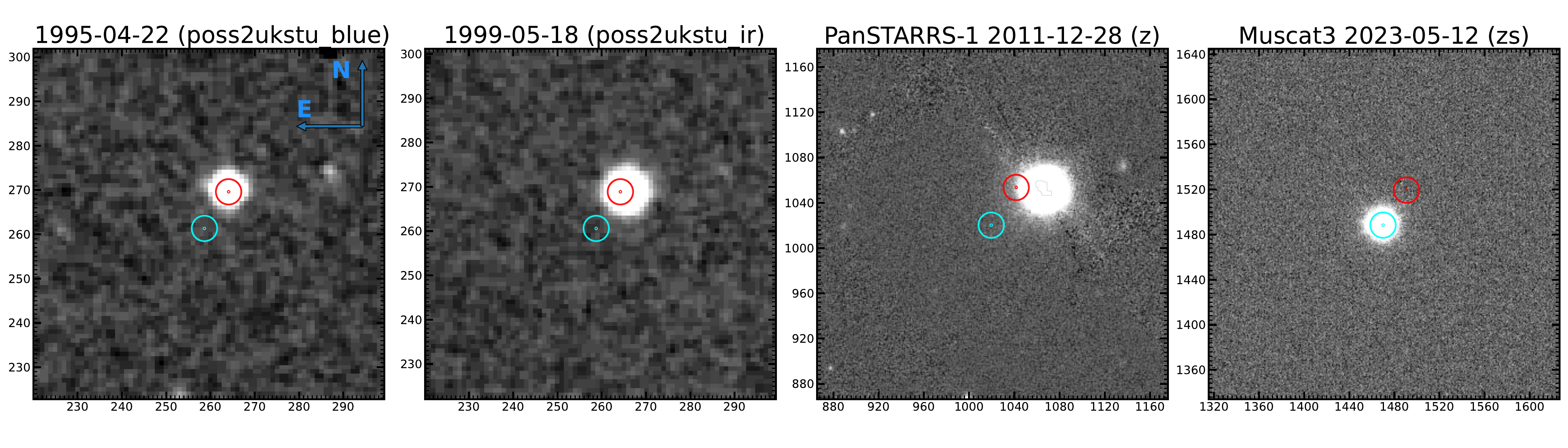}
    \includegraphics[scale=0.25]{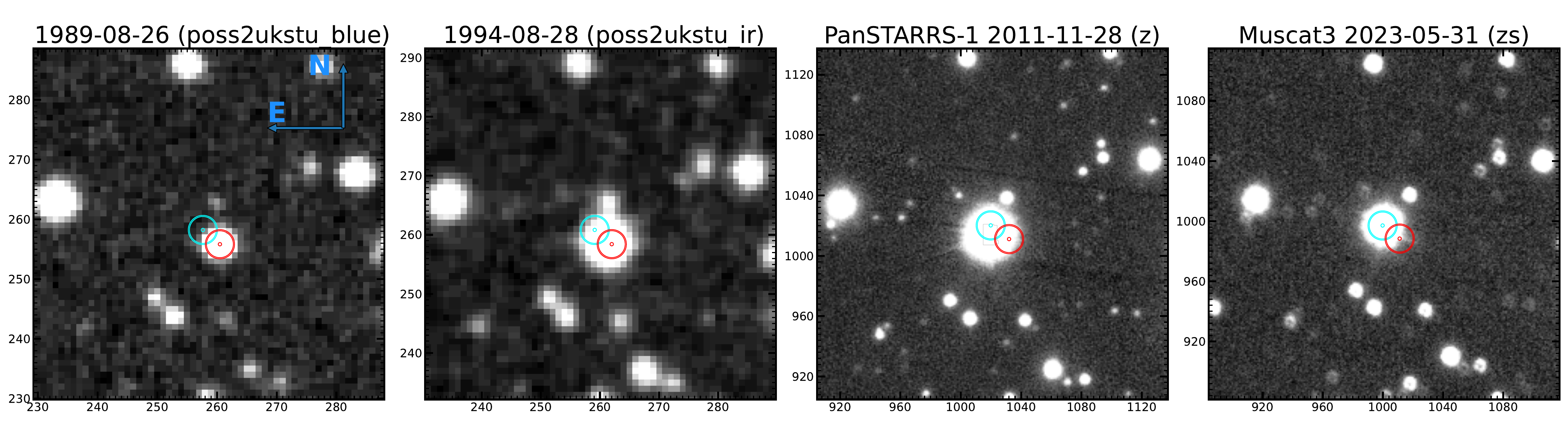}
    \includegraphics[scale=0.25]{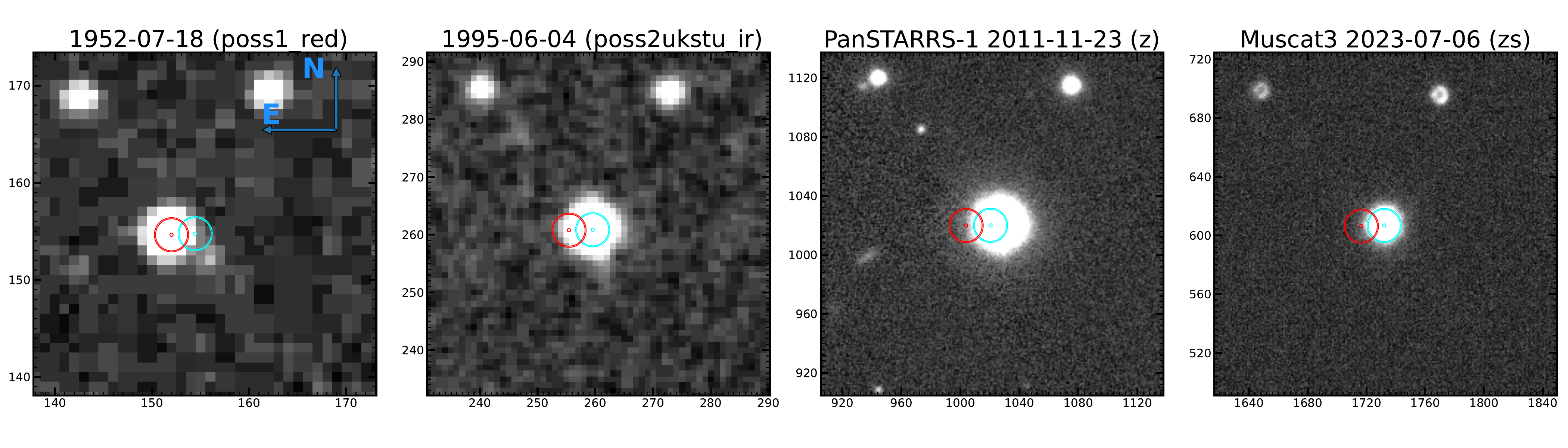}
	\caption{Archive images of TOI-5720 (top panel), TOI-6008 (middle panel), and TOI-6086 (bottom panel). The previous and current positions of the targets are shown in red and cyan, respectively. 
    {\it Top row (TOI-5720):} 1995 blue image from POSS-II/DSS, 1999 infrared image from POSS-II/DSS2, 2011 $z_s$ image from PanSTARRS1, and 2023 $z_s$ image from MuSCAT3 (shown from left to right). 
    {\it Middle row (TOI-6008):} 1989 blue image from POSS-II/DSS, 1994 infrared image from POSS-II/DSS2, 2011 $z_s$ image from PanSTARRS1, and 2023 $z_s$ image from MuSCAT3 (shown from left to right).
    {\it Bottom row (TOI-6086):} 1952 red image from POSS-I/DSS, 1995 infrared image from POSS-II/DSS2, 2011 $z_s$ image from PanSTARRS1, and 2023 $z_s$ image from MuSCAT3 (shown from left to right).} 
	\label{Archive_images}
\end{figure*}

TOI-5720 has a high proper motion of 350.04~$mas/yr$. We obtained images from POSS-II/DSS \citep{1963POSS-I} in 1995, POSS-II/DSS \citep{1996DSS_POSS-II} in 1999, and PanSTARRS-1 \citep{2016_Pan-STARRS1} in 2011 in the blue, infrared, and $z_s$ bands, respectively, and spanning 28 years with our current observations. 
TOI-6008 has a relatively low proper motion of 158.3~$mas/yr$. We obtained images from POSS-II/DSS (in 1989 in blue band), POSS-II/DSS (in 1994 in the infrared band), and PanSTARRS-1 (in 2011 in the $z_s$ band), and spanning 34 years with our current observations.
TOI-6086 also has a relatively low proper motion of 110.6~$mas/yr$. We also obtained images from POSS-I/DSS (in 1952 in red band), POSS-II/DSS (in 1995 in the infrared band), and PanSTARRS-1 (in 2011 in the $z_s$ band), and spanning 71 years with our current observations. 
TOI-5720, TOI-6008, and TOI-6086 have only moved  by $\sim 9.8$, $\sim 5.4$, and $\sim 7.8$~arcsec between 1995 and 2023, 1989 and 2023, and 1952 and 2023, respectively. There is no source in the current day positions of TOI-5720, TOI-6008, and TOI-6086 in any of these archival images (see \autoref{Archive_images}).

\subsection{Statistical validation}
We also employed the \texttt{TRICERATOPS} pipeline \citep{Giacalone2021} to assess the likelihood of false-positive scenarios. \texttt{TRICERATOPS} calculates the probability that a light curve originates from a transiting planet or from a wide range of other scenarios in a Bayesian framework. In particular, it returns the false-positive probability (FPP), which is the probability that the observed transit is not caused by a planet (e.g., an eclipsing binary), and the nearby false-positive probability (NFPP), which is the probability that the signal comes from a nearby resolved star. Moreover, \texttt{TRICERATOPS} also allows us to implement, if available, the high-contrast imaging observations in order to better constrain the calculation. Typically, a planet candidate is considered statistically validated when it returns an FPP$<0.01$ and a NFPP $<0.001$ \citep{Morton2012, Morton2015, Giacalone2021}. We ran \texttt{TRICERATOPS} using the light curves of TOI-5720.01, TOI-6008.01, and TOI-6086.01 phase-folded on the orbital periods obtained by the photometric modelling, as we will show in Sect.\ref{Global_modelling}. We constrained the overall calculation by implementing the high-contrast imaging observations mentioned in Sect~\ref{sec:archival_imaging}. Using ground-based observations (see Sect.\ref{follow_photometric}), we detected the source of a transit signal on-target, and thus we excluded other nearby sources, which clearly do not contaminate the transit signal, from our FPP calculations. We therefore obtained NFPP $= 0$ for the three candidates.
We obtained FPP $=(9.6 \pm 0.4)\times 10^{-3}$ for the signal associated with TOI-5720.01,  FPP $=(0.2 \pm 0.05)\times 10^{-3}$ for TOI-6008.01, and  FPP $=(5.00 \pm 0.05)\times 10^{-3}$ for TOI-6086.01.

\subsection{Ground-based follow-up photometric validation} \label{follow_photometric}
The ground-based photometric follow-up  was carried out on the basis of two objectives: {\it i}) identify the source of the transit event, and {\it ii})  chromaticity check for the transit depth in different wavelengths. 
The closest neighboring star to TOI-5720 is TIC~903650250 at $54.7\arcsec$ with a $T_{\rm mag}$ of 18.16, and a $\Delta T_{\rm mag}$ of 6.56. 
The closest neighboring stars to TOI-6008 are  TIC~286201110 at $7.3\arcsec$ with a $T_{\rm mag}$ of 17.32 and a $\Delta T_{\rm mag}$ of 5.02, TIC~286201086 at $12.4\arcsec$ with a $T_{\rm mag}$ of 17.43 and a $\Delta T_{\rm mag}$ of 5.13, TIC~286201079 at $14.5\arcsec$ with a $T_{\rm mag}$ of 16.98, and a $\Delta T_{\rm mag}$ of 4.68, and TIC~286201077 at $16.7\arcsec$ with a $T_{\rm mag}$ of 17.59 and a $\Delta T_{\rm mag}$ of 5.29.
The closest neighboring stars to TOI-6086 are TIC~18318294 at $25.3\arcsec$ with a $T_{\rm mag}$ of 16.23, and a $\Delta T_{\rm mag}$ of 3.83, and TIC~18318295 at $33.0\arcsec$ with a $T_{\rm mag}$ of 16.81 and a $\Delta T_{\rm mag}$ of 4.41 (see the TESS FOV in \autoref{Target_pixel}).

\begin{figure}[!ht]
	\centering
	\includegraphics[width=1.\columnwidth, height=0.6\textheight, keepaspectratio]{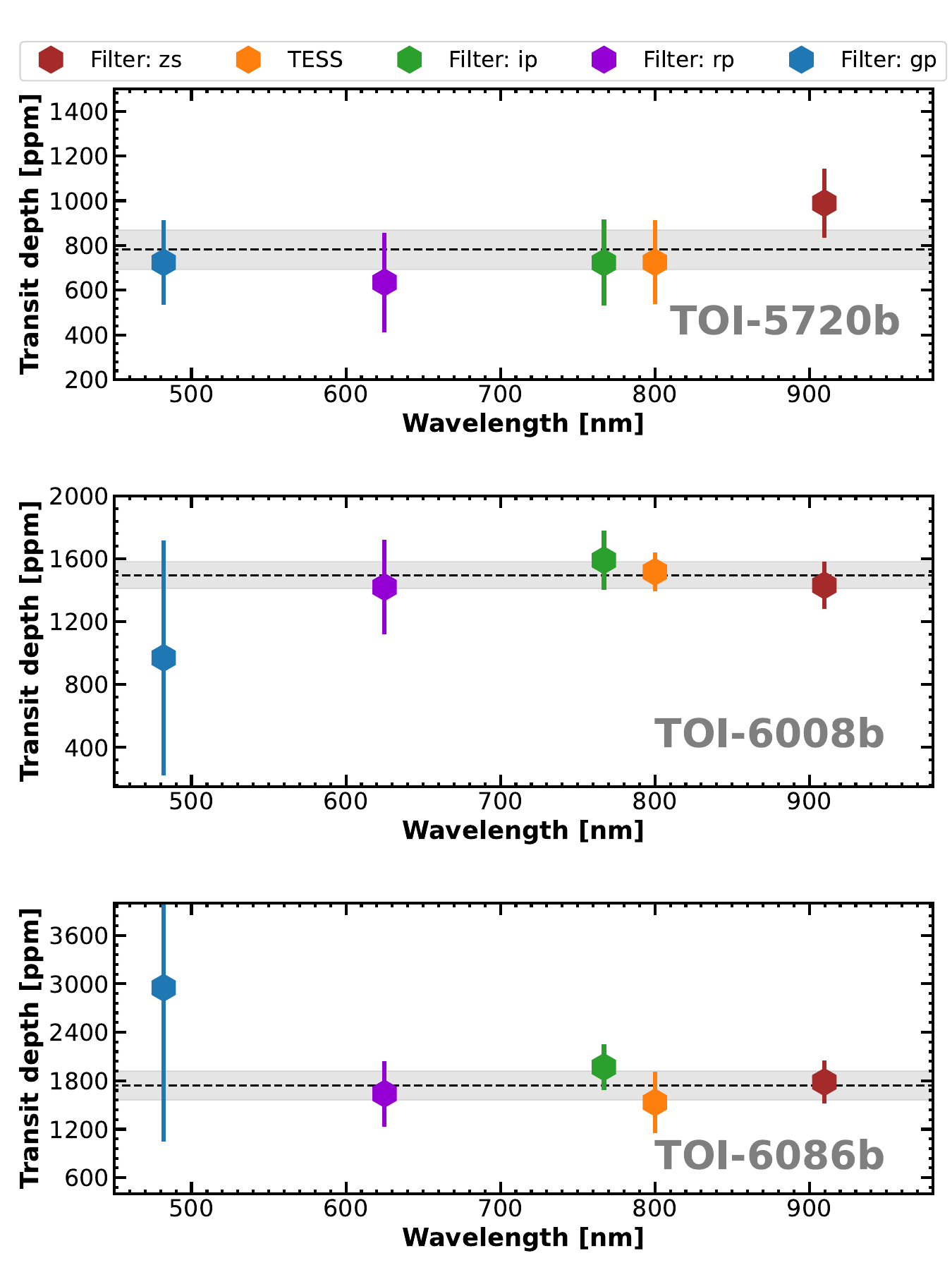}
	\caption{Measured transit depths in different filters (colored dots with error bars) obtained in the global analysis for TOI-5720\,b (top panel), TOI-6008\,b (middle panel), and TOI-6086\,b (bottom panel). The dashed line  corresponds to the depth obtained from the achromatic fit in the global analysis with a $1\sigma$ error bar (shaded region). All measurements agree with the common transit depth at $1\sigma$. }
	\label{Transit_depths}
\end{figure}

We used uncontaminated photometric aperture sizes of only a few arcseconds using ground-based telescopes in order to confirm the transit events on the expected target stars for TOI-5720\,b, TOI-6008\,b and TOI-6086\,b.
We collected photometric observations for all targets in different bands (\textit{zs}, Sloan-\textit{i'}, Sloan-\textit{r'}, and Sloan-\textit{g'}), spanning from 477 to 1100\,nm. We measured a matching transit depth within $1\sigma$ in all bands. The transit depths obtained in different bands for TOI-5720\,b, TOI-6008\,b, and TOI-6086\,b are presented in \autoref{Transit_depths} and \autoref{tois_mcmc_params}.

\section{Planet searches using TESS photometry} \label{search}

\begin{figure}[!ht]
	\centering
	\includegraphics[width=1.\columnwidth, height=0.6\textheight, keepaspectratio]{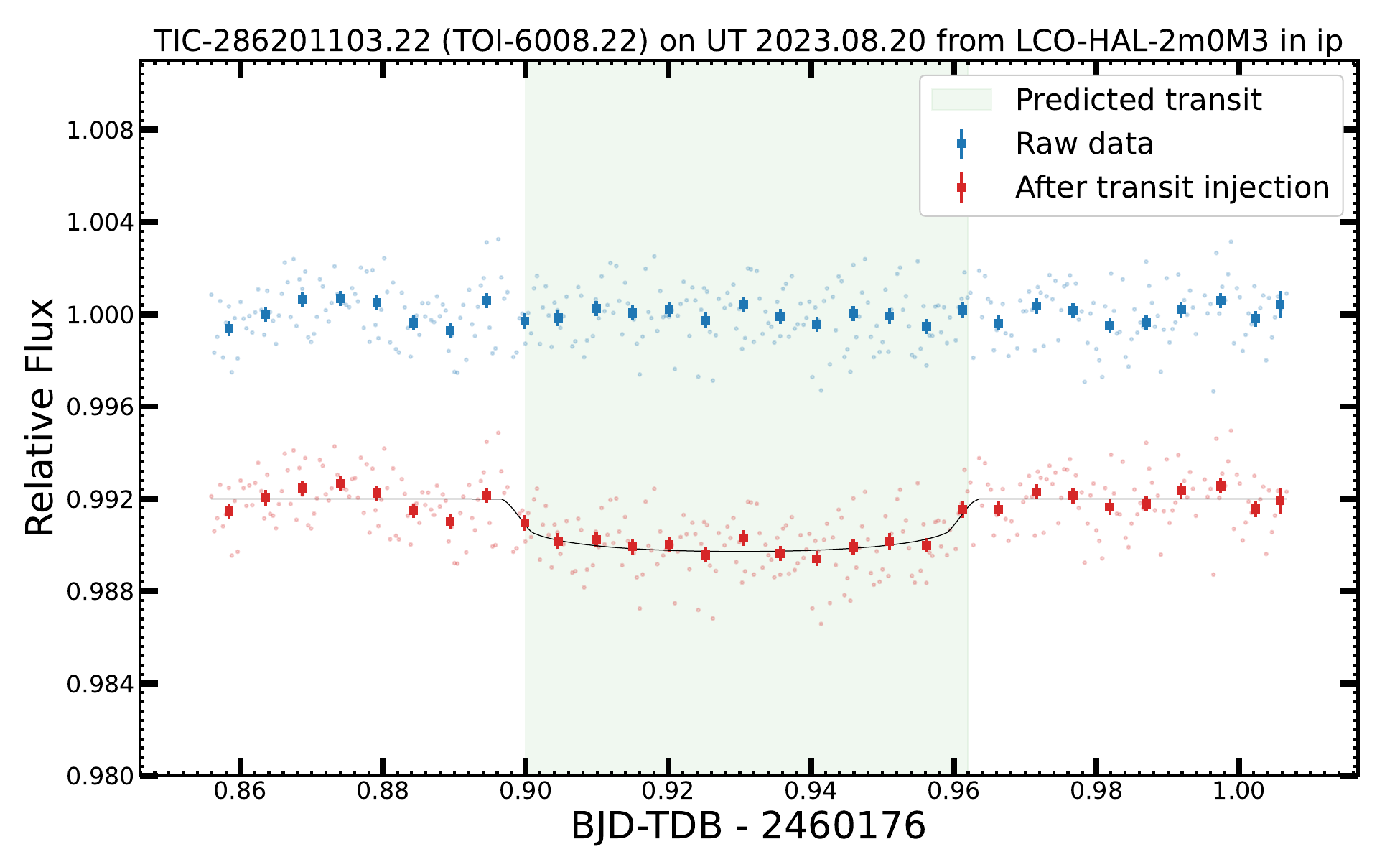} 
	\caption{LCO-HAL-2m0/MuSCAT photometry of TOI-6008.22 collected on UT August 20, 2023 in the Sloan-$i'$ filter with an exposure time of 30s. The blue data points show the raw light curve. The red data points show the light curve after the transit injection of an Earth-sized planet of $\sim 1.12$~R$_\oplus$ with the best-fitting model  superimposed in black. The green region shows the predicted ingress and egress of the transit. The light curves are shifted along the y-axis for clarity. }
	\label{LC_toi6008.22_M3}
\end{figure}

Using the TESS data described in Section~\ref{tess_phot} and our \texttt{SHERLOCK} package \citep{pozuelos2020,demory2020}, we aimed to ({\it i}) recover the candidates TOI-5720.01, TOI-6008.01, and TOI-6086.01 and ({\it ii}) search for additional transiting planets in these systems that may remain unnoticed due to detection thresholds of the SPOC. The \texttt{SHERLOCK} pipeline is especially set up for searching low-S/N transit-like features that might be attributable to planets. This pipeline is used to explore space-based data produced by the TESS and Kepler/K2 missions; for example, it is used in the SPECULOOS project \citep{sebastian2021}, where it searches for transiting planets orbiting ultracool dwarfs, and the FATE project \citep{vangrootel2021}, where it searches for planetary remnants orbiting hot subdwarfs \citep{thuillier2022}. In particular, it uses a multidetrend approach applying a biweight filter several times with different window sizes and the \texttt{TLS} algorithm \citep{hippke2019} to optimize the transit search by converging to the signals with the highest S/N and signal-detection-efficiency (SDE) \citep[see, e.g.,][]{Delrez2022, Pozuelos_2023AA}. In all the cases, we first found the signal corresponding to the planetary candidate to which SPOC alerted; that is, we successfully recovered all the TOIs studied in this work. In subsequent runs, for TOI-5720 and TOI-6086, we found no other signal that might be attributable to planets. On the other hand, for TOI-6008, we found a weak signal corresponding to an Earth-sized planet of $\sim$1.12~R$_{\oplus}$ with an orbital period of 16.35~days. The semiautomatic vetting conducted by \texttt{SHERLOCK} did not find any indication of systematics that might  produce the signal; however, the statistical validation (automatically carried out by \texttt{SHERLOCK} but relaying in TRICERATOPS) yielded an NFPP and FPP of 0.175 and 0.356, which means that it is a likely false positive \citep{Giacalone2021}. We still decided to conduct ground-based observations to fully confirm the false-positive character of this signal. We conducted three observations with MuSCAT2 on August 3 2023 and one observation with MuSCAT3 on August 19 2023. We found no evidence of a transit feature in the data, and therefore  conclude that this signal is indeed a false positive. Moreover, if the planet were present, we would see a clear transit in the MuSCAT3 light curve (See \autoref{LC_toi6008.22_M3}) with a signal-to-noise ratio of $S/N\sim 6$.
Finally, 
we also computed  the Lomb–Scargle periodogram \citep{Lomb_1976ApSS,Scargle_1982ApJ}, which showed  no indications of flaring activity or stellar modulation  in the TESS data of these three systems. This implies that the rotational periods of the stars are probably longer than the TESS observation window for a single sector.

\section{Global modeling: Photometry and radial velocities} \label{Global_modelling}

 We performed a global fit of all available transit light curves from the TESS, MuSCAT2, and MuSCAT3 telescopes and of all radial velocity measurements from the Subaru-8.2m/IRD spectrograph  described in Sections~\ref{photometric_observation}, using the Metropolis-Hastings \citep{Metropolis_1953,Hastings_1970} method implemented in {\tt TRAFIT}, a revised version of the Markov chain Monte Carlo (MCMC) code described in \cite{Gillon2010AA,Gillon2012,Gillon2014AA}. 
 The RVs collected by the Subaru/IRD spectrograph were modeled with a two-body Keplerian model \citep{Murray_2010exopbook}, while the transit light curves  were modeled using the \cite{Mandel2002} quadratic limb-darkening model, multiplied by a transit baseline, with the aim to correct for external systematic effects (time, airmass, x- and y- position of the star on the detector, background, and FWHM of the PSF).

 The baseline model for each transit was selected by minimizing the Bayesian information criterion (BIC; \citet{schwarz1978}), and photometric error bars were rescaled using the  correction factor $CF = \beta_{w} \times \beta_{r}$, where $\beta_{r}$ is the red noise and $\beta_{w}$ is the white noise \citep{Gillon2012}.  The error bars of TOI-6008 RV measurements were quadratically rescaled using the jitter noise $= (<Error_{RVs}>^2 - \sigma^2_{RVs})^{1/2}$, where $<Error_{RVs}>$ is the  mean error of the RV measurements, and  $\sigma_{RVs}$ is the standard deviation of the best-fit residuals.

 For each system, the jump parameters  were sampled by the MCMC: the transit epoch, the planet orbital period, the impact parameter, the transit depth, the total transit duration, and the combination $q_1 = (u_1 + u_2)^2$ and $q_2 = 0.5u_1(u_1 + u_2)^{-1}$ \citep{Kipping_2013MNRAS.435.2152K} of its quadratic limb-darkening coefficients $u_1$ and  $u_2$, were calculated from \cite{Claret_2012AA,Claret_2018AandA}. 

 For each target, we applied a Gaussian prior distribution on the stellar parameters derived from the SED and spectroscopy analysis ($R_\star$: stellar radius, $M_\star$: stellar mass, $T_{\rm eff}$: stellar effective temperature, $[Fe/H]$: stellar metallicity, and  $\log g_\star$: stellar surface gravity). Two MCMC analyses were performed for each system, the first assuming a circular orbit (i.e., $ecc=0$), and the second assuming an eccentric orbit. Based on the Bayes factor $BC = \exp{(-\Delta BIC/2)}$, our results are well compatible with a circular orbit solution.

 For each transit light curve, we performed a preliminary analysis composed of  one Markov chain with 100,000 steps to compute the $CF$. In the second step, we performed a global MCMC analysis composed  of three Markov chains with 100,000 steps to infer the stellar and planetary physical parameters for each system. 
We used the \cite{Gelman1992} statistical test in order to confirm the convergence of each Markov chain. 

The results for TOI-5720, TOI-6008, and TOI-6086 are presented in Tables \ref{stellarpar}, \ref{tois_mcmc_params}, and \ref{tois_LDs_param}.

\section{Results and discussion} \label{Result_discuss}

 We present the validation and discovery of TOI-5720\,b, TOI-6008\,b and TOI-6086\,b, by the TESS mission, which were confirmed using follow-up photometric observations obtained with the MuSCAT2, MuSCAT3, and LCOGT-1m0 telescopes (see \autoref{TOI_6008_5720_TESS_LC}), and radial velocity measurements obtained with the Subaru-8.2m/IRD spectrograph (see RVs curve in \autoref{RVs_curve_TOI6008}). 
We characterized the host-stars by combining spectroscopic observations obtained with Shane/Kast (see Section~\ref{Sec_Shane_Kast}) and IRTF/SpeX (see Section~\ref{Sec_IRTF_SpeX}), the spectral energy distribution (SED), and stellar evolutionary models (see Section~\ref{SED_evol_analys}).  We performed a global analysis of the TESS data and ground-based photometric and radial velocity observations in order to constrain the physical parameters for each system (see Section~\ref{Global_modelling}). \autoref{stellarpar} shows the stellar properties of the host stars TOI-5720, TOI-6008, and TOI-6086. The derived stellar and planetary physical parameters for each system are shown in Tables \ref{stellarpar}, \ref{tois_LDs_param}, and \ref{tois_mcmc_params}.
Figures \ref{corner_TOI5720}, \ref{corner_TOI6008}, and \ref{corner_TOI6086} show the  posterior distribution parameters for TOI-5720, TOI-6008, and TOI-6086, respectively. 

\subsection{TOI-5720\,b, TOI-6008\,b, and TOI-6086\,b}

The host star TOI-5720 is a nearby (36~pc) $K_{\rm mag} = 9.54$ M3.5-type star, with an effective temperature of $T_{\rm eff} = 3325 \pm 75$~K, a surface gravity of $\log g_\star = 5.01 \pm 0.02$~dex, a radius of $R_\star = 0.383 \pm 0.018~R_\odot$, a mass of $M_\star = 0.383 \pm 0.019~M_\odot$ (derived from  the SED analysis), and a metallicity of $[Fe/H] = -0.19 \pm 0.12$ (deduced from the near-infrared spectrum).
TOI-5720\,b is a short-period Earth-sized planet orbiting its host star every $P = 1.4344555 \pm 0.0000036$~days, which has a radius of $R_p = 1.09 \pm 0.07~R_\oplus$, an equilibrium temperature of $T_{\rm eq} = 708 \pm 19$~K (assuming an albedo of 0), an incident flux of $S_p = 41.7 \pm 4.5$ times that of Earth, and an upper limit on the  mass of $M_p < 4~M_\oplus$ based on two TRES RV measurements with a $3\sigma$ error bar (see Section~\ref{TRES_spectr}). This confirmed the nonstellar transiting companion.   Moreover, the predicted mass using the \cite{Chen_Kipping_2017ApJ} mass-radius relation implemented in the {\tt forecaster}\footnote{forecaster: \url{https://github.com/chenjj2/forecaster}} package it is found to be $M_p = 1.32^{+0.92}_{-0.47}$~M$_\oplus$.

TOI-6008 is an M5-type ($K_{\rm mag} = 9.29$) star at a distance of 23~pc, with an effective temperature of $T_{\rm eff} = 3075 \pm 75$~K, a surface gravity of $\log g_\star = 5.01 \pm 0.04$~dex, a radius of $R_\star = 0.242 \pm 0.013~R_\odot$, a mass of $M_\star = 0.230 \pm 0.011~M_\odot$ (derived from  the SED analysis), and a metallicity of $[Fe/H] = -0.21 \pm 0.12$ (deduced from the near-infrared spectrum).
TOI-6008\,b is a $1.03 \pm 0.05~R_\oplus$ short-period planet with an orbital period of $P = 0.8574347 \pm 0.0000424$~day, an equilibrium temperature of $T_{\rm eq} = 707 \pm 19$~K (assuming an albedo of 0), and an incident flux of $S_p = 41.5 \pm 4.5$ times that of Earth. We collected 15 RV measurements of TOI-6008 using the Subaru/IRD spectrograph. The resulting mass is found to be $M_p = 1.1 ^{+3.3}_{-0.9}~M_\oplus$ with a $3\sigma$ error bar. This value ruled out a stellar transiting companion, and confirmed the nature of the planetary companion. Additionally, the predicted planetary mass using the mass-radius  relation implemented in the {\tt forecaster} package, is found to be $M_p = 1.10^{+0.73}_{-0.36}$~M$_\oplus$. 

TOI-6086 is a $K_{\rm mag} = 9.99$ M3-type star at a distance of 32~pc from the Sun, with an effective temperature of $T_{\rm eff} = 3200 \pm 75$~K, a surface gravity of $\log g_\star = 5.01 \pm 0.02$~dex, a radius of $R_\star = 0.259 \pm 0.013~R_\odot$, a mass of $M_\star = 0.254 \pm 0.011~M_\odot$ (derived from  the SED analysis), and a metallicity of $[Fe/H] = -0.2 \pm 0.3$ (deduced from the near-infrared spectrum).
TOI-6086\,b is a short-period ($P = 1.3888725 \pm 0.0000827$~days) Earth-sized planet ($R_p = 1.18 \pm 0.07~R_\oplus$) orbiting its host star, with an equilibrium temperature of $T_{\rm eq} = 634 \pm 16$~K (assuming an albedo of 0), an incident flux of $S_p = 26.8 \pm 2.7$ times that of Earth. Using the mass-radius relation implemented in the {\tt forecaster} package, we find that TOI-6086\,b has a predicted mass of $M_p = 1.68^{+1.20}_{-0.55}$~M$_\oplus$. 

\begin{figure}[!ht]
	\centering
	\includegraphics[width=1.\columnwidth, height=0.6\textheight, keepaspectratio]{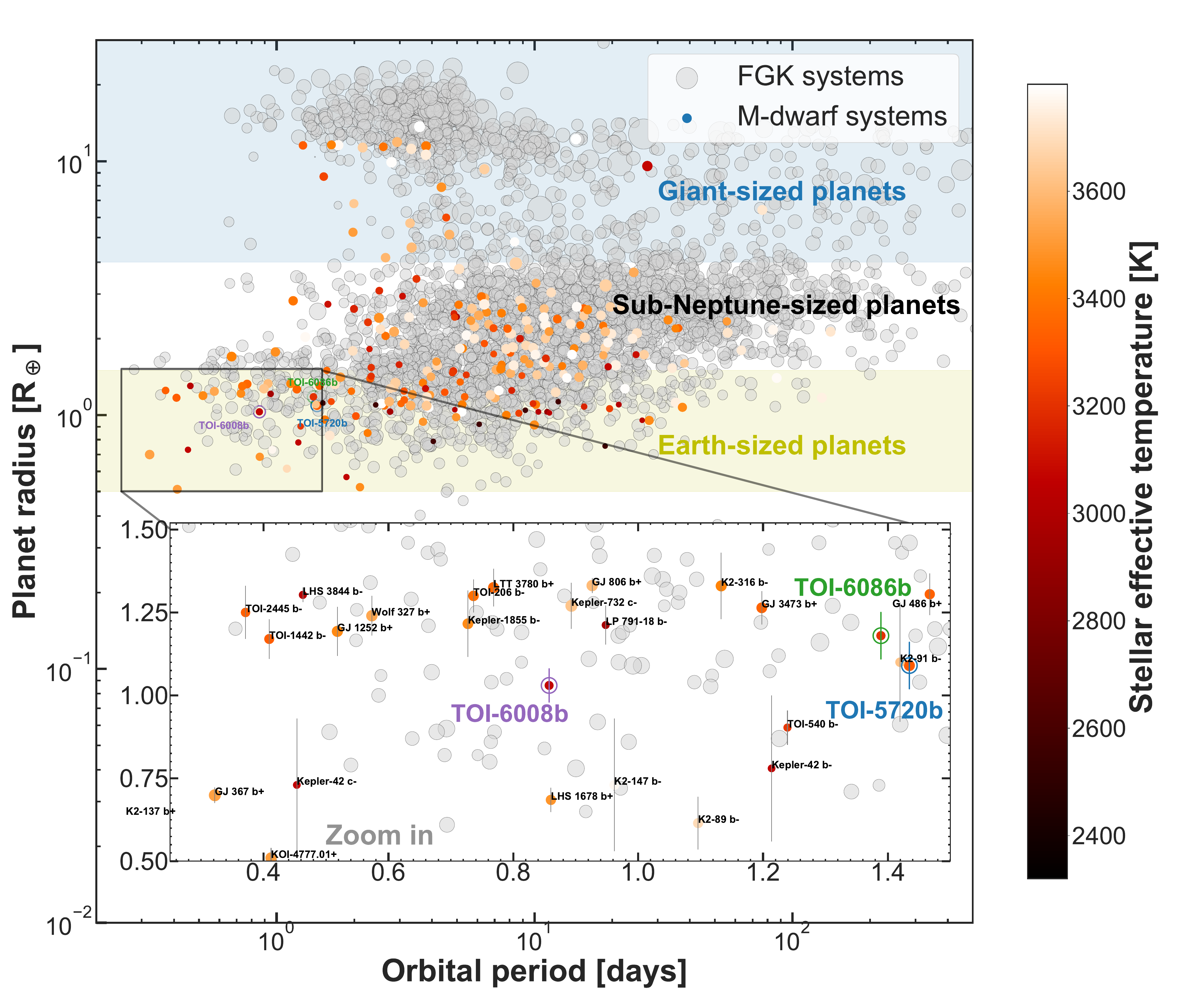} \\
 \includegraphics[width=\columnwidth, height=0.4\textheight, keepaspectratio]{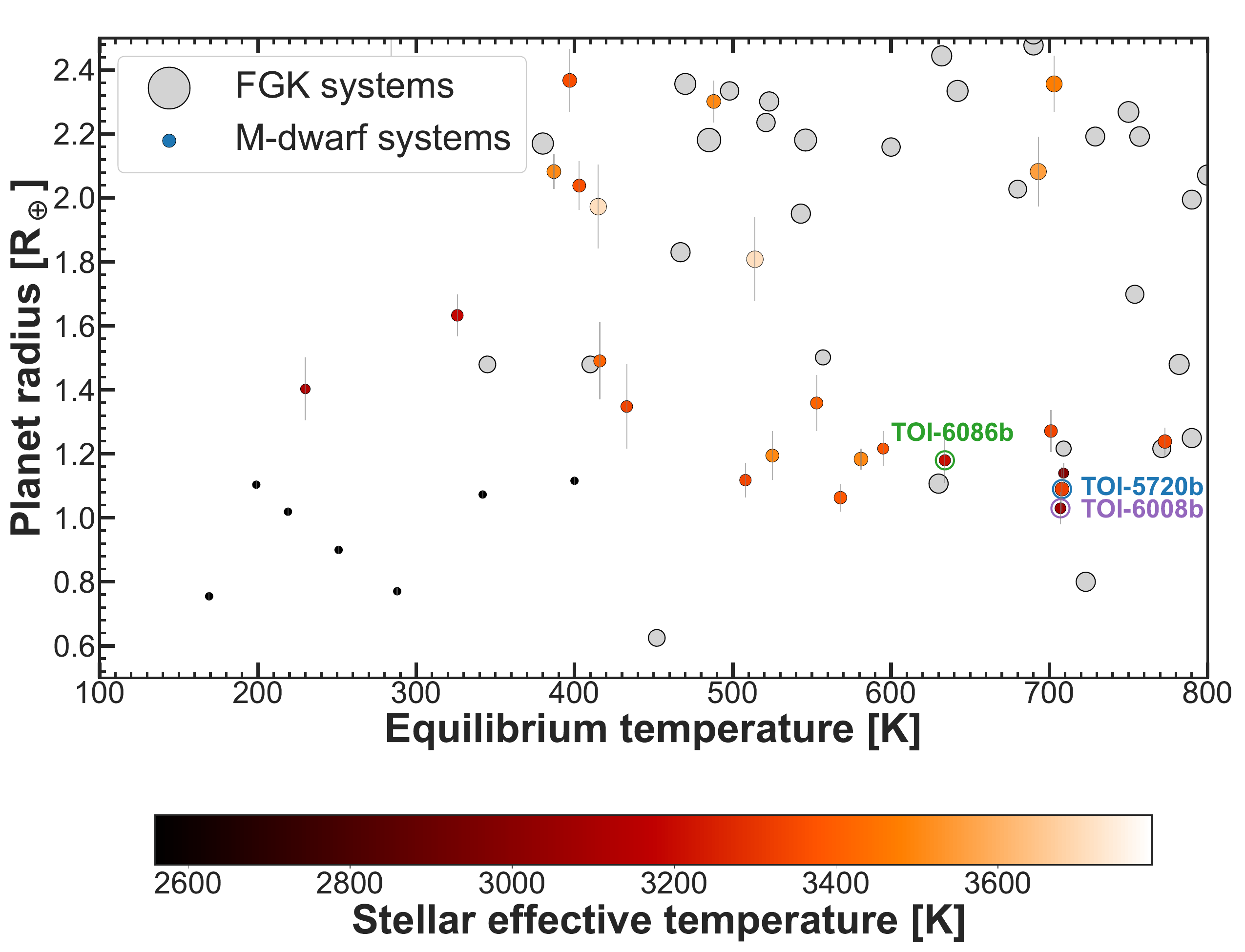} 
	\caption{Planetary radius {\it vs.}  orbital period (top panel) and Planetary radius {\it vs.} planetary equilibrium temperature (bottom panel) diagrams of known transiting exoplanets from {\it NASA Archive of Exoplanets}. Grey circles show the FGK planetary systems, and colored points show M-dwarfs systems. M-dwarfs planets are colored by the host star effective temperature. The size of the points scale according to the stellar radius.  All short-period ($<$1.5~days) Earth-sized planets orbiting around M-dwarfs are highlighted with error bars, and planets with and without mass measurements are highlighted with \textbf{+} and \textbf{-}, respectively. TOI-5720\,b, TOI-6008\,b and TOI-6086\,b are circled in blue, purple and green, respectively.}
	\label{Rp_Period_Teq_diagram}
\end{figure}

Figures \ref{Rp_Period_Teq_diagram} shows the distribution of our planets versus orbital period and equilibrium temperature. 
These systems are suitable targets for high-precision radial velocity facilities in order to provide the planetary mass measurements, and bulk densities and to search for additional outer planets in the system. 
When we assume a rocky composition for these planets, the predicted RV semi-amplitudes using the mass-radius relation by \citet{Chen_Kipping_2017ApJ}, are found to be $K_{\rm TOI-5720} = 1.4^{+1.0}_{-0.5}$~m/s for TOI-5720, $K_{\rm TOI-6008} = 2.0^{+1.3}_{-0.6}$~m/s for TOI-6008, and $K_{\rm TOI-6086} = 2.4^{+1.6}_{-0.8}$~m/s for TOI-6086. A combination of the predicted RV semi-amplitudes with the relative faintness of the host stars ($V_{\rm mag}$ of 13.9 to 16.0) would require high-precision spectroscopic observations using  stabilized spectrographs such as MAROON-X \citep{Seifahrt_MaroonX} mounted on the 8.1m Gemini-North telescope. For exposure times of 1800s and good observing conditions, the MAROON-X exposure-time calculator (ETC) predicts an RV precision of  0.6~m/s, 1.1~m/s, and 1.0~m/s for TOI-5720, TOI-6008, and TOI-6086, respectively.
A few RV measurements for each target will allow us to measure the planetary masses, bulk densities, and the eccentricities of the orbits with a precision 10~$\sigma$.

\begin{figure}[!ht]
	\centering
	\includegraphics[width=\columnwidth, height=0.4\textheight, keepaspectratio]{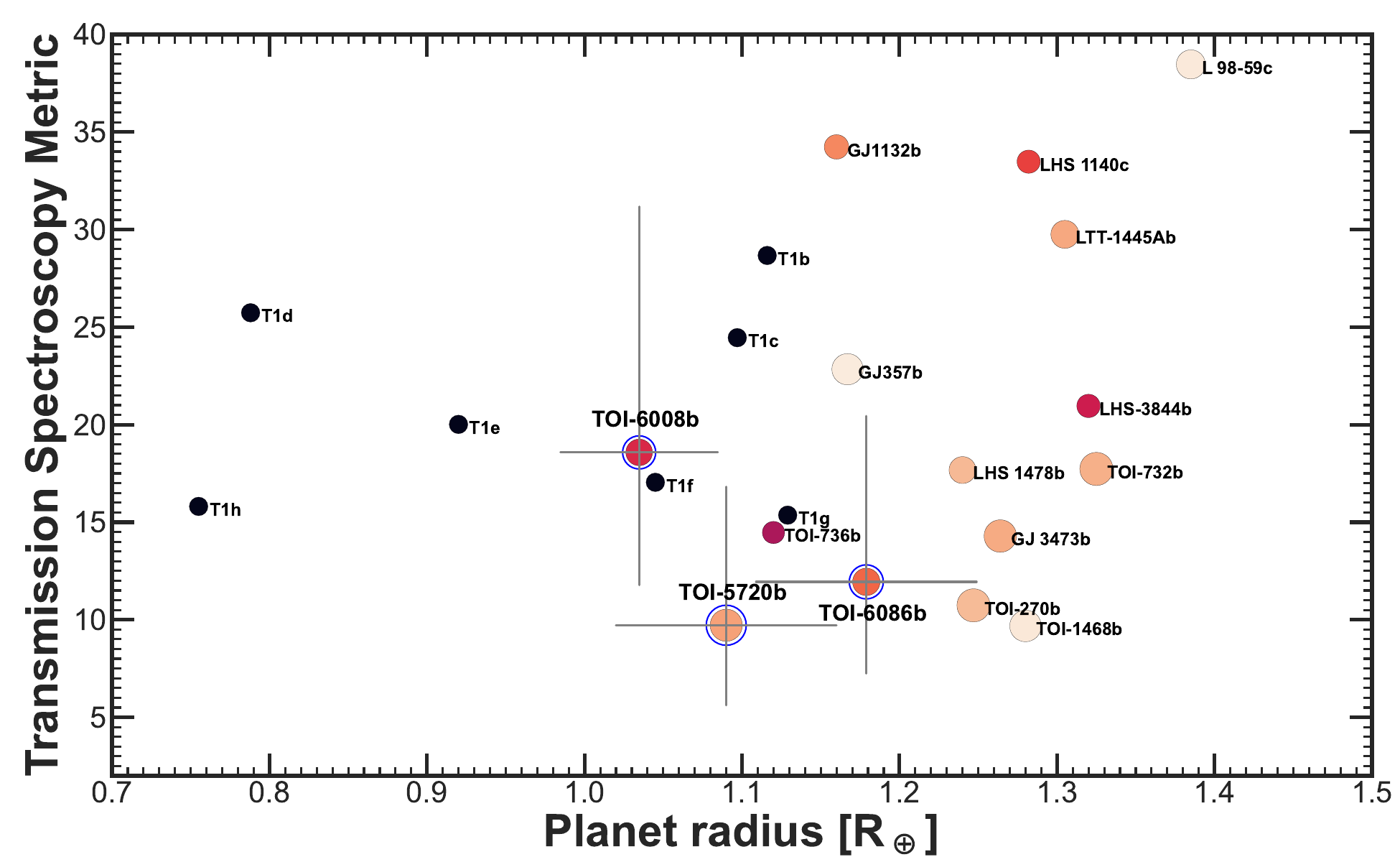} 
    \includegraphics[width=\columnwidth, height=0.4\textheight, keepaspectratio]{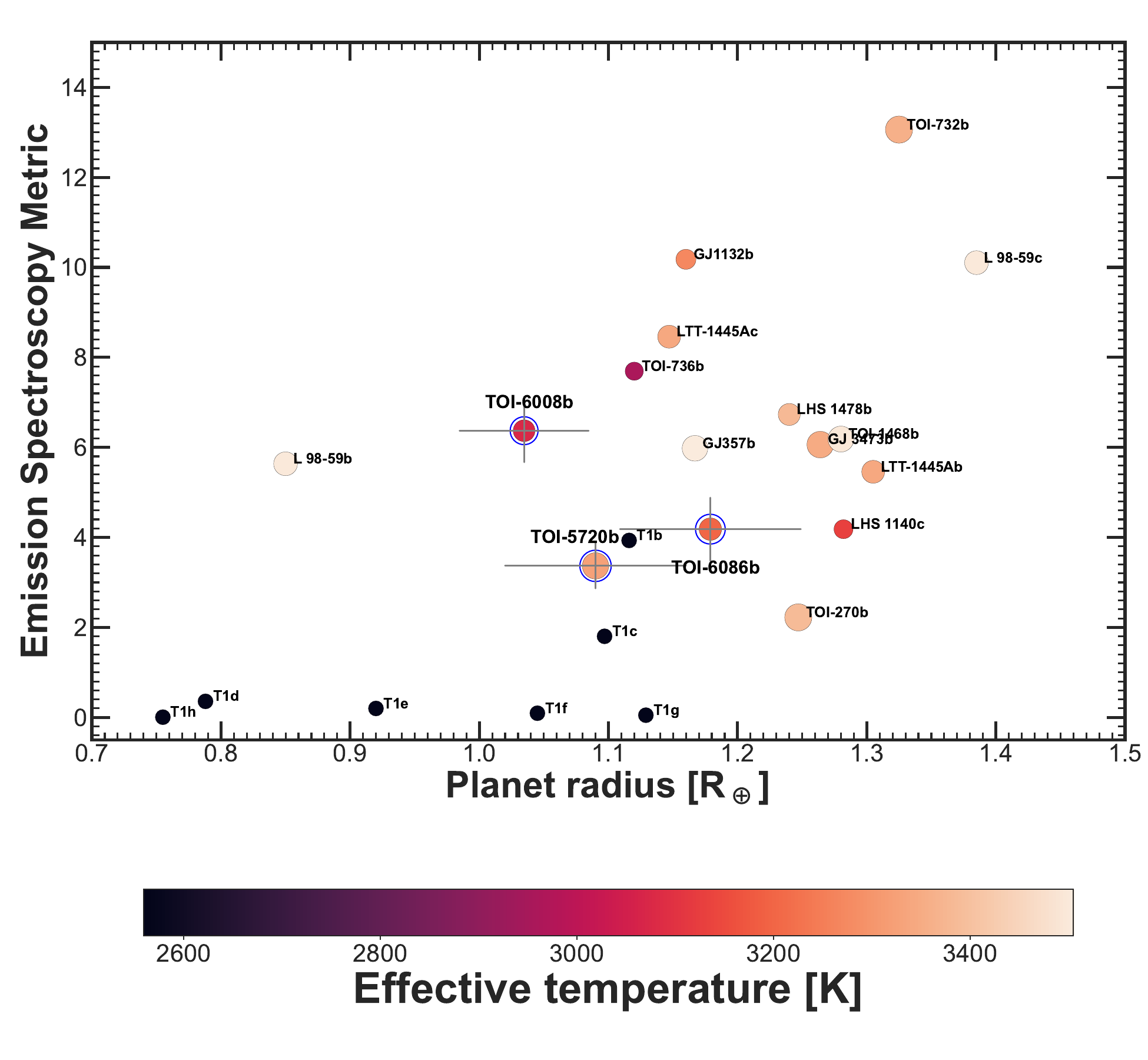} 
	\caption{ Feasibility of TOI-5720\,b, TOI-6008\,b, and TOI-6086\,b for transmission and emission spectroscopy studies.
    Transmission spectroscopy metric (TSM, top panel) and emission spectroscopy metric (ESM, bottom panel)  as a function of planetary radius. Our planets are compared to the Earth-sized ($R_p < 1.5$~R$_\oplus$) known transiting exoplanets observed or scheduled on \emph{JWST} around M dwarfs ($T_{\rm eff} < 3800$). The size of the points scales with the stellar radius. The points are colored according to the effective stellar temperature. 
    The blue circles highlight our planets TOI-5720\,b, TOI-6008\,b, and TOI-6086\,b.} 
	\label{TSM_ESM_diagram}
\end{figure}

The combination of the equilibrium temperature, the high incident flux, the infrared brightness, and small sizes of the host stars, make all three planets  suitable for transmission and emission spectroscopy studies with \emph{JWST}.
We calculated the transmission spectroscopy metrics (TSM) using equation~1 from \cite{kem}. Our predicted TSM are  $9.7 ^{+7.1}_{-4.1} $ for TOI-5720\,b, $18.8^{+12.6}_{-6.8}$ for TOI-6008\,b, and $12.0^{+8.5}_{-4.7}$ for TOI-6086\,b.
We also calculated their emission spectroscopy metrics (ESM) using equation~4 from \cite{kem}, and we find that ESM to be $3.4 \pm 0.5$ for TOI-5720\,b, $6.4 \pm 0.7$ for TOI-6008\,b, and $4.2 \pm 0.7$ for TOI-6086\,b.
\autoref{TSM_ESM_diagram} shows the planet positions of TSM and ESM versus planetary radius diagrams of Earth-sized ($R_p<1.5$~R$_\oplus$) transiting exoplanets around M dwarfs ($T_{\rm eff} < 3800$) observed or scheduled on \emph{JWST}. 
Based on the TSM and ESM parameters, TOI-6008\,b is one of favourable terrestrial planet for atmospheric characterization (emission, transmission, and phase curve photometry) compared to other systems  that have been observed with \emph{JWST}
(e.g., TRAPPIST-1\,b, TRAPPIST-1\,c, TRAPPIST-1\,e, and TRAPPIST-1\,g; \citet{Gillon_2017Natur},
L 98-59\,b; \citet{Kostov_2019_L98-59}, 
GJ 357\,b; \citet{Luque_2019AA_Gl357},
GJ 3473\,b; \citet{Kemmer_2020AA_GJ3473},
LHS 1140\,c; \citet{Dittmann_2017Natur_LHS1140},
LTT 1445A\,b; \citet{Winters_2019AJ_LTT1445b},
LTT 1445A\,c; \citet{Winters_2022AJ_LTT1445Ac},
TOI-1468\,b; \citet{Chaturvedi_2022AA_TOI-1468b},
LHS 1478\,b; \citet{Soto_2021AA_LHS1478b},
TOI-732\,b; \citet{Nowak_2020AA_TOI732b},
LP 791-18\,b; \citet{Crossfield_2019_LP791-18}). The relatively high equilibrium temperatures of the planets, their  small sizes, and the brightness of the stars, also make TOI-5720\,b and TOI-6086\,b  good targets for  emission spectroscopy and phase-curve photometry compared with TRAPPIST-1\,b and TRAPPIST-1\,c.

\subsection{Prospects for atmospheric characterization}

We provide a thorough examination of the simulated transmission spectra of planets using PLanetary Atmospheric Transmission for Observer Noobs \citep[\texttt{PLATON};][]{2019PASP..131c4501Z,2020ApJ...899...27Z}. Specifically, we employed \texttt{PandExo} \citep{2017PASP..129f4501B} to generate synthetic data, mimicking observations that would be made with the James Webb Space Telescope (\emph{JWST}). Our simulation method encompasses various parameters such as planetary and stellar radii, effective temperatures, stellar metallicity, stellar gravity, transit duration, orbital period, and inclination. Through meticulous adjustment of these inputs, we created mock datasets to replicate transmission spectra. We ensured a minimum of three transit observations, spanning between $3$-$4$ hours of observation time, with at least one-half of the time allocated for out-of-transit observations for each target to acquire the necessary simulated data. The resultant synthetic transmission spectra offer insights into the atmospheric properties of these planets. Through detailed simulation and analysis of these spectra, our goal is to contribute valuable information regarding the composition and characteristics of exoplanetary atmospheres. \autoref{fig:STS} presents the synthetic spectra for TOI-5720\,b, TOI-6008\,b, and TOI-6086\,b.

The James Webb Space Telescope (\emph{JWST}) is equipped with four scientific instruments \citep{2006SSRv..123..485G}, delivering high-resolution and highly sensitive observations of celestial targets. The \emph{JWST} NIRSpec \citep{2003AAS...20312407R} employs fixed-slit (FS) spectroscopy, offering high-sensitivity single-object spectroscopy for a wide wavelength range from 0.6~$\mu$m to 5.3~$\mu$m. The \emph{JWST} NIRCam \citep{2002AAS...20115103H}, in its wide-field slitless spectroscopy mode, uses grisms and filters covering wavelengths between 2.4 $\mu$m to 5.0 $\mu$m, achieving a resolving power of $1,600$ at 4 $\mu$m. Our study used the F322W2 and F444W filters, spanning wavelengths from 2.7 $\mu$m to 4 $\mu$m and 4 $\mu$m to 5 $\mu$m, respectively. The \emph{JWST} Near Infrared Imager and Slitless Spectrograph (NIRISS; \citep{2012cosp...39..478D}) provides slit-less spectroscopy across wavelengths ranging from 0.6~$\mu$m to 5.0~$\mu$m. The Mid-Infrared Instrument (MIRI; \citep{2003ESASP.539..355B}) extends our observational capabilities into the mid-infrared spectrum, covering spectral wavelengths from 4.9 $\mu$m to 28.8 $\mu$m. 

In our investigation of the three validated planets, we used various grism and filter combinations available for each instrument. These combinations act as instrumental tools facilitating a comprehensive simulation of the transmission spectra for each planet.

\subsubsection{PLATON}
\texttt{PLATON} \citep{2019PASP..131c4501Z,2020ApJ...899...27Z} is a Python package that computes transmission spectra for exoplanets and derives atmospheric characteristics based on observed spectra. The software accommodates key atmospheric parameters, including temperature, metallicity, C/O ratio, cloud-top pressure, and scattering slope. Employing an equilibrium chemistry model computed with \texttt{GGchem} \citep{2018AA...614A...1W}, \texttt{PLATON} allows users to input the metallicity and C/O ratio. From these inputs, \texttt{PLATON} calculates the abundances of $34$ chemical species in each atmospheric layer. GGchem computes these abundances across a grid of metallicity, C/O ratio, temperature, and pressure, assuming equilibrium chemistry with or without condensation. The parameter ranges are as follows: a metallicity from $0.1$ to $1000$ times solar, a C/O ratio from $0.2$ to $2$, a temperature from $300$ to $3000$~K, and a pressure from $10^{-4}$ to $10^{8}$~Pa. To compute a transmission spectrum, we adopted the assumption of an isothermal atmosphere with equilibrium chemistry. \texttt{PLATON} divided the atmosphere into 500 equally spaced layers in $\log{P}$ from $10^{-4}$ to $10^{8}$. The physical depth of each layer was determined by applying the hydrostatic equation. The opacity of each layer was subsequently calculated, taking into account factors such as gas absorption, collisional absorption, and Rayleigh scattering. The radiative transfer calculation was executed by considering a ray passing through the atmosphere with an impact parameter. The optical depth experienced by the ray was then computed, and the transit depth was derived by summing the cross-sectional areas of each layer, with weights assigned based on the amount of light that was permitted to pass through. Subsequently, corrections for unocculted starspots were implemented, which were deemed crucial for active stars due to the distinct spectra of  hot, unspotted regions and colder starspots. The correction involved adjusting the transit depth based on factors such as the spot fraction, the spot temperature, the temperature of unspotted regions, and the spectrum of the stellar surface. The composition of each layer was determined by considering equilibrium chemistry, which facilitated the calculation of the mean molecular weight. The fitted parameters included the mixing ratios of these molecules, the atmospheric mean molecular weight, the surface pressure, and the planet radius. The specified radius for the planet was taken at a reference pressure of $1$~bar, which was chosen for its proximity to the standard atmospheric pressure on Earth. Our model assumed a clear sky without clouds or haze.

\subsubsection{PandExo}
\texttt{PandExo} \citep{2017PASP..129f4501B} serves as both an online interface and an open-source Python package, which we employed to model instrumental noise for exoplanet transit time-series spectroscopy modes with \emph{JWST} (NIRISS, NIRCam, NIRSpec, and MIRI LRS) and HST (WFC3). Drawing on the foundations of the Space Telescope Science Institute (STScI) Exposure Time Calculator, Pandeia \citep{2016SPIE.9910E..16P}, \texttt{PandExo} exhibits excellent agreement, with deviations well within $10\%$ not only for HST WFC3, but also across the entire spectrum of the \emph{JWST} instruments. The input parameters encompass critical factors, including a stellar SED model sourced from the Phoenix stellar atlas \citep{2013A&A...553A...6H}, the apparent magnitude of the star, details of the planet spectrum (primary or secondary), the transit duration, the ratio of time dedicated to in-transit versus out-of-transit observations, the number of observed transits, the saturation threshold expressed as a percentage of the full-well capacity, and the user-defined noise floor. To compute the atmospheric transmission data of the planet, we used the built-in molecular line grid in \texttt{PandExo}. Our approach involved selecting the planetary temperature closest to the computed temperature of each of the eight validated planets. We considered equilibrium chemistry with TiO and medium Rayleigh scattering. It is essential to note that our modeling assumed conditions devoid of clouds and hazes, representing clear skies for these analyses.

\subsection{Results}
We discuss about the synthetic spectra we obtained using \texttt{PandExo} and the best-fit model we found using \texttt{PLATON}. In Table \ref{tab:ExpSNR} we provide the details of exposure time and S/N values.

\subsubsection{TOI-5720\,b}
Our model spectra using \texttt{PLATON} align well with the simulated data from \texttt{PandExo}. $\rm H_2O$ lines can be found at 1, 1.3, 1.8, and 2.8 $\mu$m using NIRSpec, NIRISS, and NIRCam. Observation using MIRI may also add a contribution to detecting subtle $\rm H_2O$ signatures at higher wavelengths. We also found a $\rm CO_2$ signature at 4.3 $\mu$m, depending on the metallicity and C/O chosen for our study. Our fitting efforts yield a best-fit model with an estimated metallicity of the atmosphere of TOI-5720b of 1.26 times solar metallicity and C/O of 1.01. We also find $\rm CH_4$ lines at 1.2, 1.4, 1.7, 2.3, and 3.3 $\mu$m with NIRSpec, NIRISS, and NIRCam (see top panel of \autoref{fig:STS}). $\rm NH_3$ may also be present in a small amount.

\subsubsection{TOI-6008\,b}
Our model spectra using \texttt{PLATON} align well with the simulated data from \texttt{PandExo}. $\rm H_2O$ lines can be found at 0.8, 1.2, 1.4, 1.8, and 2.7 $\mu$m using NIRSpec, NIRISS, and NIRCam. Observation using MIRI may also add a contribution to detecting subtle $\rm H_2O$ signatures at higher wavelengths. We also found $\rm CO_2$ signatures at 4.3 $\mu$m and $\rm CO$ at 4.8 $\mu$m, depending on the metallicity and C/O chosen for our study. Our fitting efforts yield a best-fit model with an estimated metallicity of the atmosphere of TOI-6008b of 1.26 times solar metallicity and C/O of 0.5 (see middle panel of \autoref{fig:STS}).

\subsubsection{TOI-6086\,b}
\begin{figure}
    \centering
    \includegraphics[scale=0.15]{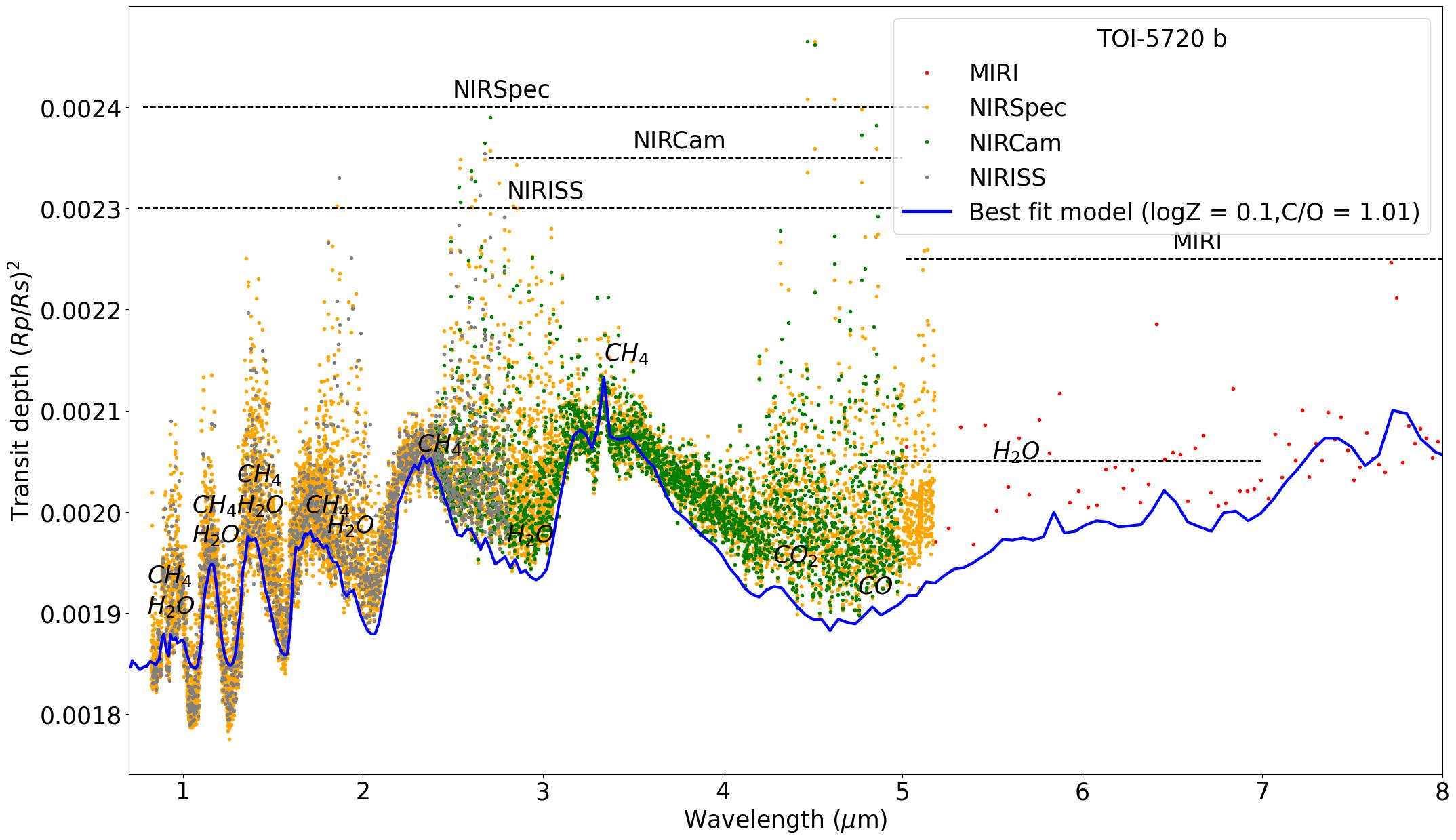}
    \includegraphics[scale=0.15]{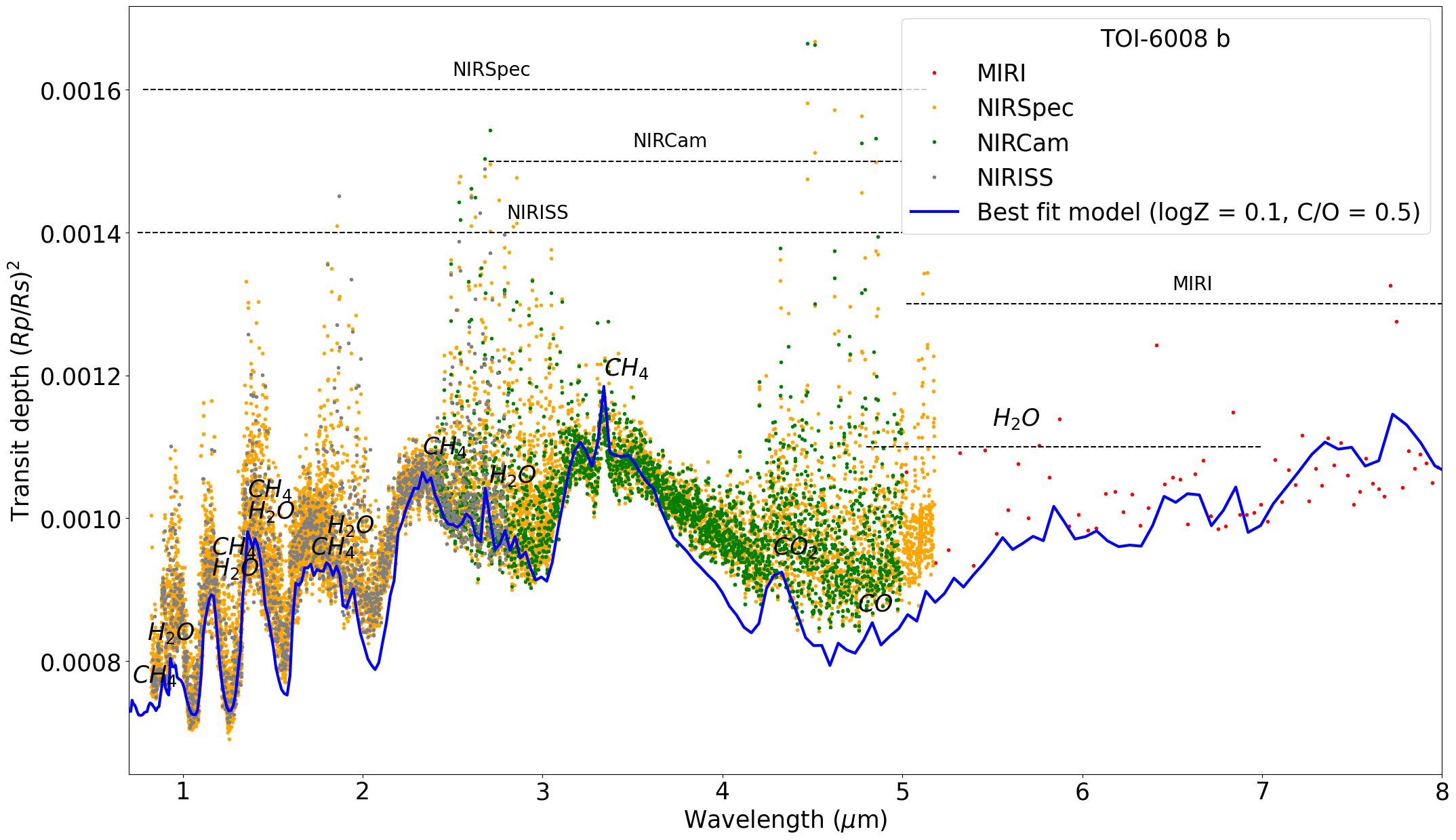}
    \includegraphics[scale=0.15]{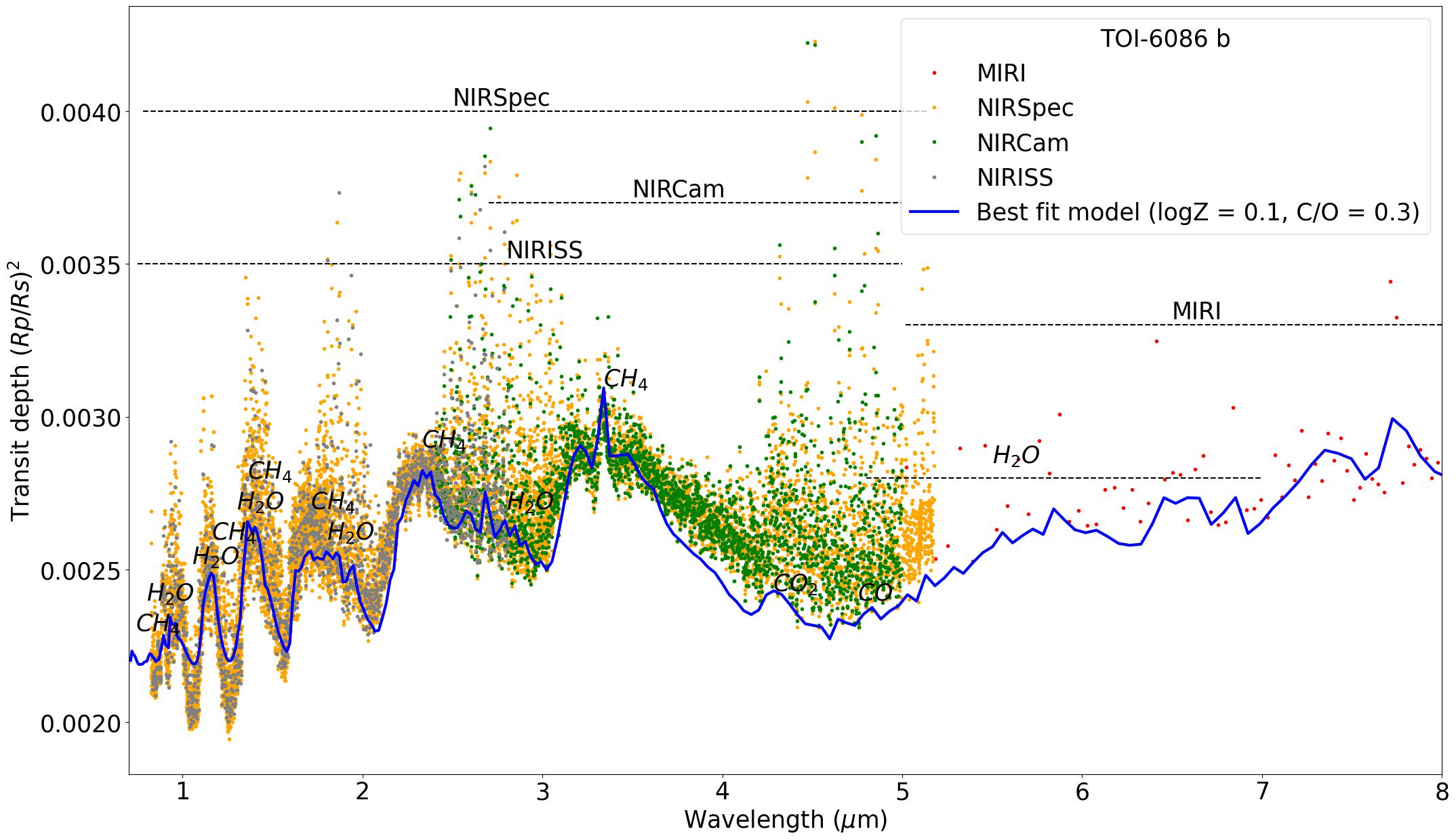}
    \caption{Synthetic transmission spectrum in \emph{JWST} for TOI-5720\,b (top panel), TOI-6008\,b (middle panel) and TOI-6086\,b (bottom panel) generated with the help of the \texttt{PLATON} and \texttt{PandExo} tools. The red, yellow, green, and grey points are simulated data using Jthe WST/MIRI, JWST/NIRSpec, JWST/NIRCam, and JWST/NIRISS instruments, and the best-fitting model is superimposed in blue.}
    \label{fig:STS}
\end{figure}

Our model spectra using \texttt{PLATON} align well with the simulated data from \texttt{PandExo}. $\rm H_2O$ lines can be found at about 0.8, 1.0, 1.3, 1.8, and 3 $\mu$m using NIRSpec, NIRISS, and NIRCam. Observation using MIRI may also add a contribution to detecting subtle $\rm H_2O$ signatures at higher wavelengths.  We also found $\rm CO_2$ signatures at 4.3 $\mu$m and $\rm CO$ at 4.8 $\mu$m, depending on the metallicity and C/O chosen for our study. Our fitting efforts yield a best-fit model with an estimated metallicity of the atmosphere of TOI-6086b of 1.26 times solar metallicity and C/O of 0.3. $\rm NH_3$ may also be present in a small amount of  present. We also find $\rm CH_4$ lines at 0.8, 1.2, 1.4, 1.7, 2.3, and 3.3 $\mu$m with NIRSpec, NIRISS, and NIRCam (see bottom panel of \autoref{fig:STS}).

\begin{table}[]
    \centering
    {\renewcommand{\arraystretch}{1.25}
    \begin{tabular}{c c c c}
         \hline
         Instrument & Mode $^a$ & Exposure & S/N \\
                    &           & Time     &     \\
         \hline
         \hline
         \multicolumn{4}{c}{\bf TOI-5720\,b} \\
         MIRI & Slit-less Spectroscopy & 7003 s & 3095.56 \\
         NIRSpec & G140H/F070LP & 7179 s & 3906.26 \\
                 & G140H/F100LP & 7168 s & 3713.27 \\
                 & G235H/F170LP & 7170 s & 4072.01 \\
                 & G395H/F290LP & 7177 s & 3693.03 \\
         NIRCam & F322W2 & 7221 s & 3514.38 \\
                & F444W  & 7177 s & 2894.01 \\
         NIRISS & SUBSTRIP256 & 7211 s & 5645.12 \\
         & & & \\
         \multicolumn{4}{c}{\bf TOI-6008\,b} \\
         MIRI & Slit-less Spectroscopy & 8370 s & 2513.81 \\
         NIRSpec & G140H/F070LP $^{b}$ & 8526 s & 3744.90 \\
                 & G140H/F100LP & 8510 s & 4142.91 \\
                 & G235H/F170LP & 8534 s & 4258.71 \\
                 & G395H/F290LP & 8518 s & 3555.67 \\
         NIRCam & F322W2 & 8635 s & 3536.38 \\
                & F444W  & 8500 s & 2755.55 \\
         NIRISS & SUBSTRIP96 $^{c}$ & 8664 s & 6833.68 \\
         & & & \\
         \multicolumn{4}{c}{\bf TOI-6086\,b} \\
         MIRI & Slit-less Spectroscopy & 8222 s & 1934.40 \\
         NIRSpec & G140H/F070LP & 8303 s & 3115.41 \\
                 & G140H/F100LP & 8315 s & 3345.80 \\
                 & G235H/F170LP & 8350 s & 3214.50 \\
                 & G395H/F290LP & 8376 s & 3037.43 \\
         NIRCam & F322W2 & 8431 s & 2903.43 \\
                & F444W  & 8516 s & 2342.96 \\
         NIRISS & SUBSTRIP256 & 8386 s & 5107.16 \\
         \hline
    \end{tabular}}
    $^a$ Details of the different modes, resolving power and wavelength can be found at: \url{https://jwst-docs.stsci.edu/}\\
    $^b$ Disperser-filter combinations. We chose combinations that can provide a high resolution power of $~$2700. G = Grism, F = Filter.\\
    $^c$ Size of the subarray for single-object slitless spectroscopy. \\
    \caption{Exposure times and S/N obtained for each synthetic specrum of JWST.}
    
    \label{tab:ExpSNR}
\end{table}

\section{Software}
This work made use of many publicly available packages and tools for which the authors are immensely grateful. In addition to the software and pipelines cited throughout the paper, we also used
\texttt{Matplotlib} \citep{Matplotlib}, 
\texttt{Numpy} \citep{Numpy},
\texttt{Astropy} \citep{Astropy},
\texttt{SciPy} \citep{Scipy},
\texttt{Wotan} \citep{Hippke2019AJ_wotan}, and
\texttt{Photutils} \citep{larry_bradley_2020_4044744}.

\section{Acknowledgements}
The postdoctoral fellowship of KB is funded by F.R.S.-FNRS grant T.0109.20 and by the Francqui Foundation.
This paper made use of data collected by the TESS mission and are publicly available from the Mikulski Archive for Space Telescopes (MAST) operated by the Space Telescope Science Institute (STScI). Funding for the TESS mission is provided by NASA’s Science Mission Directorate. We acknowledge the use of public TESS data from pipelines at the TESS Science Office and at the TESS Science Processing Operations Center. Resources supporting this work were provided by the NASA High-End Computing (HEC) Program through the NASA Advanced Supercomputing (NAS) Division at Ames Research Center for the production of the SPOC data products.
This work is partly supported by JSPS KAKENHI Grant Numbers JP18H05439, JP 18H05442, JP21K20376, and JST CREST Grant Number JPMJCR1761. This article is based on observations made with the MuSCAT2 instrument, developed by ABC, at Telescopio Carlos Sánchez operated on the island of Tenerife by the IAC in the Spanish Observatorio del Teide. This paper is based on observations made with the MuSCAT3 instrument, developed by the Astrobiology Center and under financial supports by JSPS KAKENHI (JP18H05439) and JST PRESTO (JPMJPR1775), at Faulkes Telescope North on Maui, HI, operated by the Las Cumbres Observatory.
(Some of the) Observations in the paper made use of the NN-EXPLORE Exoplanet and Stellar Speckle Imager (NESSI). NESSI was funded by the NASA Exoplanet Exploration Program and the NASA Ames Research Center. NESSI was built at the Ames Research Center by Steve B. Howell, Nic Scott, Elliott P. Horch, and Emmett Quigley. We acknowledge financial support from the Agencia Estatal de Investigaci\'on of the Ministerio de Ciencia e Innovaci\'on MCIN/AEI/10.13039/501100011033 and the ERDF “A way of making Europe” through project PID2021-125627OB-C32, and from the Centre of Excellence “Severo Ochoa” award to the Instituto de Astrofisica de Canarias.
This work makes use of observations from the LCOGT network. Part of the LCOGT telescope time was granted by NOIRLab through the Mid-Scale Innovations Program (MSIP). MSIP is funded by NSF.
This research has made use of the Exoplanet Follow-up Observation Program (ExoFOP; DOI: 10.26134/ExoFOP5) website, which is operated by the California Institute of Technology, under contract with the National Aeronautics and Space Administration under the Exoplanet Exploration Program.
MG is F.R.S.-FNRS Research Director. KAC and CNW acknowledge support from the TESS mission via subaward s3449 from MIT.
B.V.R. thanks the Heising-Simons Foundation for Support.
This material is based upon work supported by the National Aeronautics and Space Administration under Agreement No.\ 80NSSC21K0593 for the program ``Alien Earths''.
The results reported herein benefited from collaborations and/or information exchange within NASA’s Nexus for Exoplanet System Science (NExSS) research coordination network sponsored by NASA’s Science Mission Directorate.
F.J.P acknowledges financial support from the Severo Ochoa grant CEX2021-001131-S funded by MCIN/AEI/10.13039/501100011033 and through the project PID2022-137241NB-C43. 

\bibliographystyle{aa}
\bibliography{aa.bib}

\begin{appendix}

\section{Discovery TESS light curves for TOI-5720, TOI-6008, and TOI-6086}

\begin{figure*}[!h]
	\centering
	   \includegraphics[scale=0.44]{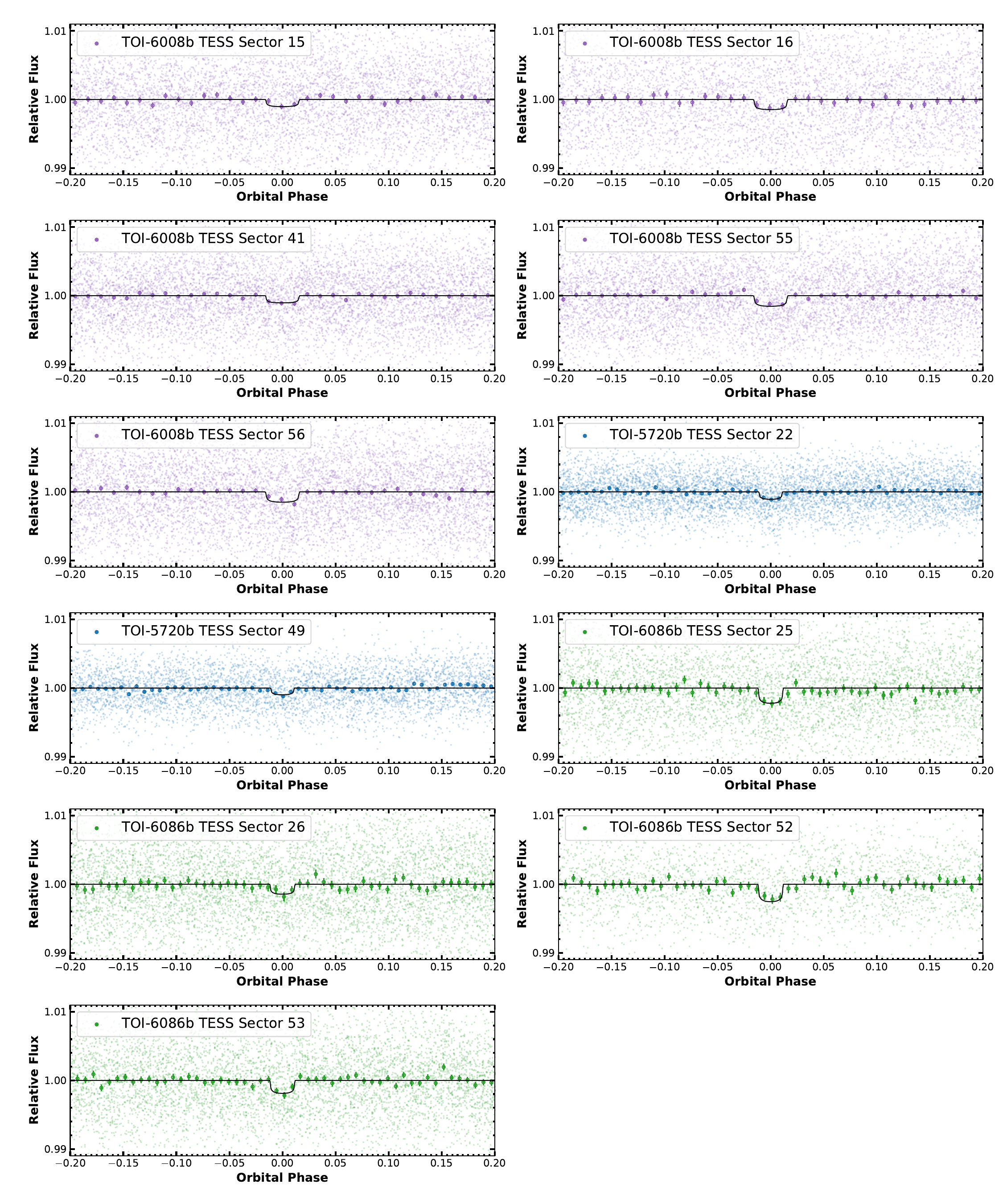}
	\caption{TESS phase-folded photometric data of TOI-5720\,b (blue), TOI-6008\,b (purple), and TOI-6086\,b (green). The best-fitting transit model is superimposed.}
	\label{TESS_LCs}
\end{figure*}

\newpage
\section{High-resolution imaging: WIYN 3.5~m/NESSI}
We observed TOI-5720 on UT 2023 February 5 using the NN-EXPLORE Exoplanet Stellar Speckle Imager (NESSI; \cite{Scott_2018PASP_NESSI}), a speckle imager employed at the WIYN 3.5~m telescope on Kitt Peak.  NESSI was used to obtain simultaneous speckle imaging in two filters with central wavelengths of $\lambda_c = 562$ and 832~nm. The observation consisted of a set of nine 1000-frame 40~ms exposures. The NESSI field of view was set by a $256\times256$ pixel subarray readout, which resulted in a field of $4.6\times4.6$~arcsecond.  However, our speckle measurements were confined to an outer radius of 1.2~arcseconds from the target star.  Speckle imaging of a point source standard star was also taken in conjunction with the observation of the TOI.   The standard observation consisted of a single 1000-frame image set and was used to calibrate the underlying PSF.

\begin{figure}[!ht]
    \centering
    \includegraphics[width=\columnwidth, height=0.4\textheight, keepaspectratio]{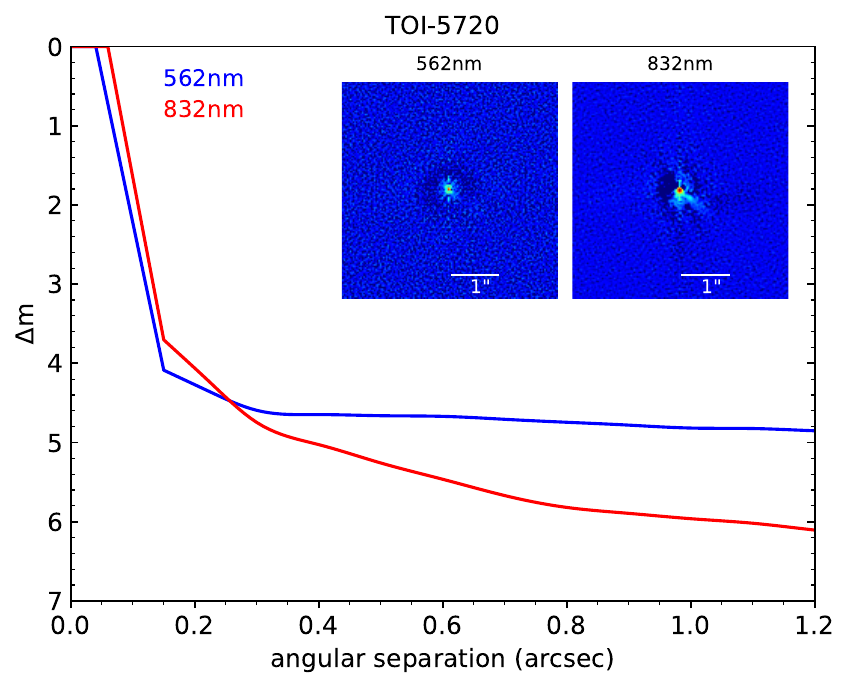}
    \caption{ High-resolution imaging of TOI-5720 obtained with the WIYN-3.5m/NESSI on UT 2023 Feb 5 in two filters with central wavelengths $\lambda_c = 562$ and 832~nm. No close companion sources were detected near the target.}
    \label{toi5720_WIYN}
\end{figure}

These speckle data were reduced using a pipeline described by \citet{Howell_2011AJ_Speckle}.  The pipeline products include a reconstructed image of the field around TOI-5720 in each filter.  The reconstructed image was used to measure a contrast curve setting the detection limits on any point source in close proximity to TOI-5720.  The contrast level was obtained by measuring fluctuations in the noise-like background level as a function of separation from the target star. No companion sources were detected for TOI-5720 in the NESSI data (see \autoref{toi5720_WIYN}). 

\section{High-resolution imaging: Robo AO}

\begin{figure}[!ht]
    \centering
    \includegraphics[width=\columnwidth, height=0.4\textheight, keepaspectratio]{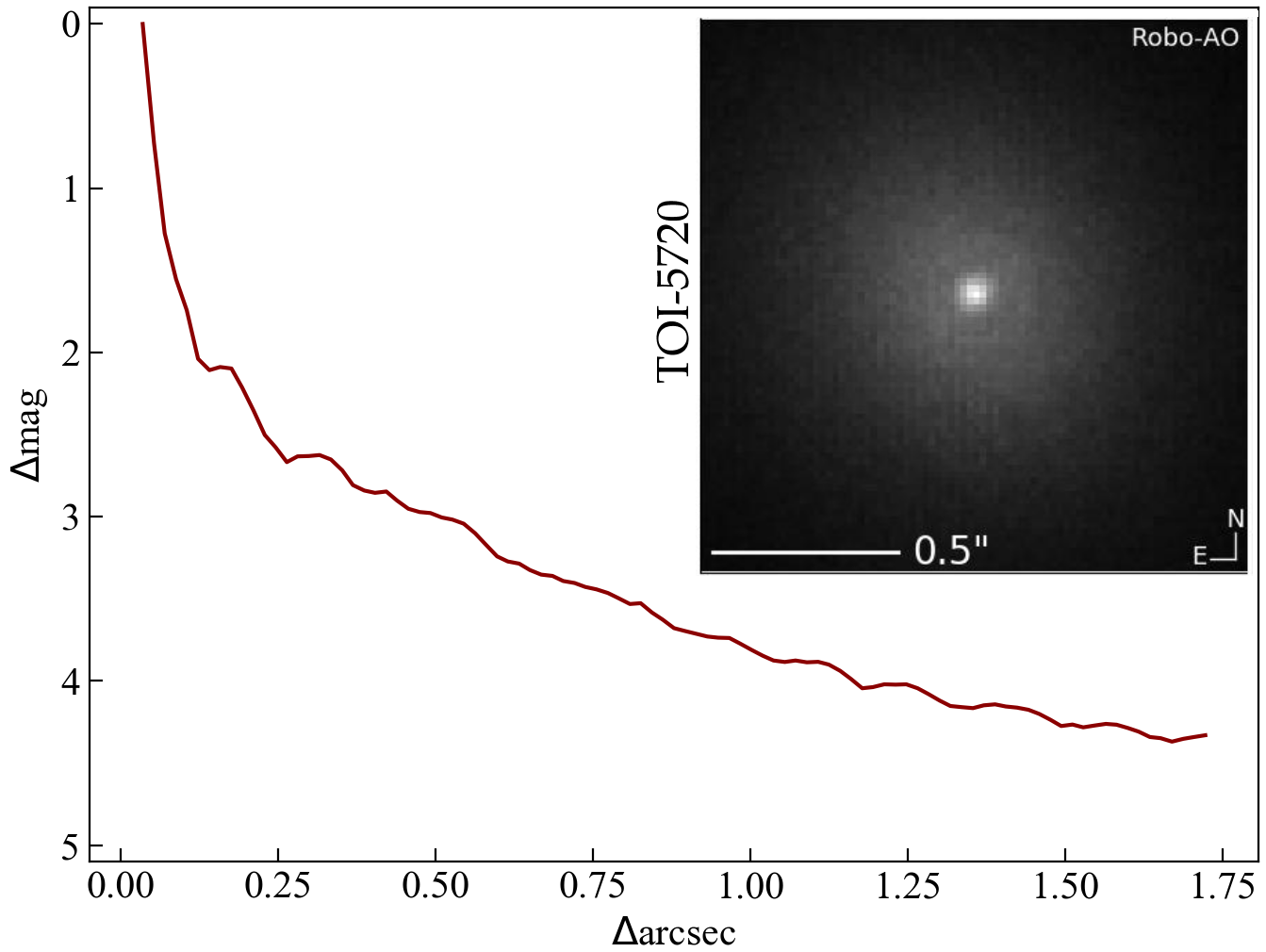}
    \includegraphics[width=\columnwidth, height=0.4\textheight, keepaspectratio]{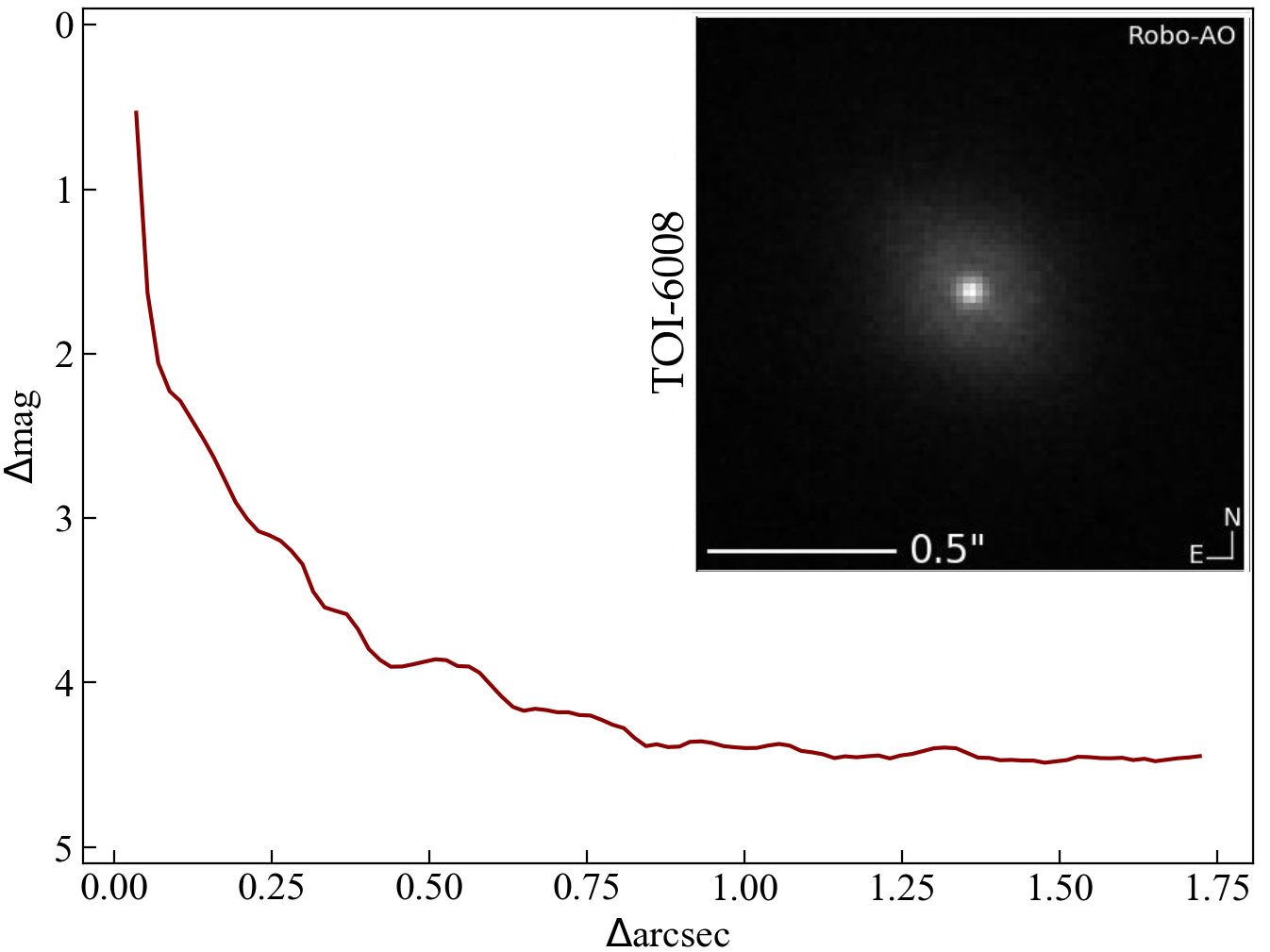}
    \caption{Robo-AO contrast curves and images of TOI-5720 (top) and TOI-6008 (bottom). These observations were taken via an automated laser-guided adaptive optics system on the 2.1m Kitt Peak telescope. No stellar companions were detected in these observations \citep{Lamman_2020AJ}.}
    \label{toi5720_6008_RoboAO}
\end{figure}

We used additional subarcsecond imaging taken with Robo-AO, an autonomous laser-guided adaptive optics (AO) system \citep{Baranec_2014}. The Robo-AO M-dwarf multiplicity catalog contains AO imaging of TOI-5720 and TOI-6008, taken on 2016 February 18 and 2016 June 18, respectively, with the 2.1m Kitt Peak telescope \citep{Lamman_2020AJ}.  The images were taken with an Andor iXon DU-888 camera in the $i'$ filter with an exposure time of 90s. Thee median seeing at the telescope was 1.44", which resulted in an FWHM of about 0.12". Observations were then processed with automatic pipelines, including one pipeline that was optimized for images with either high or low signal-to-noise ratios and a high-contrast imaging pipeline that creates a locally optimized PSF-subtracted image around the target using reference stars observed on the same night \citep{Jensen_2018AJ}. The contrast curves of the observations are shown in \autoref{toi5720_6008_RoboAO};  there is no indication that any additional close stellar companions were detected above these contrast curves.

\section{Posterior probability distribution for the systems TOI-5720, TOI-6008, and TOI-6086.}

\begin{figure*}[h!]
	\centering
	\includegraphics[scale=0.35]{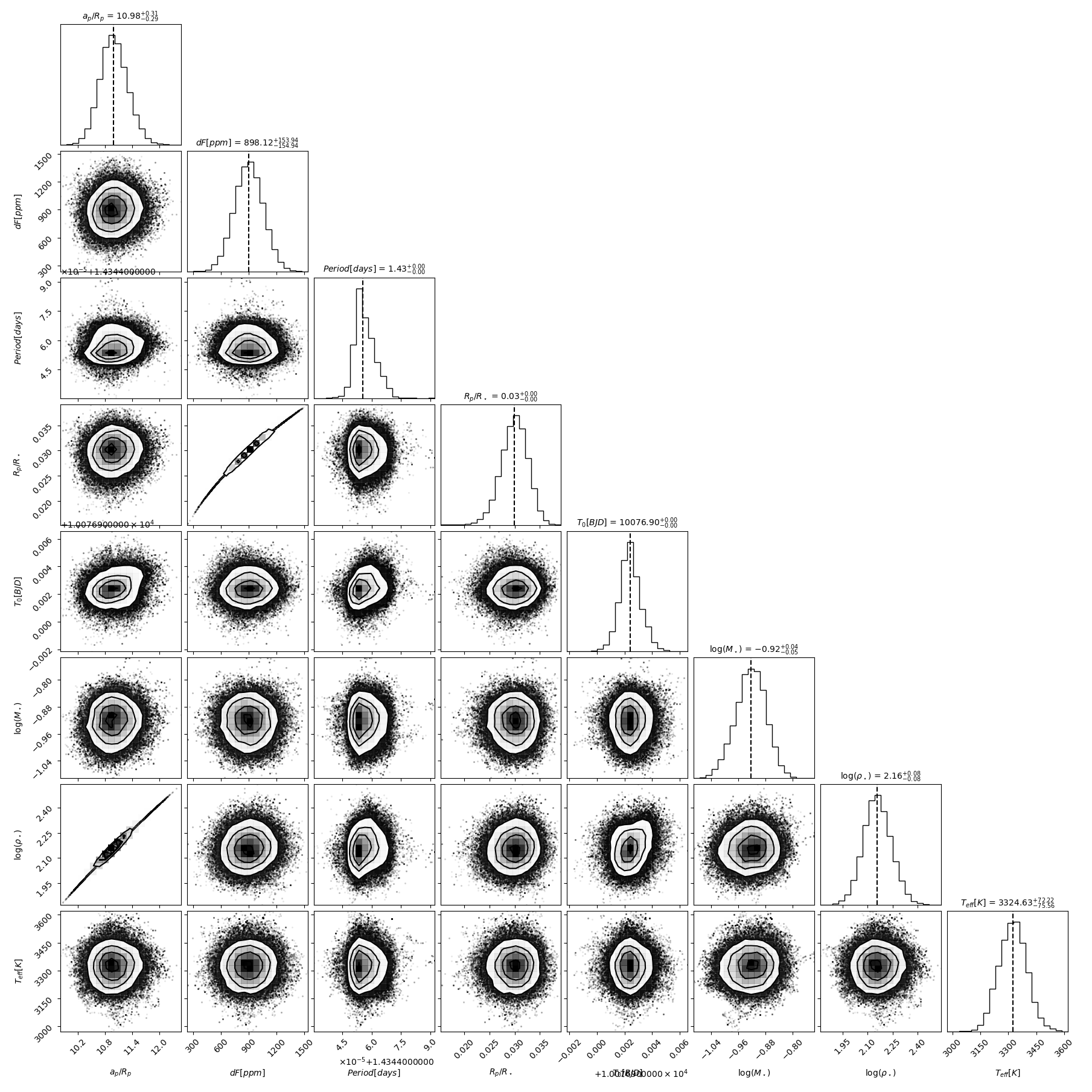}
	\caption{ Posterior probability distribution for the TOI-5720 system parameters derived from our global MCMC analysis.  The median value for each parameter is represented by the vertical dashed lines.}
	\label{corner_TOI5720}
\end{figure*}

\begin{figure*}[h!]
	\centering
	\includegraphics[scale=0.35]{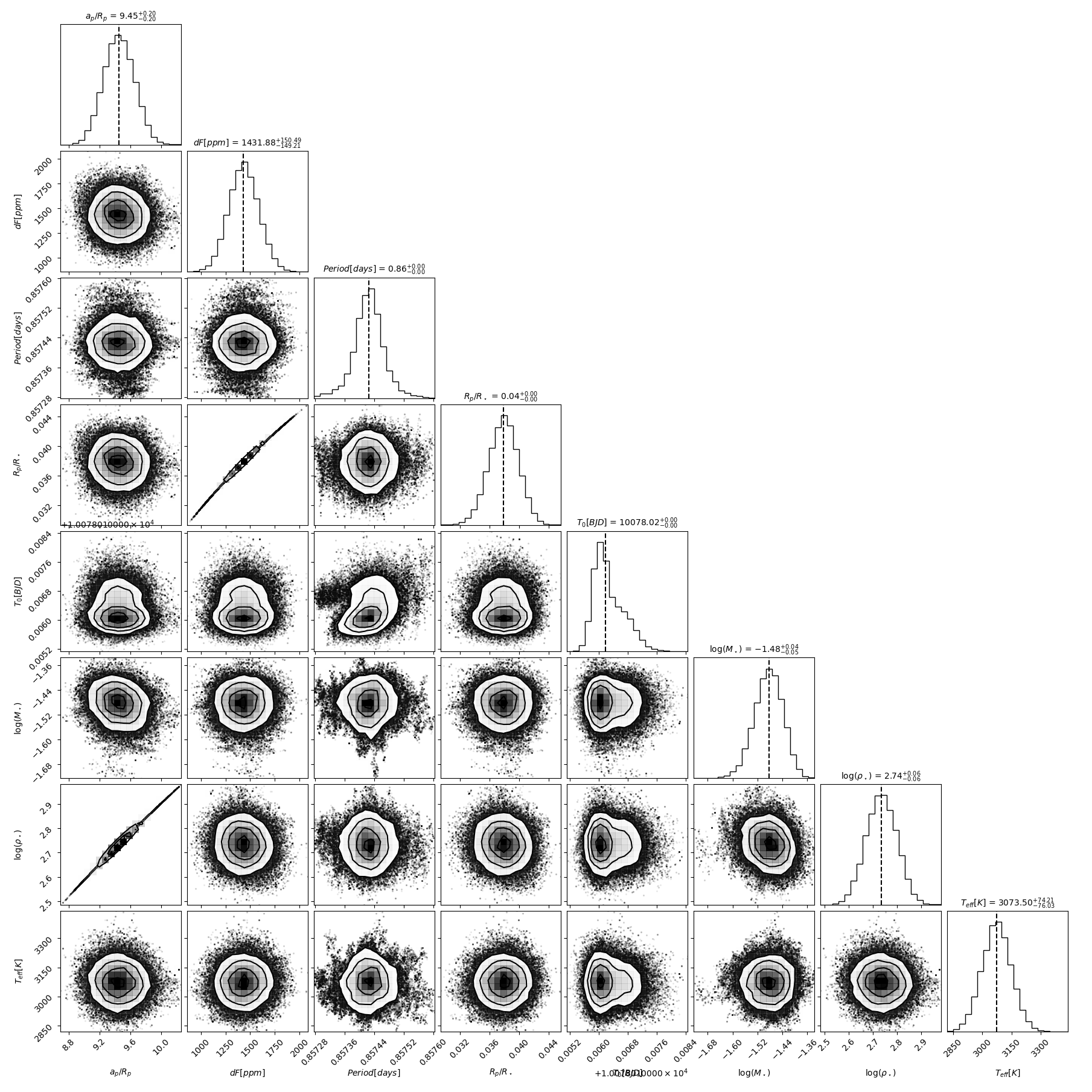}
	\caption{Posterior probability distribution for the TOI-6008 system.}
	\label{corner_TOI6008}
\end{figure*}

\begin{figure*}[h!]
	\centering
	\includegraphics[scale=0.35]{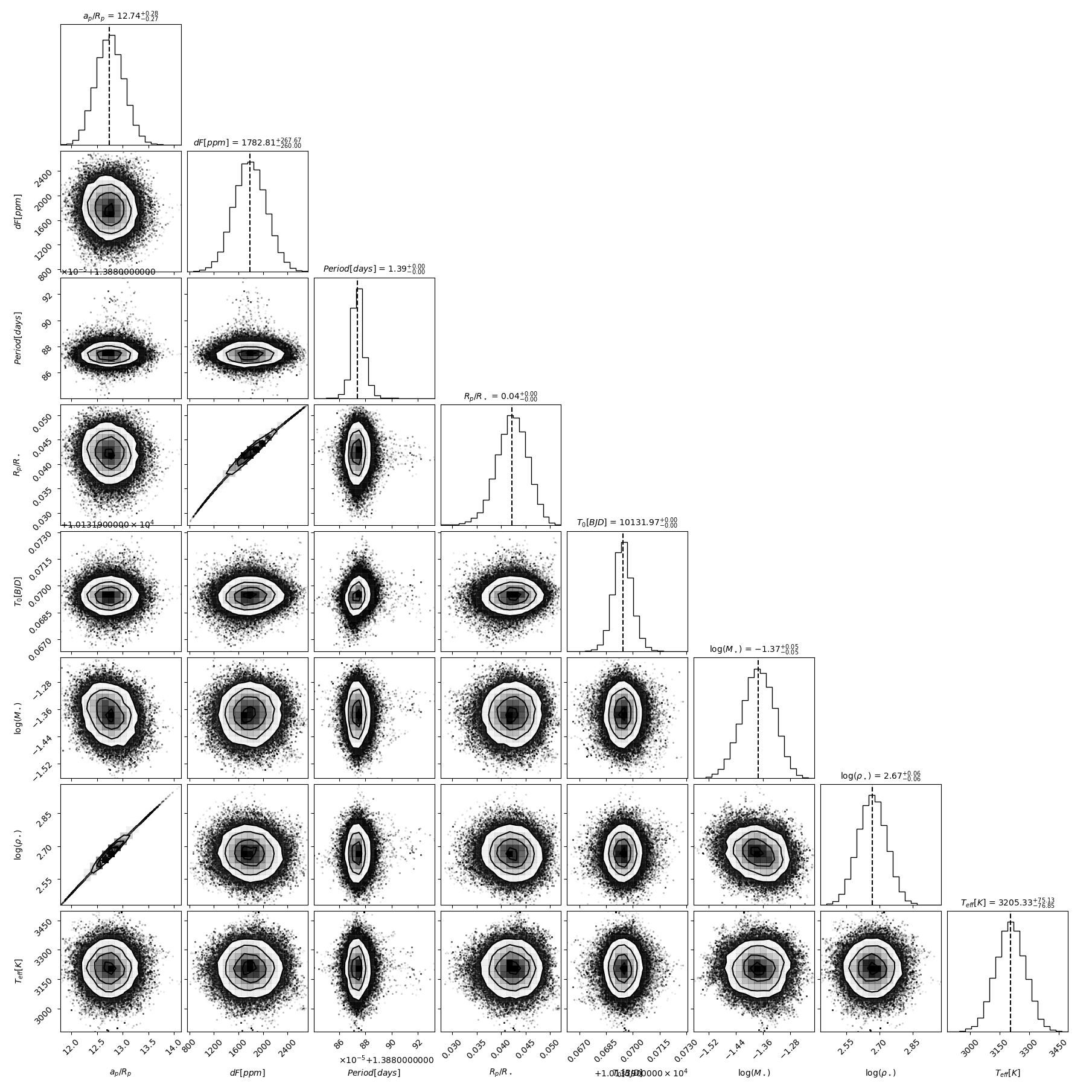}
	\caption{Posterior probability distribution for the TOI-6086 system.}
	\label{corner_TOI6086}
\end{figure*}

\end{appendix}

\end{document}